%% file: main.tex
\documentclass{aastex631}

\usepackage{newtxtext,newtxmath,amsmath,float,xspace,threeparttable,bm,amsmath,booktabs,enumitem}
\usepackage{rotating}

\input{mycommands.tex}

\newcommand{\NAGN}{858}  
\newcommand{\Nobscuredall}{393} 
\newcommand{\Nbroadexcluded}{359} 
\newcommand{\Nbeamexcluded}{106} 
\newcommand{\Nobscureddr}{359} 
\newcommand{\Nbonusvdisp}{125} 
\newcommand{\Nvdisp}{484}  
\newcommand{\Nvdispsec}{158} 

\newcommand{\Nvdispmeas}{956}
\newcommand{\Nvdispspec}{642}
\newcommand{\Nfirst}{281}
\newcommand{\Nlowmass}{6}

\newcommand{\Nspecfitall}{960}
\newcommand{\Ntert}{30}
\newcommand{\Nfitfail}{265}
\newcommand{\Nresfail}{45}
\newcommand{\Nexcludeallmeas}{340}

\accepted{February 16, 2022} 

\submitjournal{ApJS}

\shorttitle{BASS XXVI: DR2 Stellar Velocity Dispersions}
\shortauthors{Koss et al.}

\graphicspath{{./}{figures/}}

\begin{document}

\title{BASS XXVI: DR2 Host Galaxy Stellar Velocity Dispersions}

\correspondingauthor{Michael Koss}
\email{mike.koss@eurekasci.com}

\author[0000-0002-7998-9581]{Michael J. Koss}
\affiliation{Eureka Scientific, 2452 Delmer Street Suite 100, Oakland, CA 94602-3017, USA}
\affiliation{Space Science Institute, 4750 Walnut Street, Suite 205, Boulder, CO 80301, USA}

\author[0000-0002-3683-7297]{Benny Trakhtenbrot}
\affiliation{School of Physics and Astronomy, Tel Aviv University, Tel Aviv 69978, Israel}

\author[0000-0001-5231-2645]{Claudio Ricci}
\affiliation{N\'ucleo de Astronom\'ia de la Facultad de Ingenier\'ia, Universidad Diego Portales, Av. Ej\'ercito Libertador 441, Santiago 22, Chile}
\affiliation{Kavli Institute for Astronomy and Astrophysics, Peking University, Beijing 100871, People's Republic of China}

\author[0000-0002-5037-951X]{Kyuseok Oh}
\affiliation{Korea Astronomy \& Space Science institute, 776, Daedeokdae-ro, Yuseong-gu, Daejeon 34055, Republic of Korea}
\affiliation{Department of Astronomy, Kyoto University, Kitashirakawa-Oiwake-cho, Sakyo-ku, Kyoto 606-8502, Japan}
\affiliation{JSPS Fellow} 

\author[0000-0002-8686-8737]{Franz E. Bauer}
\affiliation{Instituto de Astrof\'{\i}sica  and Centro de Astroingenier{\'{\i}}a, Facultad de F\'{i}sica, Pontificia Universidad Cat\'{o}lica de Chile, Casilla 306, Santiago 22, Chile}
\affiliation{Millennium Institute of Astrophysics (MAS), Nuncio Monse{\~{n}}or S{\'{o}}tero Sanz 100, Providencia, Santiago, Chile}
\affiliation{Space Science Institute, 4750 Walnut Street, Suite 205, Boulder, CO 80301, USA}

\author[0000-0003-2686-9241]{Daniel Stern}
\affiliation{Jet Propulsion Laboratory, California Institute of Technology, 4800 Oak Grove Drive, MS 169-224, Pasadena, CA 91109, USA}

\author[0000-0002-9144-2255]{Turgay Caglar}
\affiliation{Leiden Observatory, PO Box 9513, 2300 RA, Leiden, the Netherlands}

\author[0000-0002-8760-6157]{Jakob S. den Brok}
\affiliation{Institute for Particle Physics and Astrophysics, ETH Z{\"u}rich, Wolfgang-Pauli-Strasse 27, CH-8093 Z{\"u}rich, Switzerland}
\affiliation{Argelander Institute for Astronomy, Auf dem H{\"u}gel 71, D-53231, Bonn, Germany}

\author[0000-0002-7962-5446]{Richard Mushotzky}
\affiliation{Department of Astronomy, University of Maryland, College Park, MD 20742, USA}
\affiliation{Joint Space-Science Institute, University of Maryland, College Park, MD 20742, USA}

\author[0000-0001-5742-5980]{Federica Ricci}
\affil{Dipartimento di Fisica e Astronomia, Università di Bologna, via Gobetti 93/2, I-40129 Bologna, Italy}
\affil{INAF Osservatorio Astronomico di Bologna, via Gobetti 93/3, I-40129 Bologna, Italy}

\author[0000-0001-8450-7463]{Julian E. Mej\'ia-Restrepo}
\affiliation{European Southern Observatory, Casilla 19001, Santiago 19, Chile}

\author[0000-0003-3336-5498]{Isabella Lamperti}
\affiliation{Centro de Astrobiología (CAB), CSIC–INTA, Cra. de Ajalvir Km. 4, E-28850 Torrejón de Ardoz, Madrid, Spain}

\author[0000-0001-7568-6412]{Ezequiel Treister}
\affiliation{Instituto de Astrof{\'i}sica, Facultad de F{\'i}sica, Pontificia Universidad Cat{\'o}lica de Chile, Casilla 306, Santiago 22, Chile}

\author[0000-0001-5481-8607]{Rudolf E. B\"{a}r}
\affiliation{Institute for Particle Physics and Astrophysics, ETH Z{\"u}rich, Wolfgang-Pauli-Strasse 27, CH-8093 Z{\"u}rich, Switzerland}

\author{Fiona Harrison}
\affiliation{Cahill Center for Astronomy and Astrophysics, California Institute of Technology, Pasadena, CA 91125, USA}

\author[0000-0003-2284-8603]{Meredith C. Powell}
\affiliation{Institute of Particle Astrophysics and Cosmology, Stanford University, 452 Lomita Mall, Stanford, CA 94305, USA}

\author[0000-0003-3474-1125]{George C. Privon}
\affiliation{National Radio Astronomy Observatory, 520 Edgemont Road, Charlottesville, VA 22903, USA}
\affiliation{Department of Astronomy, University of Florida, P.O. Box 112055, Gainesville, FL 32611, USA}

\author[0000-0002-1321-1320]{Rog\'erio Riffel}
\affiliation{Departamento de Astronomia, Instituto de F\'\i sica, Universidade Federal do Rio Grande do Sul, CP 15051, 91501-970, Porto Alegre, RS, Brazil}

\author[0000-0003-0006-8681]{Alejandra F. Rojas}
\affiliation{Centro de Astronom\'{i}a (CITEVA), Universidad de Antofagasta, Avenida Angamos 601, Antofagasta, Chile} 

\author[0000-0001-5464-0888]{Kevin Schawinski}
\affiliation{Modulos AG, Technoparkstrasse 1, CH-8005 Zurich, Switzerland}

\author[0000-0002-0745-9792]{C. Megan Urry}
\affiliation{Yale Center for Astronomy \& Astrophysics, Physics Department, PO Box 208120, New Haven, CT 06520-8120, USA}


\begin{abstract}
We present new central stellar velocity dispersions for \Nvdisp\ Sy\,1.9 and Sy\,2 from the second data release of the Swift/BAT AGN Spectroscopic Survey (BASS DR2). This constitutes the largest study of velocity dispersion measurements in X-ray-selected, obscured AGN with \Nvdispmeas\ independent measurements of the \Cahk\ and \mgb\ region (3880--5550\,\AA) and the calcium triplet region (8350--8730\,\AA) from \Nvdispspec\ spectra mainly from VLT/X-Shooter or Palomar/DoubleSpec. Our sample spans velocity dispersions of 40--360\,\kmpssh, corresponding to 4--5 orders of magnitude in black hole mass ($\mbh=10^{5.5-9.6}$\,\Msun), bolometric luminosity ($\lbol\sim10^{42-46}$\,\ergps), and Eddington ratio ($\lledd\sim10^{-5}$--2).  For \Nfirst\ AGN, our data and analysis provide the first published central velocity dispersions, including \Nlowmass\ AGN with low-mass black holes (\mbh=$10^{5.5-6.5}$\,\Msun), discovered thanks to high spectral resolution observations ($\sigma_{\rm inst}\sim25$\,\kmpssh).   The survey represents a significant advance with a nearly complete census of velocity dispersions of hard X-ray-selected obscured AGN with measurements for 99\% of nearby AGN ($z{<}0.1$) outside the Galactic plane ($\lvert b\rvert{{>}}10^{\circ}$).     The BASS AGN have much higher velocity dispersions than the more numerous optically selected narrow-line AGN (i.e., $\sim$150 vs. $\sim$100\,\kmpssh), but are not biased towards the highest velocity dispersions of massive ellipticals (i.e., $>$250\,\kmpssh).  Despite sufficient spectral resolution to resolve the velocity dispersions associated with the bulges of small black holes ($\sim10^{4-5}$\,\Msun), we do not find a significant population of super-Eddington AGN.    Using estimates of the black hole sphere of influence from velocity dispersion, direct stellar and gas black hole mass measurements could be obtained with existing facilities for more than $\sim$100 BASS AGN.  

\end{abstract}


\section{Introduction} \label{sec:intro}

Over the last two decades, dedicated surveys have established several scaling relations between supermassive black hole (SMBH) mass (\mbh) and host galaxy properties, including the bulge luminosity and mass ($M_{\rm bulge}$), stellar velocity dispersion inferred from spectral absorption lines (\sigs), concentration index, and Sersic index \citep[see, e.g., review by][and references therein]{Kormendy:2013:511}.  
These scaling relations strongly support a picture in which SMBHs and their host galaxies ``coevolve'', in the sense that the growth histories of the two components are physically interlinked, perhaps through some form of SMBH-related feedback mechanism \citep[see, e.g., the review by][and references therein]{Fabian:2012:455a}.

Observationally, the tightest and perhaps most robust link between SMBHs and their hosts is the \mbh-\sigs\ scaling relation \citep[e.g., $<0.5$ dex;][]{Marsden:2020:61a}.  Calibrating this relation has relied on a rather modest number \citep[e.g., $N$=145,][]{Sahu:2019:155,Sahu:2019:10} of direct \mbh\ measurements with varying precision from stellar or gas dynamics,
kinematics of megamasers, proper motion, or recent direct imaging techniques.   The spatially resolved measurements of stellar and/or gas dynamics comprise the majority of these measurements and typically require very high spatial resolutions that are possible only for the nearest galaxies \citep[e.g.,][]{McConnell:2013:184}. Encouragingly, the high resolution offered by the Atacama Large Millimeter/submillimeter Array (ALMA) is increasingly being used to this end for nearby galaxies, tracking the dynamics of the circumnuclear gas dynamics as it is strongly affected by the SMBH gravitational field \citep[e.g.,][]{Cohn:2021:77}.  
Other direct measurements such as using H$_2$O megamasers are only possible for a small number of galaxies with favorable alignments \citep[][]{Greene:2016:L32}.  

Using these direct measurements, various studies have suggested systematic environmental differences or possible selection biases in the \mbh-\sigs\ relation in elliptical (typically higher mass) and spiral (typically lower mass) morphological galaxy types, barred galaxies, and pseudobulges \citep[e.g.,][]{Graham:2008:167,Greene:2016:L32, Shankar:2017:4029}.  Other studies have suggested variations of \sigs\ due to spectral regions probing different stellar populations \citep[e.g.,][]{Riffel:2015:2823}.
Notwithstanding these limitations and possible biases, the $\mbh-\sigs$ relation has been found to extend unbroken down to $\mbh\sim10^5$\,\Msun\ \cite[although with increasing scatter; e.g.,][]{Greene:2020:257} and seems to be the most fundamental relation between SMBHs and their host galaxies  \citep[e.g.,][]{vandenBosch:2015:10,Shankar:2019:1278}.
This, in turn, makes it useful for inferring \mbh\ in much larger samples of galaxies. 

Active galactic nuclei (AGN) provide the best tracer of SMBH accretion throughout cosmic time.  Although technically challenging in the presence of (optical) AGN contamination \cite[e.g.,][]{Greene:2006:117}, velocity dispersion measurements have been obtained for various AGN populations, including unobscured broad-line AGN (``type 1'' AGN, hereafter ``Sy 1'') to calibrate so-called ``virial'' \mbh\ estimators \citep[e.g.,][]{Grier:2013:90a,Woo:2013:49}; AGN in dwarf galaxies, which probe the low-mass end of the ``coevolutionary'' picture \citep[e.g.,][]{Martin-Navarro:2018:L20,Baldassare:2020:L3}; and obscured narrow-line AGN \citep[``type 2'', hereafter ``Sy\,2'', e.g.,][]{Garcia-Rissmann:2005:765}.
Among the many ways to survey the AGN population, hard X-rays (${{>}}2$\,keV) provide the most complete census for distant, strongly accreting AGN \citep[see, e.g.,][]{Brandt:2005:827}, as a large fraction--indeed the majority--of the AGN population is obscured \citep[see review by, e.g.,][]{Hickox:2018:625}. 
ultrahard X-ray emission (${>}10$\,keV) provides an even more complete tracer of the radiation from obscured AGN ({i.e. an equivalent neutral hydrogen absorbing column, $\NH {>} 10^{22}\,\cmii$), probing into the Compton-thick (CT) regime (i.e., $\NH {>} 10^{24}\,\cmii$; see, e.g., \citealp{Ricci:2015:L13,Koss:2016:85}).  
Ultimately, studying the stellar velocity dispersion of a large sample of nearby, ultrahard X-ray-selected AGN is crucial to get good constraints on \sigs\ in obscured AGN and thus the SMBH mass distribution among this population and serves as a critical nearby benchmark for high-redshift AGN, where host galaxy velocity dispersions are difficult to measure.

Despite the importance of central (e.g. $\lessapprox$\,kpc) velocity dispersions for inferring \mbh\ in obscured AGN, relatively few large studies (i.e., ${>}100$ systems) have been performed, mostly within the context of large-scale surveys of galaxies such as the Sloan Digital Sky Survey \citep[SDSS; e.g.,][]{Greene:2005:721,Thomas2013}.  Compared to obscured (narrow-line) AGN selected via strong line ratio diagnostics from the SDSS \citep[e.g., \OIII/\Hbeta\ vs.~\NII/\Halpha,][]{Baldwin:1981:5,Veilleux:1987:295,Kewley:2001:37}, ultrahard X-ray-selected AGN are sometimes missed due to being in significantly more dusty host galaxies \citep[e.g.,][]{Koss:2017:74} with higher star formation rates \citep[e.g.,][]{Koss:2013:L26,Koss:2021:29}.  Thus, surveying the velocity dispersions of ultrahard X-ray-selected AGN host galaxies offers an important complement to optical surveys.   

The goal of the BAT AGN Spectroscopic Survey{\footnote{\href{https://www.bass-survey.com/}{https://www.bass-survey.com/}}} (BASS) is to generate the largest available optical spectroscopic dataset for the sample of Swift/BAT ultrahard X-ray (14--195\,keV) detected AGN.  As part of this effort, the first data release (DR1) of BASS \cite[][]{Koss:2017:74} used mostly archival optical spectra for 641 AGN from the 70 month BAT catalog \citep{Baumgartner:2013:19} to derive central velocity dispersions for 202 AGN.  The DR1 found that BAT AGN tend to have larger central velocity dispersions than SDSS-selected, narrow-line Sy\,2 AGN.   Notably, almost all of the DR1 observations were obtained with smaller (1.5--2.5m) telescopes spread across various surveys and other past studies, each using bespoke reduction routines, leading to substantial inhomogeneity in quality and parameter constraints. Importantly, many BASS DR1 spectra also had spectral resolutions that were too low ($R{<}1000$) to robustly measure the stellar velocity dispersion in low-mass AGN ($\mbh\lesssim10^7$\,\Msun).


In this paper, part of the BASS DR2 Special Issue, we greatly improve upon the DR1 results by presenting and analyzing targeted observations of central velocity dispersions for a nearly complete sample of AGN ($\geq95\%$) drawn from the 70 month BAT catalog with higher spectral resolution and higher signal-to-noise ratios (S/Ns).  
We focus on a complete sample of Sy\,1.9 and Sy\,2, where these measurements can be made in the absence of significant AGN contamination based on a large number of new, high-quality spectra.  A more detailed study of the nearest ($z{<}0.01$) 19 local luminous AGN within the 70 month Swift/BAT catalog is provided in \citet[][]{Caglar:2020:A114}. 
Another complementary DR2 study \citep{Caglar_DR2_Msigma} will focus on the velocity dispersions of the broad-line Sy 1 AGN (with broad \Hbeta, e.g. FWHM, $>1000$\,\kmpssh) in the same parent sample.  

An overview of the BASS DR2 spectroscopic sample used in this work is provided in \citet{Koss_DR2_overview}, while full details of the \NAGN\ AGN, including revised counterparts, classifications, observations, and reductions are in \citet{Koss_DR2_catalog}. Broad emission line measurements for the BASS DR2 sample are presented in \citet{Mejia_Broadlines}, while narrow-line measurements are presented in \citet{Oh_DR2_NLR}.  
\citet{Ananna_DR2_XLF_BHMF_ERDF} uses the highly complete set of BASS DR2 measurements to derive the intrinsic X-ray luminosity function (XLF), black hole mass function (BHMF) and Eddington ratio distribution function (ERDF), for both obscured and unobscured low-redshift AGN, relying on the \sigs\ measurements presented here (for narrow-line systems).  
Completing the DR2 Special Issue, the details of the DR2 near-infrared (NIR) spectroscopy are provided in \citet{denBrok_DR2_NIR}, with an investigation of NIR coronal lines. The NIR emission from broad-line regions and associated virial SMBH mass estimates are studied in \citet{Ricci_DR2_NIR_Mbh}. Finally, \citet{Pfeifle:2022:3} investigated the relationship between mid-IR colors and X-ray column density.


Throughout this work, we adopt $\Omega_{\rm M} {=} 0.3$, $\Omega_\Lambda {=} 0.7$, and $H_0 {=} 70\,\kms\,{\rm Mpc}^{-1}$.  To determine the extinction due to Milky Way foreground dust, we use the maps of \citet{Schlegel:1998:525} and the extinction law derived by \citet{Cardelli:1989:245}. For consistency with the BASS DR1 \citep{Koss:2017:74}, we define Sy 1 as AGN with broad \hbeta\ line emission, Sy\,1.9 as having narrow \hbeta\ and broad \halpha, and Sy\,2 AGN as having both narrow \hbeta\ and narrow \halpha\ (this latter category includes small numbers of LINERs and AGN in star formation-dominated galaxies).  


\section{AGN Spectroscopic Sample and Data} \label{sec:agnsample}
The goal of the BASS DR2 is to provide a complete sample of SMBH mass estimates and multiwavelength ancillary measurements using targeted observations for all AGN in the 70 month Swift/BAT survey.  The optical spectroscopy component focuses on either broad emission lines (mostly Balmer lines) or (host galaxy) stellar velocity dispersion measurements to obtain SMBH estimates while also covering the broadest possible spectral range (i.e., within the accessible visual range, 3200--10000\,\AA) for emission line measurements for the entire catalog of \NAGN\ AGN.  We discuss the obscured AGN sample along with some aspects of the parent optical spectra sample below and refer the reader to the DR2 overview  \citep{Koss_DR2_overview} and the detailed catalog paper \citep{Koss_DR2_catalog} for more details.

\subsection{AGN sample}
The AGN sample starts with the \NAGN\ AGN listed in the 70 month BAT catalog that comprise the BASS DR2 \citep{Koss_DR2_catalog}. As stellar velocity dispersions are difficult (or impossible) to measure when the spectrum is dominated by AGN continuum emission, we excluded from our sample \Nbeamexcluded\ beamed and/or lensed AGN and \Nbroadexcluded\ AGN with broad \hbeta, based on the classifications from \cite[][]{Mejia_Broadlines} as these have broad-line based mass estimates.  A separate investigation of velocity dispersions in a subset of BASS DR2 Sy1 systems was carried out by \citet{Caglar_DR2_Msigma}, where special care is taken to tackle the AGN contamination.   This leaves \Nobscuredall\ Sy\,1.9 and Sy\,2 AGN from the parent DR2 sample of \NAGN\ AGN (i.e., 46\%, \Nobscuredall/\NAGN).  For potential dual AGN in the sample \citep[e.g.,][]{Koss:2011:L42,Koss:2012:L22,Koss:2015:149,Koss:2016:L4}, each AGN is only included in the sample if X-ray detected and bright enough to be detected individually by Swift/BAT. From this parent sample of \Nobscuredall\ obscured DR2 AGN, we were able to successfully measure \sigs\ in \Nobscureddr\ AGN.

The velocity measurements described below were also carried out successfully on \Nbonusvdisp\ additional obscured AGN from the deeper 105 month  Swift/BAT all-sky survey \cite[e.g.,][]{Oh:2018:4}, which were collected as part of the ongoing BASS efforts.  As counterpart identification is still ongoing for the 105 month sample, we stress that this sample is neither complete nor final and does not represent a flux- or volume-complete subset of the deeper BAT data. Thus, we omit this additional bonus sample when discussing completeness measurements. So the final obscured AGN sample totals \Nvdisp\ from \Nobscureddr\ DR2 and \Nbonusvdisp\ bonus 105 month AGN.

\subsection{Sample of Spectra}
Our sample starts from 960 BASS spectra of the Sy\,1.9 and Sy\,2 AGN  from the 70 month AGN catalog \citep[for review, see,][]{Koss_DR2_catalog} and additional obscured AGN from the deeper 105 month.   The majority of the spectra (67\%, 651/960) are newly obtained from the Very Large Telescope (VLT) using the X-Shooter instrument,  a multiwavelength  (3000--25000\,\AA) echelle spectrograph  \citep{Vernet:2011:A105} or from the DoubleSpec (DBSP) instrument, mounted on the 200 inch telescope at Palomar Observatory. The VLT/X-Shooter consists of three spectroscopic arms covering the ultraviolet blue (UVB; 3000--5595\,\AA), visual (VIS; 5595--10240\,\AA), and NIR (10240--24800\,\AA).  The spectra also include significant new datasets from the Southern Astrophysical Research telescope (SOAR)  using the Goodman High Throughput Spectrograph \citep{2004SPIE.5492..331C}.  Finally, legacy SDSS spectra are used when available, as well as smaller contributions from other telescopes.     

Importantly for the construction of the sample and dataset used in the present work, we note that repeated observations of these AGN were carried out if either the S/N over the spectral features relevant for the detailed measurements and/or the spectral resolution was too low to robustly measure the stellar absorption features (i.e. $\Delta\sigs>20$\,\kmpssh). Specifically, high spectral resolution observations (i.e., $\sigma_{\mathrm{inst}}\approx25$\,\kmpssh), even if over a limited spectral range, were pursued primarily for obscured AGN (Sy\,1.9 and Sy\,2), for which stellar velocity measurements provide the only way to estimate BH masses.  Additional repeated observations of spectra were also sometimes attempted to improve the accuracy of \sigs\ for lower-quality measurements ($10$\,\kmps$<\Delta\sigs<20$\,\kmpssh).

Outside of the SDSS sample, which was in both the DR1 and DR2, almost all of the DR1 spectra were reobserved with larger telescopes at much higher quality and spectral resolution, so we do not include them in this analysis.   We do, however, include a sample of 21 high-quality spectra from the DR1 from Gemini. For sample completeness, we also include a single DR1 spectrum of M81, a very bright and nearby galaxy, which was not part of the DR2 release due to instrumental issues.  After including these samples, only 35 spectra remain with acceptable velocity dispersions ($\Delta\sigs<20$\,\kmpssh) from the DR1, all of which were reobserved in the DR2, so we do not remeasure the remaining DR1 spectra in this release.

%





Throughout the text, due to the frequent duplications of AGN spectra from different telescopes and from the same telescope using a higher-resolution setup, we use the nomenclature ``best'' to refer to the spectra with the lowest velocity dispersion error in kilometers per second in either the 3880--5550\,\AA\ region or the 8350--8730\,\AA\ fitting region.  In other words, for an individual AGN, we compare the $\Delta\sigs$ of all of the measurements from all the spectra, and the best spectrum is the one that has the single lowest measurement error in a single region.   We also include unique ``secondary'' spectra of the same AGN observed with a different telescope or instrumental setting, which have the second-lowest errors of any spectra, to enable a robust comparison between measurements to better understand issues like apertures, stellar libraries, instrumental issues, different fitting regions, etc, but we do not include these spectra for further scientific analysis in the paper.  We exclude additional ``tertiary'' or even ``quaternary'' BASS spectra of 30 AGN of even worse quality to avoid biasing the results.

From the sample of \Nspecfitall\ BASS spectra (and 22 DR1 spectra) of the \Nvdisp\ Sy\,1.9 and Sy\,2 AGN, we excluded \Nexcludeallmeas\ due to low-quality measurements or duplications. In total, \Nfitfail\ were excluded due to large errors in measurement ($\Delta\sigs>20$\,\kmpssh), \Nresfail\ due to velocity dispersions close to or below the instrumental resolution, and \Ntert\ due to tertiary or quaternary quality measurements.  This leaves \Nvdisp\ best-fit spectra (and \Nvdispsec\ secondary spectra).

A full summary of the best and secondary spectra and instrumental setups used specifically for velocity dispersion measurements is provided in \autoref{tab:bass_dr2}. A summary of the redshift distribution, slit size in kiloparsecs, and instrumental resolutions is provided in \autoref{fig:spectra_sum}. Further details of instrumental settings, reductions, and observing conditions are provided in \citet{Koss_DR2_catalog}.  In brief, the data reduction and analysis of DR2 spectra used here maintain the uniform approach described in the initial DR1 paper \citep{Koss:2017:74}.  All new spectra are processed using the standard tasks for cosmic-ray removal, one-dimensional spectral extraction, wavelength, and flux calibrations in either \iraf or the ESO/{\tt Reflex} environment for the VLT instruments. The spectra are flux calibrated using standard stars, which were typically observed two to three times per night.  The spectra are corrected for Galactic reddening. Finally, a telluric absorption correction is applied to the spectra with the software \molecfit\ (i.e., for the 8350--8730\,\AA\ region).   The instrumental resolution and line-spread function for each spectral setup were measured using the best estimate from telluric features (i.e., for the 8350--8730\,\AA\ region) when possible or Galactic stars observed as close in time as possible (i.e., for 3880--5550\,\AA\ region, where strong telluric features are not present).


\begin{deluxetable*}{lcrclccc}
\tablewidth{0pt}
\tablecaption{Summary of BASS Spectra Used \label{tab:bass_dr2}}
\tablehead{
\colhead{Telescope}& \colhead{Instrument}& \colhead{$N_{\mathrm{best}}$}& \colhead{$N_{\mathrm{sec.}}$}&\colhead{$\lambda_{\rm range}$ (\AA)}& \colhead{Slit width (\arcsec)}& \colhead{$R_{5000}$}& \colhead{$R_{8500}$}
\\
\colhead{(1)}& \colhead{(2)}& \colhead{(3)}& \colhead{(4)}&\colhead{(5)}& \colhead{(6)}& \colhead{(7)}&
\colhead{(8)}
}
\startdata
VLT&X-Shooter&163&21&2990-10200&1.5&3850&6000\\
Palomar\tablenotemark{a}&DBSP&101&61&3150-10500&1.5&1220&1730\\
&&51&11&3970-5499/8050-9600&2&2170&4720\\
APO&SDSS&91&2&3830-9180&3&1760&2490\\
SOAR&Goodman&53&27&7900-9070&1.2&\nodata&4720\\
&&1&10&4560-8690&1.2&890&\nodata\\
&&1&4&5280-7900&1.2&1450&\nodata\\
Keck&LRIS/DEIMOS&6&3&3200-10280&1&1280&1810\\
Magellan&MagE&6&1&3300-10010&1&3850\tablenotemark{b}&\nodata\\
VLT&MUSE&6&0&4800-9300&2\tablenotemark{c}&1850&3150\\
VLT&FORS2&1&0&3400-6100&2&830&\nodata\\
\hline
Gemini\tablenotemark{d}&GMOS&3&18&4000-7000&1&1050&\nodata\\
Perkins\tablenotemark{e}&Deveny&1&0&3900-7500&2&920&\nodata\\
\hline \hline
Total&&484&158&&&&\\
\enddata
\tablecomments{The columns are as follows. (1) Telescope. (2) Instrument used. (3) and (4) Whether the spectra was included in the best measurement or a secondary measurement. (5)--(8) The wavelength range, slit width, and resolving power at 5000\,\AA and 8500\,\AA, respectively.  These represent typical values for this setup and may differ by small factors for a small number of spectra. In some cases, larger or smaller slit widths (e.g., 1\farcs5 vs.~2\arcsec) were used, resulting in different resolutions.  Here, $R$ is given at 5000 and 8500\,\AA\ depending on the spectral range.   See \citet{Koss_DR2_catalog} for a detailed list of instrument setups.}
\tablenotetext{a}{There were two spectral setups used for Palomar, one at lower resolution for general observations of AGN and another in higher-resolution mode for velocity dispersions.}
\tablenotetext{b}{The CaT region was not measured in Magellan/MagE spectra due to strong instrumental fringing.}
\tablenotetext{c}{A 2\arcsec\ diameter central aperture was extracted from the VLT/MUSE observation, except for NGC 6240, a close dual AGN where a 1\arcsec\ aperture was used for each AGN.}
\tablenotetext{d}{These Gemini spectra were used from the BASS DR1 because of their typically high S/N but are not part of the DR2.}
\tablenotetext{e}{This BASS DR1 spectrum was used because the bright galaxy M81 was not part of the DR2 due to instrumental issues with sky subtraction.}
\end{deluxetable*}

\begin{figure*} 
\centering
\includegraphics[width=8.2cm]{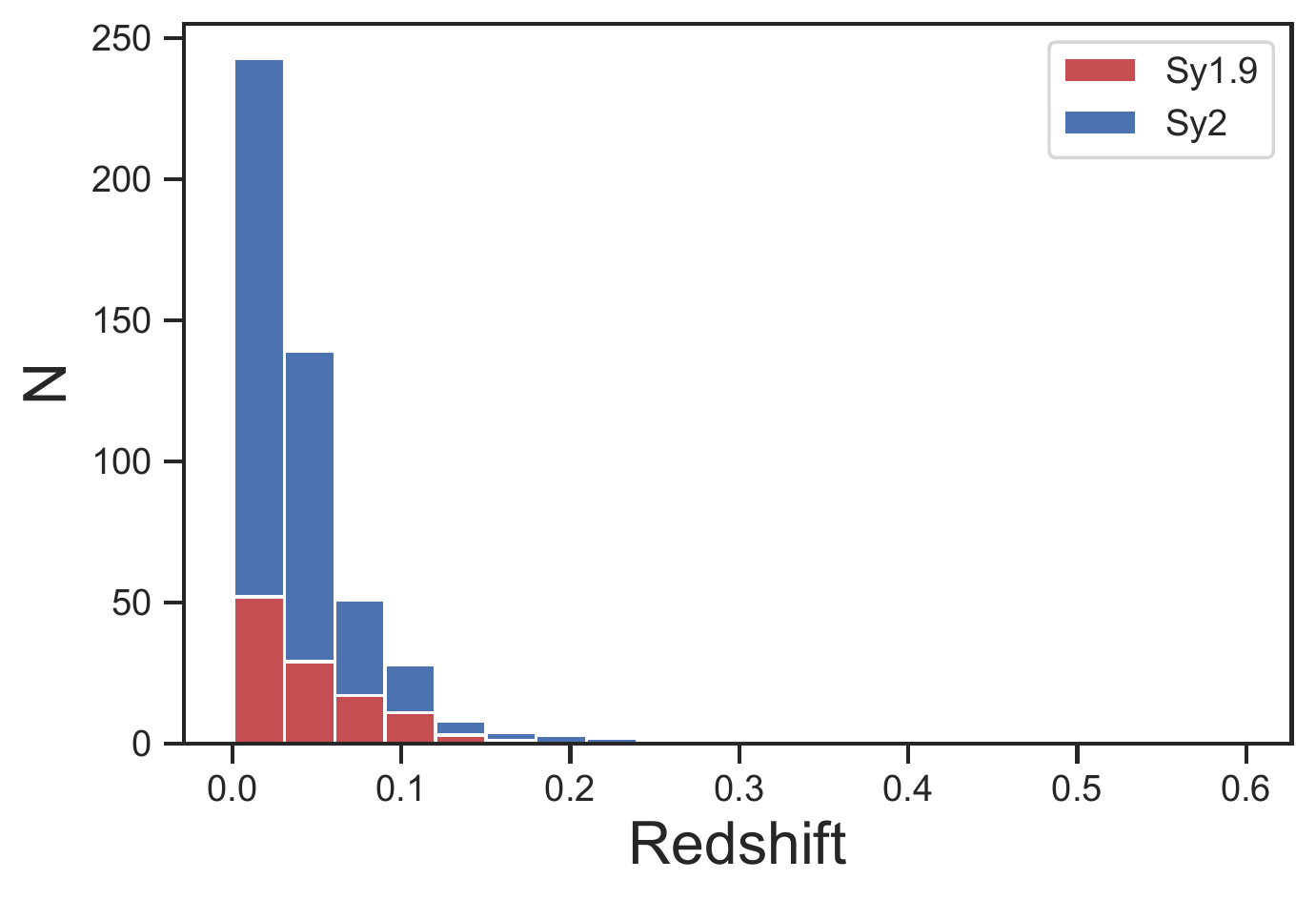}
\includegraphics[width=8.2cm]{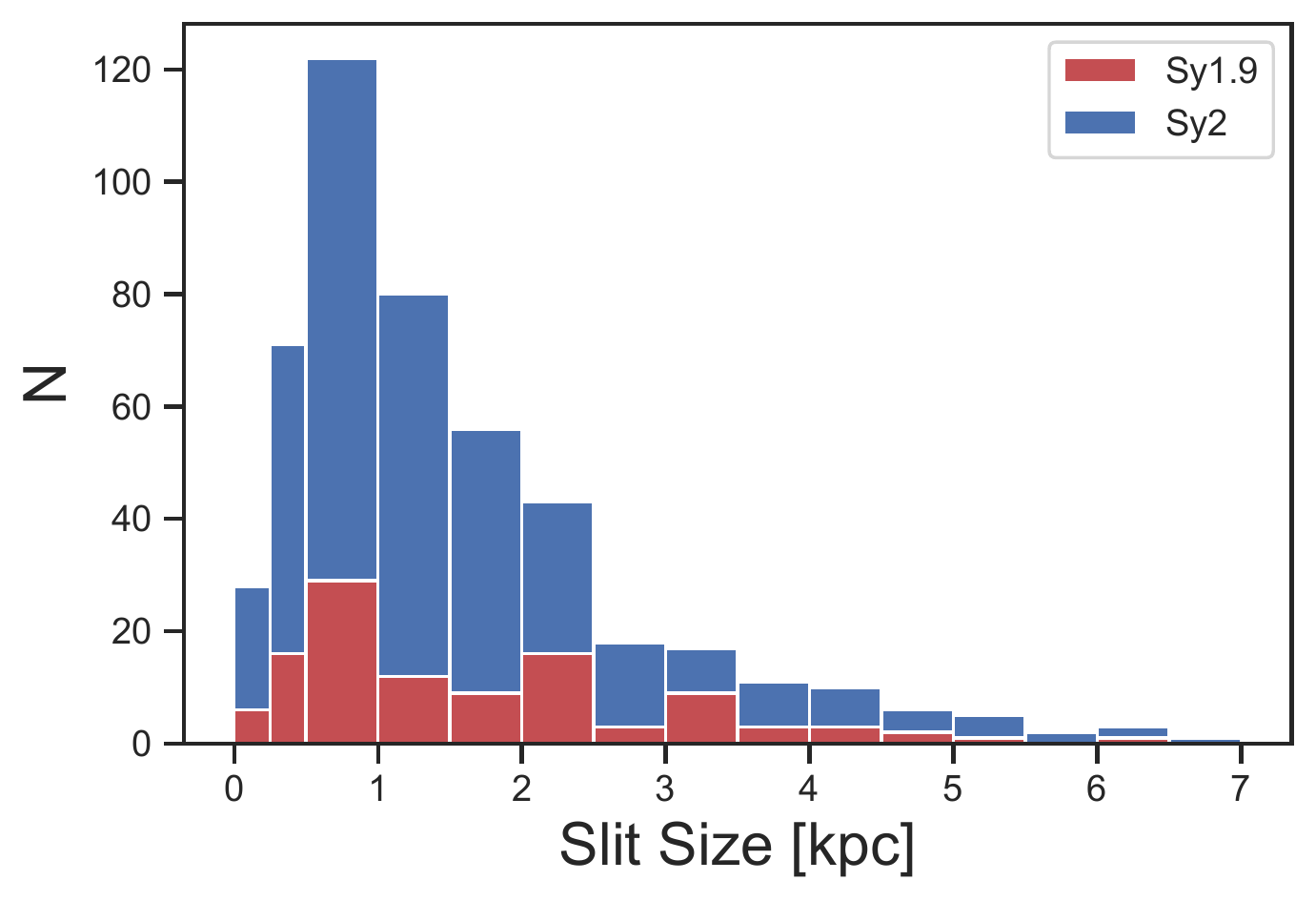}
\includegraphics[width=8.2cm]{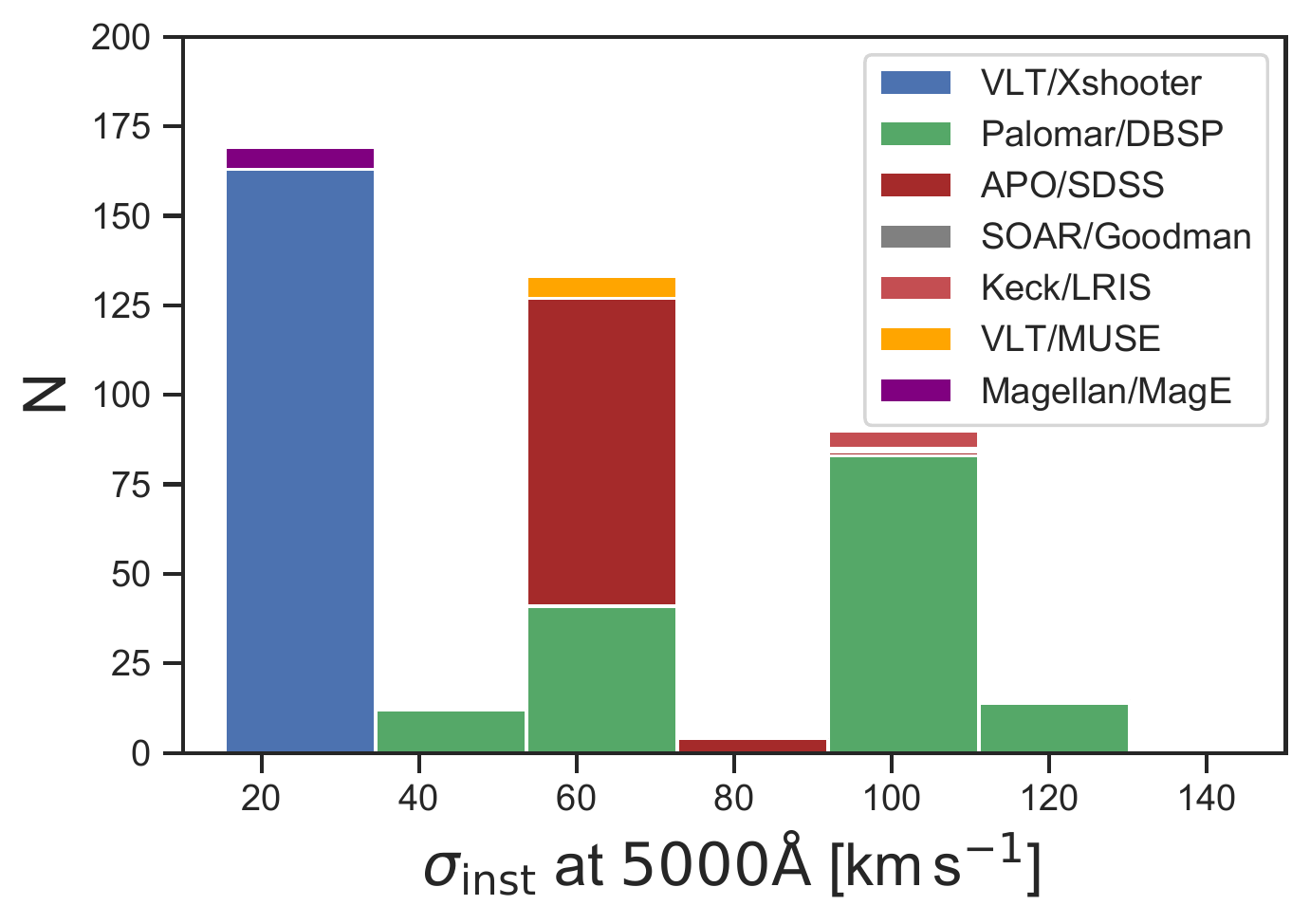}
\includegraphics[width=8.2cm]{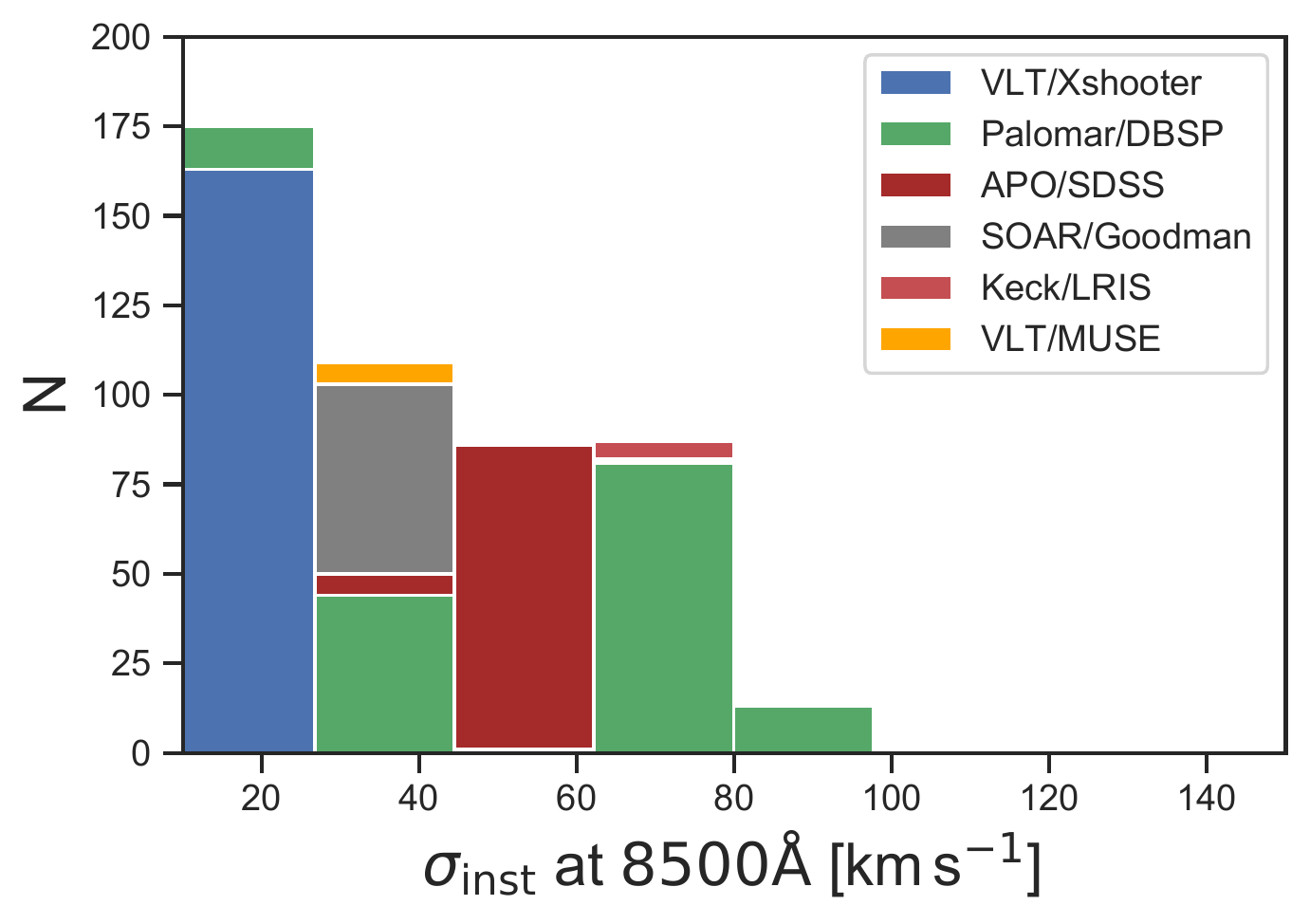}
\caption{Summary of the AGN and the different observing conditions for the best \Nvdisp\ spectra in the sample. {\it Top left:} redshift range of the AGN in the sample, split into Sy\,1.9 and Sy\,2; {\it Top right:} slit width size in kpc for our AGN (again split to Sy\,1.9 and Sy\,2 sources). {\it Bottom panels:} instrumental resolution at 5000 (left) and 8500\,\AA\ (right) for the best spectra, from which we successfully measured \sigs\ from the 3880--5550\,\AA\ or CaT spectral regions (respectively).}
\label{fig:spectra_sum}
\end{figure*}

\section{Velocity Dispersion Measurements}
\label{sec:vdisp_technique}

Here we provide an overview of the technique we use to measure stellar velocity dispersions (\sigs) for the AGN hosts in our sample.  We discuss the specifics of the fitting regions and template library, the fitting technique and software, and masked regions.

\subsection{Wavelength Regions and X-Shooter Spectral Library} \label{sec:stellarlib}

In the BASS DR1, we used stellar population synthesis models from the Miles Indo-U.S. Catalog library of stellar spectra \citep{Vazdekis:2012:157} with a spectral resolution of 2.51\,\AA\ FWHM ($R\sim$2000) and fitted the 3900--7000\,\AA\ range to measure the velocity dispersion using the \Cahk, the $G$-band (at $\sim$4300\,\AA), and \mgb\ triplet absorption lines.  We also used the 8350--8730 \AA\ region to measure the \Catrip\ triplet (hereafter CaT).  
For BASS DR2, many sources were studied at much higher spectral resolutions (i.e., $R\gtrsim5000$; particularly with VLT/X-Shooter); therefore, an updated library was needed to exploit the full instrumental resolution and to probe lower-mass systems (e.g. ${<}$65 \kmps\footnote{A 2.51\,\AA\ FWHM for a template library corresponds to $\sigma$=64\,\kmps at $\lambda$=5000\,\AA.}).  Many studies have demonstrated how template libraries have great difficulty precisely recovering a velocity dispersion below their nominal resolution  \cite[e.g.,][]{Boardman:2016:3029,Boardman:2017:4005,Gannon:2020:2582}. 
Therefore, for the BASS DR2 analysis, we used the X-Shooter Spectral Library (XSL, specifically the XSL DR2; \citealt{Gonneau:2020:A133}), which includes observations of 666 stars covering the wavelength range 3000--24500\,\AA\ at a spectral resolving power close to $R\approx10,000$.\footnote{The X-Shooter instrument has a range of spectral setups; therefore the XSL reaches a higher spectral resolution than our X-Shooter data.}



As a further improvement over BASS DR1, we have revised the fitting regions to measure \sigs\ to be the 3880--5550 and 8350--8730\,\AA\ spectral bands.  We specifically excluded the region between 5550--7000\,\AA\ due to the lack of strong stellar absorption features, residual telluric features in the spectra from the oxygen bands, and the presence of weak broad \Halpha\ in Sy\,1.9 that is difficult to mask.  Additionally, the majority of our spectra were obtained with either  VLT/X-Shooter or Palomar/DBSP, and in both cases, our spectral setups are such that the blue arm ends at $\sim$5550\,\AA. Focusing our \sigs\ measurements on two distinct spectral bands thus avoids any systematics that could have resulted from the ``stitching'' of spectra from various arms.
Our choice of spectral bands means we focus on the UVB and VIS regions of the XSL templates (regardless of the instrument used to obtain each BASS spectrum).

For the VLT/X-Shooter template sample, we began with only slit loss-corrected stars \citep{Gonneau:2020:A133}, which included 813 observations of 666 stars.  We then used only the slit loss-corrected observations available from the website,\footnote{See \url{http://xsl.u-strasbg.fr/page_dr2_all.html}} which included 628 and 718 observations for the UVB and VIS regimes, respectively.  Because the wavelength grids of the template spectra vary due to barycentric velocity corrections and cover a much wider range than necessary, we interpolated all stellar templates to a common wavelength grid with $\Delta\lambda=0.2$\,\AA, which is the native pixel sampling of X-Shooter in the UVB and VIS arms. This resampling was done while conserving flux using {\tt SpectRes} \citep{Carnall:2021:ascl:2104.019}.\footnote{\url{https://spectres.readthedocs.io/en/latest/}} 
We also limited the spectral range of the stellar templates to slightly exceed the spectral fitting regions we planned to use for velocity dispersion, to avoid any discontinuities at the edge of the spectra.
In the blue (UVB) regime, the templates are cut to $\simeq$3800--5552\,\AA\ (compared to the fitting range of 3880--5550\,\AA).
The edge at 5552\,\AA\ is the native red limit of the UVB arm of X-Shooter.  
For the red (VIS) regime, we limited the templates to cover the CaT, i.e. 8300--8900\,\AA. Excluding the rest of the spectra also helped avoid issues with correction of the water vapor bands ($\sim8100 {<} \lambda/$\AA$ {<} 8300$, $\sim8930 {<} \lambda/$\AA$ {<} 9800$).


\subsection{Spectral Template Library from Hierarchical Clustering} 
\label{sec:hc}

When fitting spectra with template libraries, the execution time needed (per observed spectrum) scales roughly with the number of templates used and the number of spectral pixels. 
Thus, to facilitate the usage of large stellar libraries to fit galaxy spectra, smaller subsamples of stars are typically used.  This is particularly important in our case, given the higher pixel sampling of some of our observations (e.g., 0.2\,\AA\ for VLT/X-Shooter) and our use of repeated bootstrap fitting for the error analysis as recommended by the penalized PiXel Fitting (\ppxf) software \citep[][see below]{Cappellari:2017:798}.    

We therefore proceeded with creating subsets of the stellar template library relying on cluster analysis, which is meant to identify groupings or clusters of similar objects (stellar spectra, in our case).  This process also enables the identification and exclusion of any emission line features from flaring late-type stars, low-S/N features in only a single spectrum in the sample, and any possible instrumental issues in the large X-Shooter library that might affect template fitting.  We follow the approach of \citet{Westfall:2019:231}, who applied a hierachical clustering approach \citep[e.g.,][]{Johnson:1967:241} to the full MILES stellar library of 985 stars to create 42 representative templates used in their study.  Their approach was shown to reduce the execution time by a factor of 25 while only slightly affecting the measurements (the median and 68th percentile confidence intervals were $\Delta \sigma_{obs}=0.9^{+3.6}_{-4.1}$ and $\Delta V=-1.3^{+2.0}_{-1.8}$).  
We note that the \citet{Westfall:2019:231} approach is particularly relevant for our needs, as it was designed to decompose the (spatially resolved) spectra of low-redshift galaxies as part of the SDSS-IV MaNGA project.

For our clustering analysis, we use the same Python software that \citet{Westfall:2019:231} used with MILES, {\tt speclus}.\footnote{Available at \url{https://github.com/micappe/speclus}}  We use as input the 628 UVB arm stellar spectra over the range of 3800--5552\,\AA.  This reduces the number of templates from 628 to 97 representative templates.  We further remove from this set 30 templates with emission lines, one template with instrumental issues, and 45 representative templates comprised solely of one individual spectrum of low S/N
 (see Appendix B for examples of the excluded templates).  The remaining 21 X-Shooter templates (\autoref{fig:xshooter_temp}) used here have features found in two up to 152 stellar spectra. While these 21 templates are half (e.g., 21/42) of what was used in the MILES-based library described in, e.g., \citet[][]{Westfall:2019:231}, we note that the fitting region, 3800--5552\,\AA, is much narrower than the one used with their MILES-based library (3525--7500\,\AA); moreover 15/42 of these MILES stars show little or no blue emission below 5552\,\AA.

In the redder (VIS) part of our library, we start the analysis with 718 spectra in the 8300--8900\,\AA\ region.  The clustering analysis reduces the number of templates from 718 down to 79 representative templates.  We further exclude 41 templates with prominent emission features in Paschen transitions or over the CaT region, which are all found in only single spectra, and 16 templates of relatively low S/N that are only comprised of a single spectrum (see Appendix B for examples of the excluded templates).  
This leaves 19 VIS templates (\autoref{fig:xshooter_temp}).

\begin{figure*} 
\centering
\includegraphics[height=4.63in]{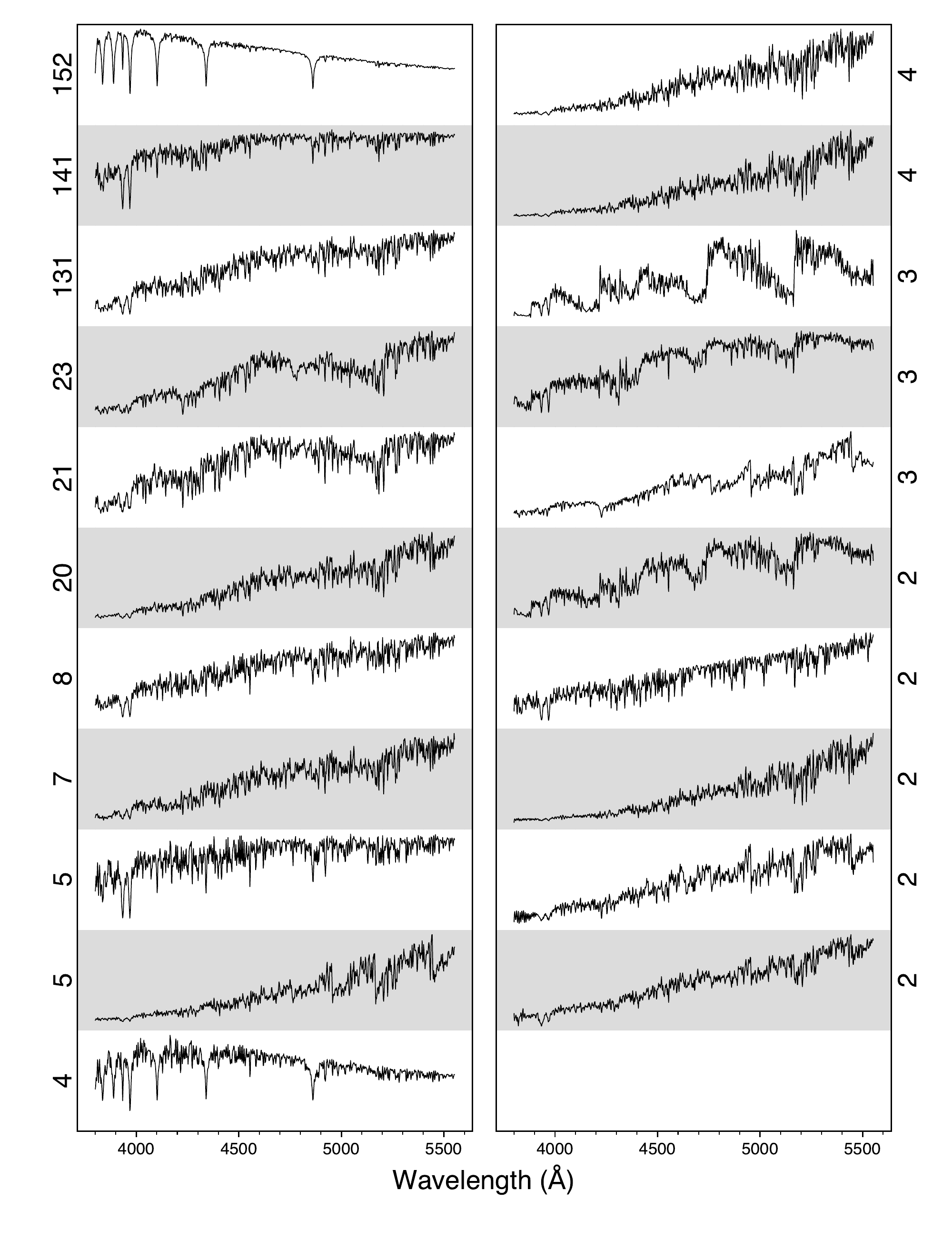}
\includegraphics[height=4.63in]{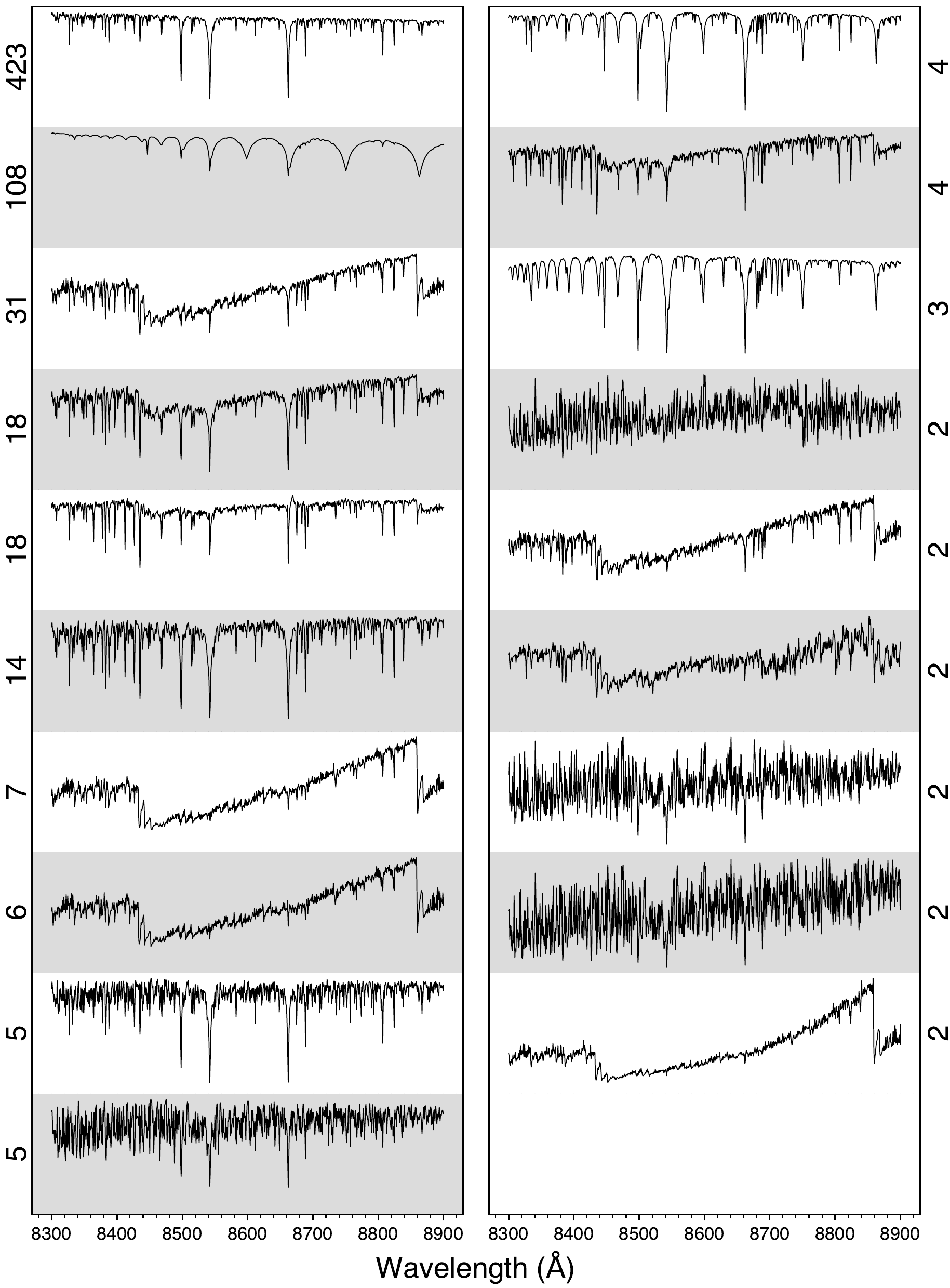}
\caption{Examples of the X-Shooter stellar templates obtained via hierarchical clustering analysis and used in our work.
{\it Left:} the 21 blue (UVB) X-Shooter templates constructed from a subset of 628 X-Shooter stellar spectra, sorted by prominence in the spectral library (see $\S$\ref{sec:hc}).  
{\it Right:} the 19 red (VIS) X-Shooter templates similarly constructed from a subset of 718 X-Shooter stellar spectra (see $\S$\ref{sec:hc}).   
The number next to the vertical axis of each template indicates the number of stellar spectra among the parent sample in which the features were found during the clustering analysis. 
All templates are shown in arbitrarily scaled units of $F_\lambda$.}
\label{fig:xshooter_temp}
\end{figure*}


\subsection{Fitting Technique} \label{sec:fitting_software}

Like in the BASS DR1, we use the \ppxf\ softwaare \citep[][]{Cappellari:2004:138} (Python version 7.4.3) to measure stellar kinematics and the central stellar velocity dispersion (\sigs). This method operates in pixel space and uses a maximum penalized likelihood approach for deriving the line-of-sight velocity distribution (LOSVD) from kinematic data \citep[][]{Merritt:1997:228}.
The \ppxf\ software is extensively used in extragalactic surveys for \sigs\ measurements, which would make our results more directly comparable with other studies.

As a first step, the \ppxf\ code creates a model galaxy spectrum $G_{\rm mod}\left(x\right)$ by convolving a template spectrum with a parameterized LOSVD. Then it determines the best-fitting parameters of the LOSVD by minimizing the value of $\chi ^2$, which measures the agreement between the model and the observed galaxy spectrum over the set of reliable data pixels used in the fitting process.  Finally, \ppxf\ uses the best-fit spectra to calculate the velocity dispersion after subtracting the instrumental resolution (in quadrature).

Within \ppxf\ we use a fitting procedure based on the example developed for the SAURON project \citep{Cappellari:2004:138} with a velocity scale ratio of 2 to sample the templates at twice the resolution of the observed spectra ({\sc velscale}=2) with the standard four moments ({\sc moments}=4).  We use additive ({\sc degree}=8) and multiplicative ({\sc mdegree}=1) polynomials to develop the best composite stellar population to fit the galaxy continuum and absorption lines, correct for inaccuracies in spectral calibration, and make the fit insensitive to a specific reddening curve. Importantly, modifying the degrees of additive polynomials (e.g., with {\sc degree}=4-12) and/or removing the multiplicative polynomial (i.e., {\sc mdegree}=0) does not significantly change the fitting results, implying that the latter are robust.

To estimate the uncertainty associated with each \sigs\ measurement, we conservatively use the error estimate by \ppxf\ or an error estimated through bootstrapping, whichever is larger.  To estimate the bootstrap error, we resample the residuals of the initial velocity dispersion fit to construct 10 mock spectra to perform 10 additional fits. In general, the error given by \ppxf\ is larger than the bootstrap errors, except for a few cases of fitting the CaT region.  In addition to the \sigs\ measurement, \ppxf\ also provides accurate, absorption features-based redshift determinations. 
These are completely independent of the emission line-based redshift measurements reported and used in other BASS (DR2) works and may be useful for studying velocity offsets between gas and stars in the (hosts of) BASS AGN.

Given the range of quality in our spectra and the presence of repeated spectroscopy for some sources, we excluded any measurements with ${>}$20\,\kmps\ error for which other velocity dispersion measurements existed within our sample.  For completeness, we included measurements with errors between 20 and 30\,\kmps\ if no other measurements existed.  

Studies have found that measurements below or near the instrumental resolution have significantly increased scatter and may also be overestimated  \cite[see][where they suggest an 18\% cut]{Scott:2018:2299a} because biases can extend above the instrumental resolution \citep[e.g., 100\,\kmps with $\sigma_\mathrm{SDSS\,inst}$=70\,\kmps][]{2019AJ....158..160B}.  Based on these studies and due to the intrinsic uncertainty in measuring instrumental resolution \citep{Koss_DR2_catalog}, we therefore conservatively exclude any measurements within 20\% above the corresponding instrumental resolution based on previous studies.

As these limitations were realized during our observational campaigns, the number of AGN covered in our work is {\it not} reduced at all, as any galaxies with \sigs\ near or below the spectral resolution were reobserved with higher-resolution setups (e.g. where $\sigma_{\rm inst}\sim25$\,\kmpssh).  Specifically, 10 AGN from 16 spectra from either Gemini/GMOS or Palomar/DBSP fell into this category, but all were reobserved using higher-resolution setups from VLT/Xshooter or Palomar/DBSP.

We chose not to apply aperture corrections \citep[e.g.,][]{Jorgensen:1995:1341} to the central velocity dispersions to place them at the effective radius or correct for inclination effects related to galaxy orientation \citep[e.g.,][]{Bellovary:2014:2667}.  At fixed aperture, an increasing part of the host bulge and disk is sampled in spectra at increasing distances (redshifts), possibly leading to larger observed velocity dispersions.   
The aperture corrections that are designed to overcome this effect act in opposite directions, depending on whether the galaxy morphology is early- or late-type \citep[e.g.,][]{Falcon-Barroso:2017:J/A+A/597/A48}.  This is extremely challenging to apply to the host galaxies in our sample, which are predominantly massive spirals with high concentration indices \cite[i.e., lenticulars;][]{Koss:2011:57}, as it is not clear whether to increase or decrease the measured \sigs\ for each galaxy under study.  Another issue is the significant fraction of mergers \citep[e.g., 24\%;][]{Koss:2010:L125,Koss:2018:214a} that may have velocity dispersions affected due to the oscillations of the stars in the two progenitor galaxies as they pass by each other and coalesce.  However, simulations suggest that during the merger, the \sigs\ values do not fall below 70\% or exceed 200\% and are much more likely to fall near the equilibrium value \citep{2012ApJ...747...33S}. However, dust attenuation associated with the merger may increase the scatter.  A detailed host morphological analysis is needed but is beyond the scope of DR2.  Additionally, LOVSD can be broadened due to orientation-dependent disk rotation, which would result in a larger dispersion outside the central spheroid component \citep[e.g.,][]{Caglar:2020:A114}.  
Finally, other studies do not find a significant offset of the SMBH masses deduced from \ppxf-measured, uncorrected \sigs\ compared to those deduced from broad emission lines (within BASS), suggesting that aperture and rotation corrections are unlikely to be a major issue, at least in a statistical sense \citep[e.g.,][]{Ricci_DR2_NIR_Mbh}.  Future studies within BASS using integral field units such as VLT/MUSE are currently ongoing \citep[e.g.,][]{2022kakkad} and will hopefully serve to address this particular issue.

\subsection{Masked Regions}

The spectral fitting procedure is significantly improved by masking out certain spectral regions that are dominated by prominent emission lines.
A full list of all the masked emission lines is given in Appendix A.  For the normal mask, the width of these masks was set to 2500\,\kms, except for the brightest \hbeta, \OIIIa, and \OIII\ lines, where the masked regions were set to 5000\,\kms, to account for any blueshifted or weak broad emission components.  We did not find that a narrower mask significantly improved the fitting.   In practice, \ppxf\ automatically applies a mask for several bright emission lines: H$\delta$, H$\gamma$, H$\beta$, \OIII, and \OI. We further masked out the regions around the Ca H $\lambda$3968 line because of overlap with the H$\epsilon$ $\lambda$3970 and \NeIIIb\ emission lines. Since the overlapping emission lines reduce the depth of the Ca\,H absorption line, we expect that \ppxf\ will also underestimate the depth and width of the Ca\,K line and consequently affect the measurement of \sigs\ \cite[see, e.g.,][]{Greene:2005:721}. Finally, we also excluded a few other bright emission lines \citep[e.g., flux ${>}5$\% \Hbeta\ in bright AGN;][]{Tran:2000:562} from the velocity dispersion fits.


To make sure that our spectral fits are not affected by any residual line emission, we visually inspected the (initial) \ppxf\ fits, and masked additional spectral regions, of weaker emission lines, motivated by past observations of bright AGN \citep[e.g.,][]{Osterbrock:1990:561,Osterbrock:1996:183}, and repeated our fits.  Here the width of the mask was set to 1000\,\kms.  In the case of the CaT region (8350--8730\,\AA), the initial fit was done with no masking of emission lines.  If emission lines were seen in the residuals, we masked the relevant weak transitions of ${\rm O}\,\textsc{i}$, $\left[{\rm Cl}\,\textsc{ii}\right]$,
$\left[{\rm Fe}\,\textsc{ii}\right]$, Pa12, and/or Pa11. Additionally, we also masked the observed-frame spectral region of 9300-9700\,\AA, which is affected by strong telluric features.  
This was particularly relevant for higher redshift AGN ($0.065{<}z{<}0.09$), where these telluric features strongly limit our ability to recover the intrinsic spectra.
At yet higher redshifts, $z{>}0.09$, no CaT measurements were made.

\section{Results}
\label{sec:results}

In this section we present the velocity dispersion fit results, compare the measurements available for some of our AGN within the BASS DR2 dataset and in other compilations to better understand uncertainties, discuss issues related to sample (in)completeness, and present the inferred estimates of BH mass for our BASS AGN.

\subsection{Velocity Dispersions}

A full catalog of \sigs\ measurements for the best spectra of \Nvdisp\ unique BASS AGN is provided in \autoref{tab:veldispbest}. In addition, \sigs\ measurements derived from secondary spectra are provided in \autoref{tab:veldispsec} (in Appendix C).
Example velocity dispersion fits are shown in \autoref{fig:Example_fits}.  Each spectrum may have a 3880--5550\,\AA\ and/or CaT spectral region fit, depending on the spectral setup and the \sigs\ measurement error.
Some instruments yielded only measurements of one spectral region. 
Specifically, spectra obtained with SOAR/Goodman cover only the CaT region, while spectra obtained with either Magellan/MagE or Gemini/GMOS only have the  3880-5550\,\AA\ region.

The distributions of the \sigs\ measurements and associated errors for all the spectra are shown in \autoref{fig:regionfitcomp}. 
The median velocity dispersions from the 3880--5550\,\AA\ region (173\,\kmpssh) are somewhat higher than those of the CaT region (144\,\kmpssh).
This is due to the higher typical redshift of the AGN with \sigs\ measured from the  3880--5550\,\AA\ region (median $z=0.042$) compared to those with a CaT-based measurement ($z=0.024$) and because of the increasing difficulty (or impossibility) of fitting the CaT region for sources at $z{>}0.065$ (due to telluric features).

In \autoref{fig:agntypecomp}, we compare the distributions of the (best) \sigs\ measurements for the Sy\,1.9 ($N=113$) and Sy\,2 ($N=371$) sources in our sample.  Overall, the two subsamples are very similar, with a median \sigs\ of 153$\pm$5 and 157$\pm3$\,\kmps for the Sy\,1.9 and Sy\,2 systems, respectively.

\begin{figure*}
\centering
\includegraphics[width=0.44\textwidth]{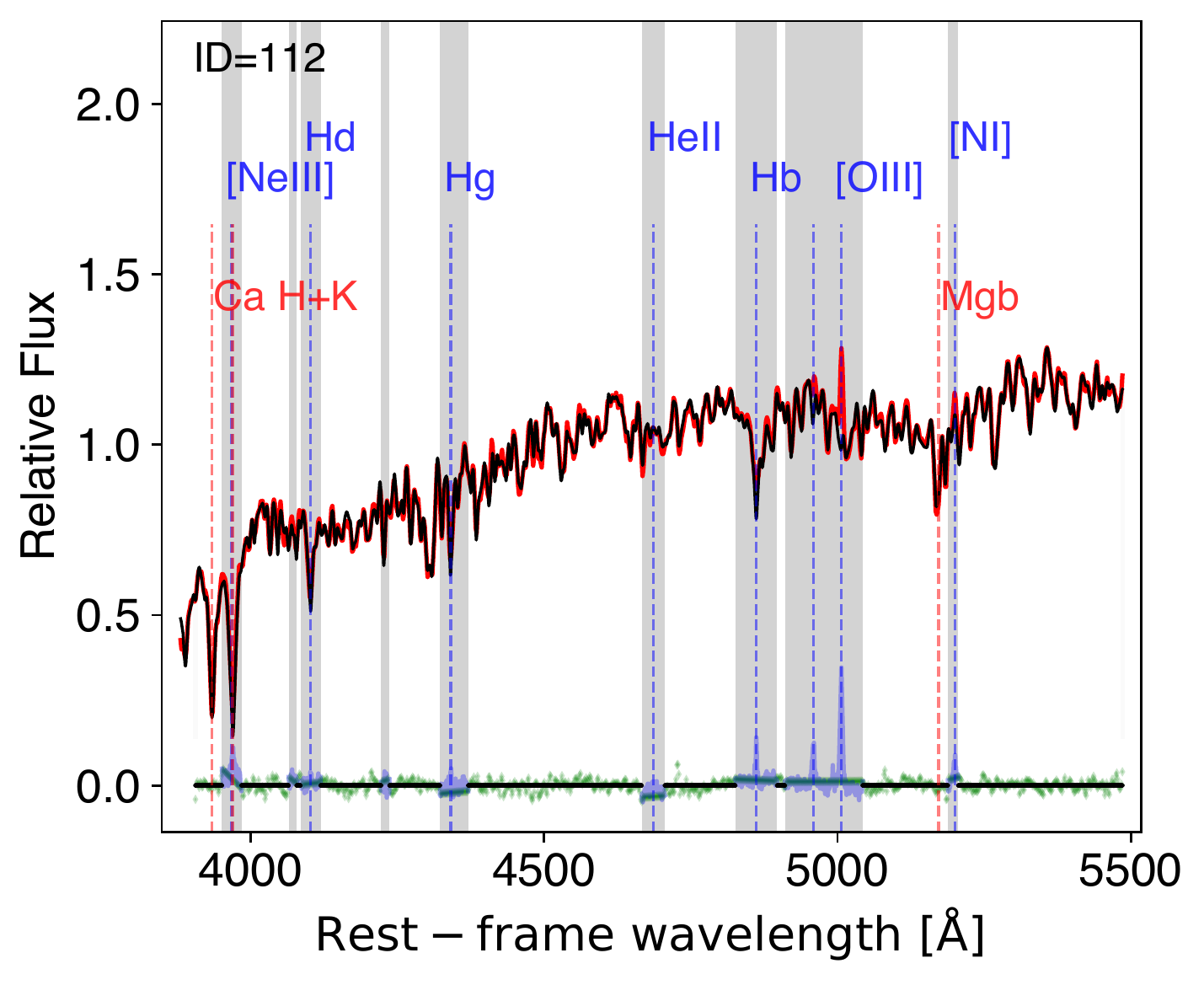}
\includegraphics[width=0.44\textwidth]{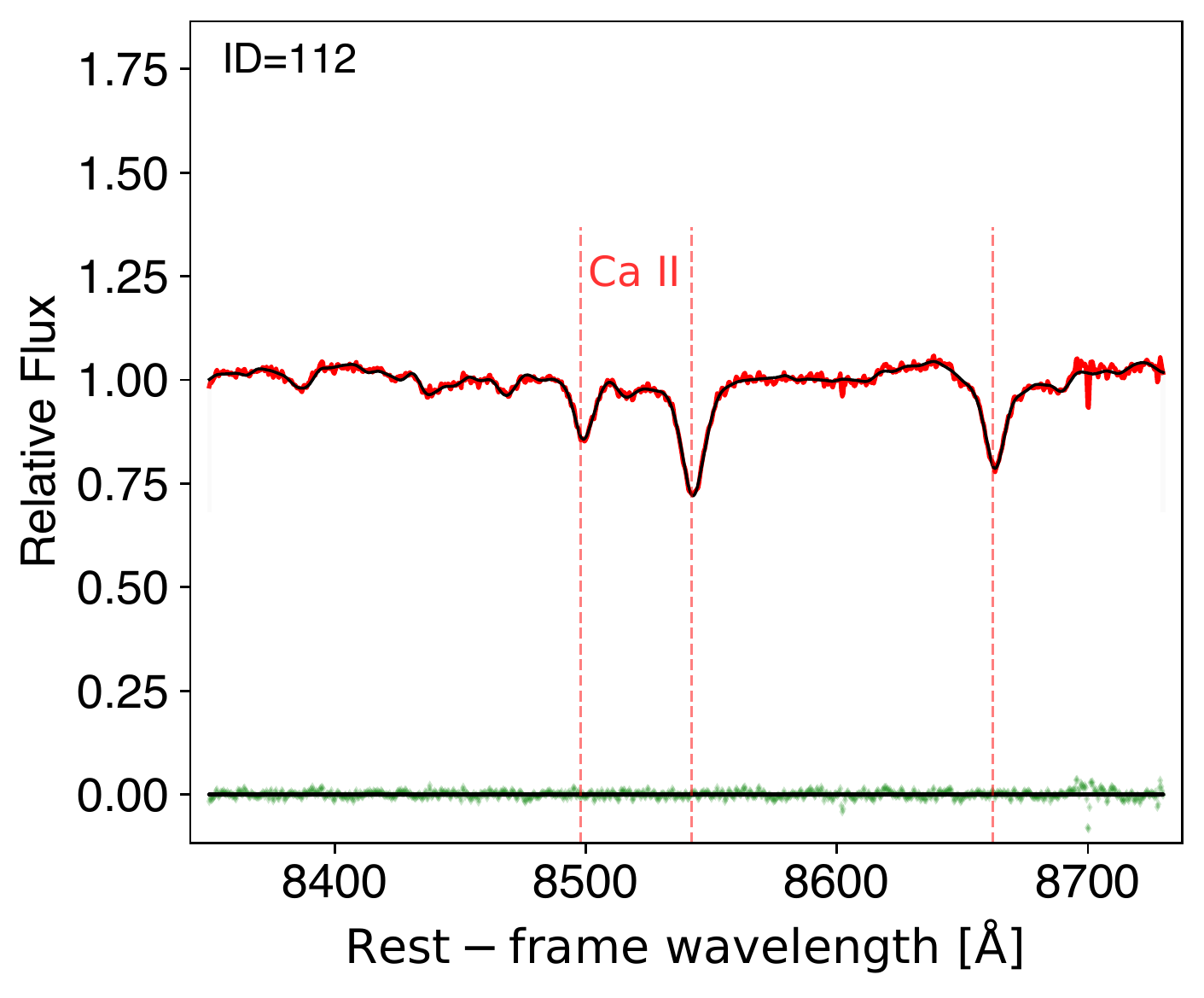}
\includegraphics[width=0.44\textwidth]{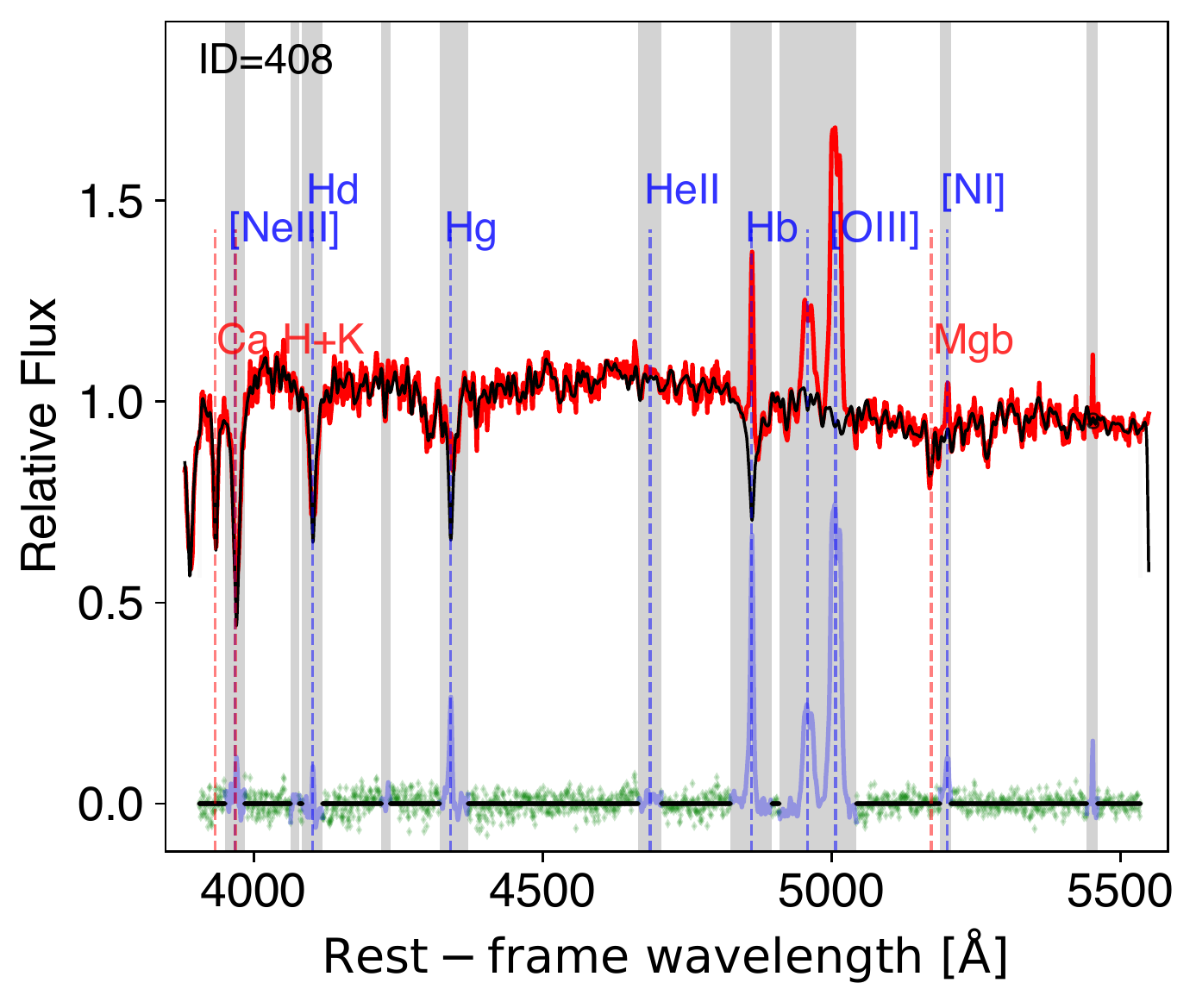}
\includegraphics[width=0.44\textwidth]{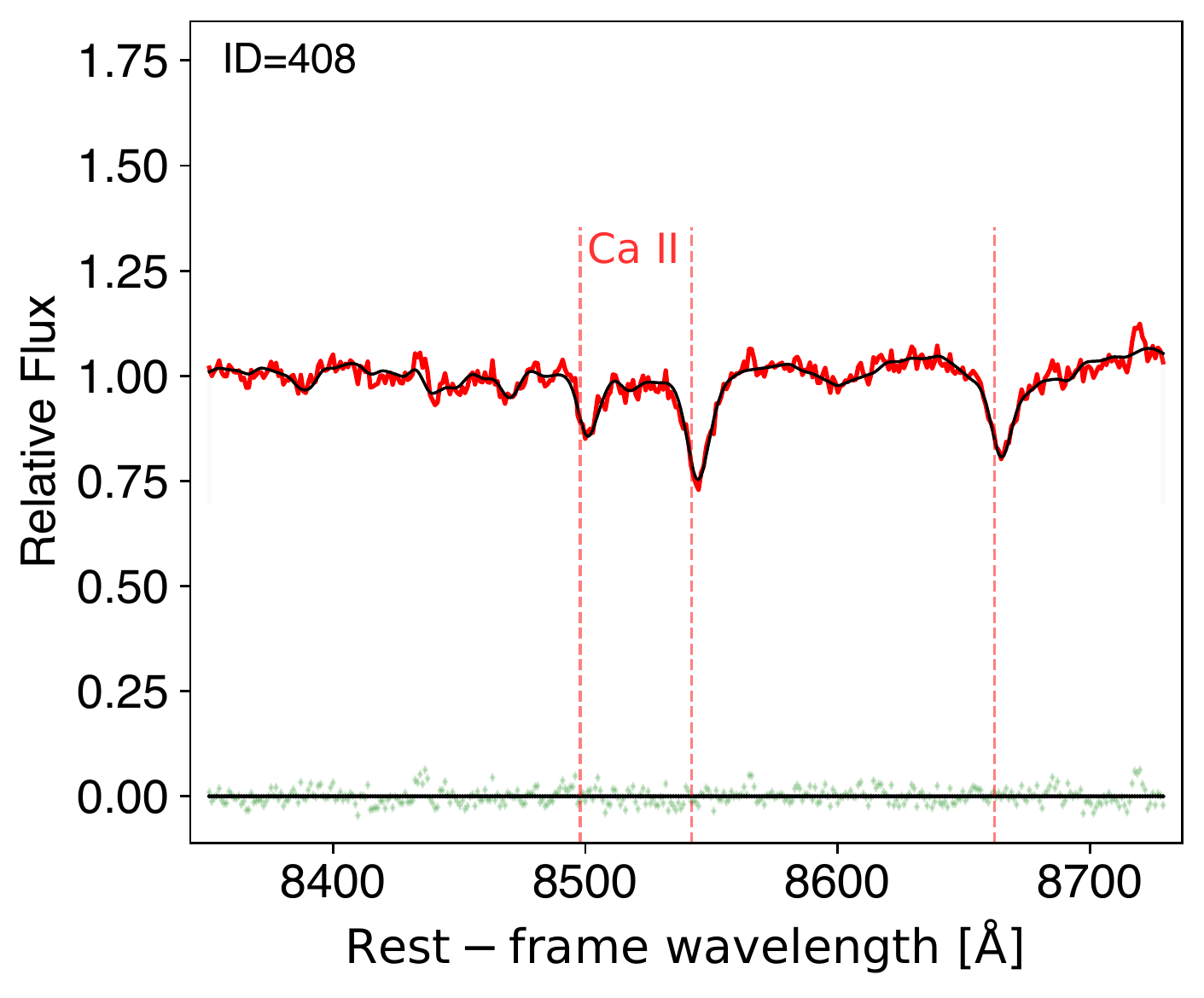}
\includegraphics[width=0.44\textwidth]{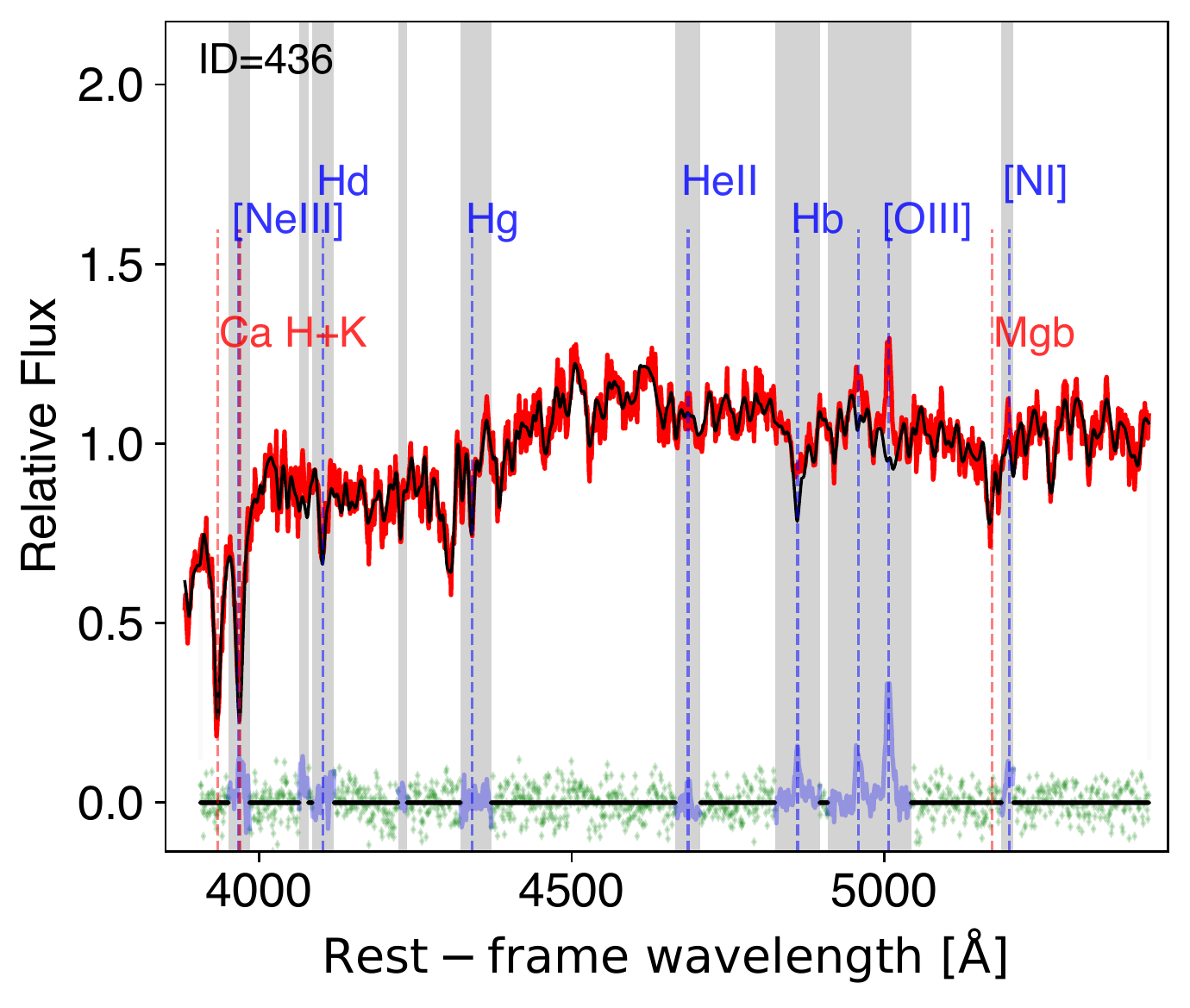}
\includegraphics[width=0.44\textwidth]{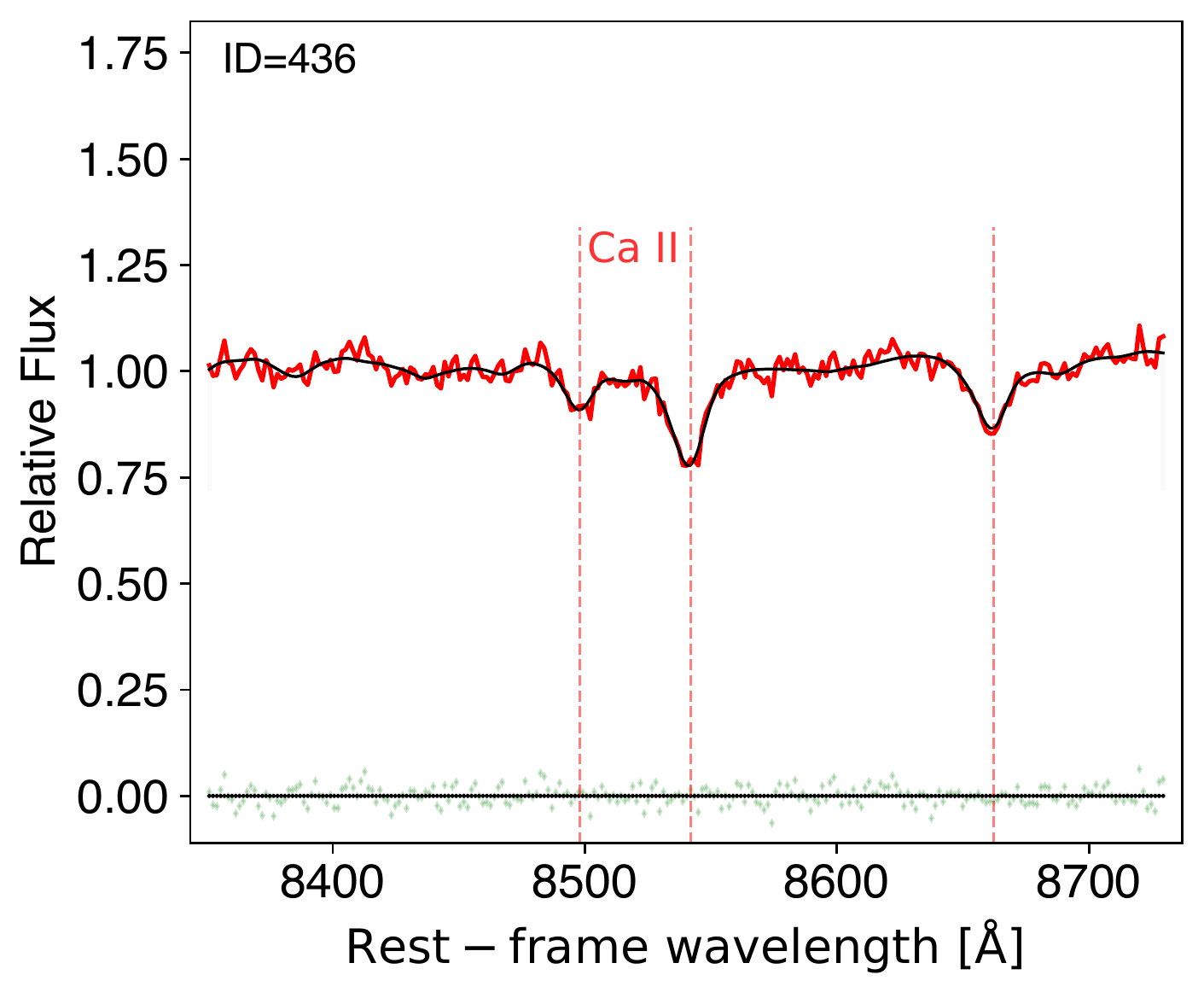}
\caption{Examples of velocity dispersion fits for three AGN spectra (from top to bottom: BAT IDs 112, 408, and 436), representative of the three most common observational setups (from top to bottom: VLT/X-Shooter, SDSS, and Palomar/DBSP).  
The three spectra shown here illustrate the fitting quality typical of the top quartile, median, and bottom quartile of the velocity dispersion errors (from top to bottom: $\Delta\sigs=$4, 7,  and 11\,\kmpssh, respectively).  In all panels, the red lines indicate the observed spectra, and the best-fitting templates are shown in black.  The red dashed vertical lines indicate important absorption features, while the blue dashed vertical lines mark possible emission features that are masked out in the gray regions.}
\label{fig:Example_fits}
\end{figure*}

\begin{figure*} 
\centering
\includegraphics[width=0.49\textwidth]{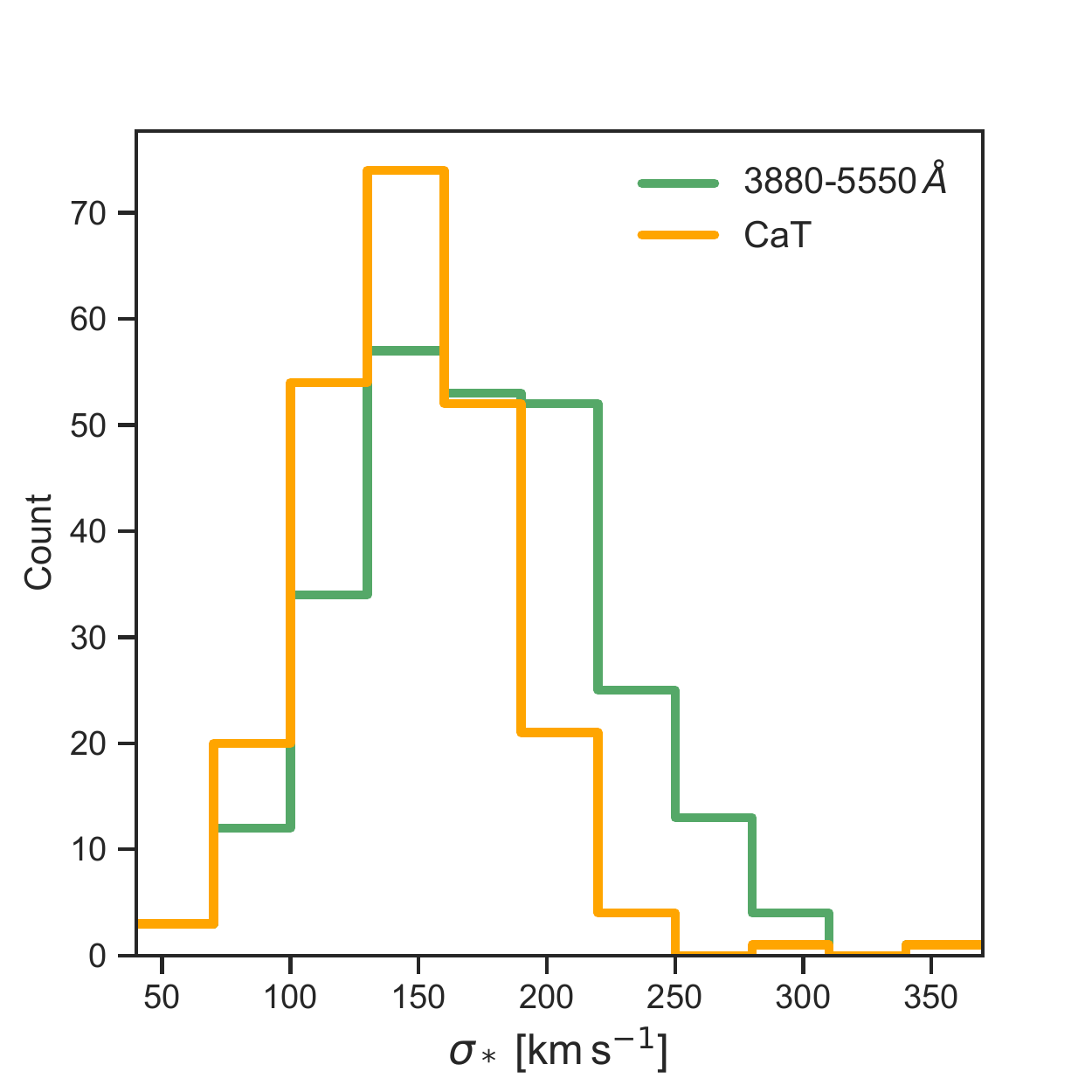}
\includegraphics[width=0.49\textwidth]{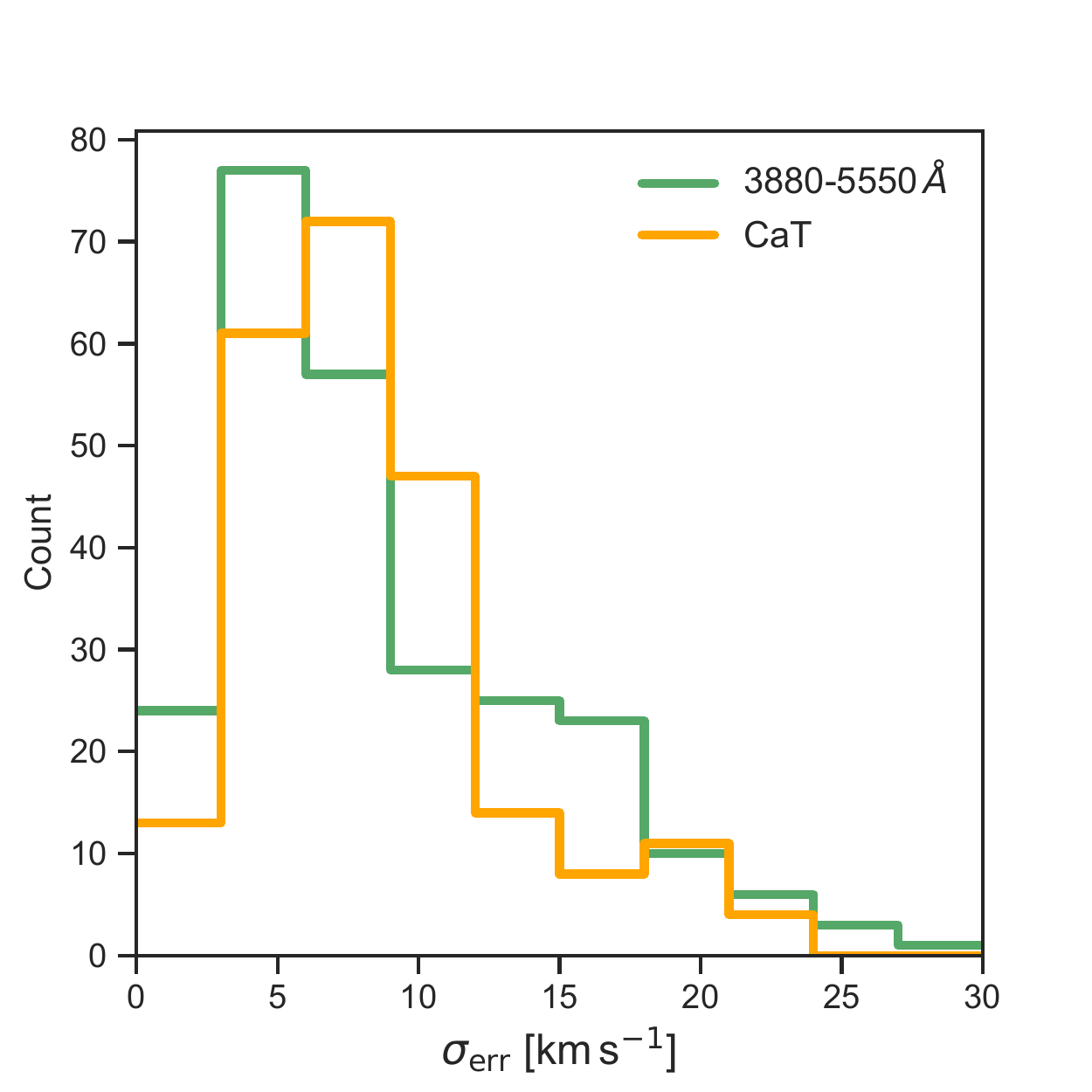}
\caption{Distributions of velocity dispersions (left) and measurement errors (right) for the 3880--5550\,\AA\ ($N=254$) and CaT ($N=230$) fitting regions from the best fits.  The median velocity dispersions from the 3880--5550\,\AA\ region are somewhat higher than those from the CaT region (173 and 144\,\kmpssh, respectively).  This is due to the typically higher redshifts of the AGN for which a fit from the 3880--5550\,\AA\ region is available (median $z=0.042$) compared to those with CaT measurements ($z=0.024$), as the latter spectral region is limited by the increasing effect of telluric features (at $z{>}0.065$).  }
\label{fig:regionfitcomp}
\end{figure*}

\begin{figure*} 
\centering
\includegraphics[width=0.49\textwidth]{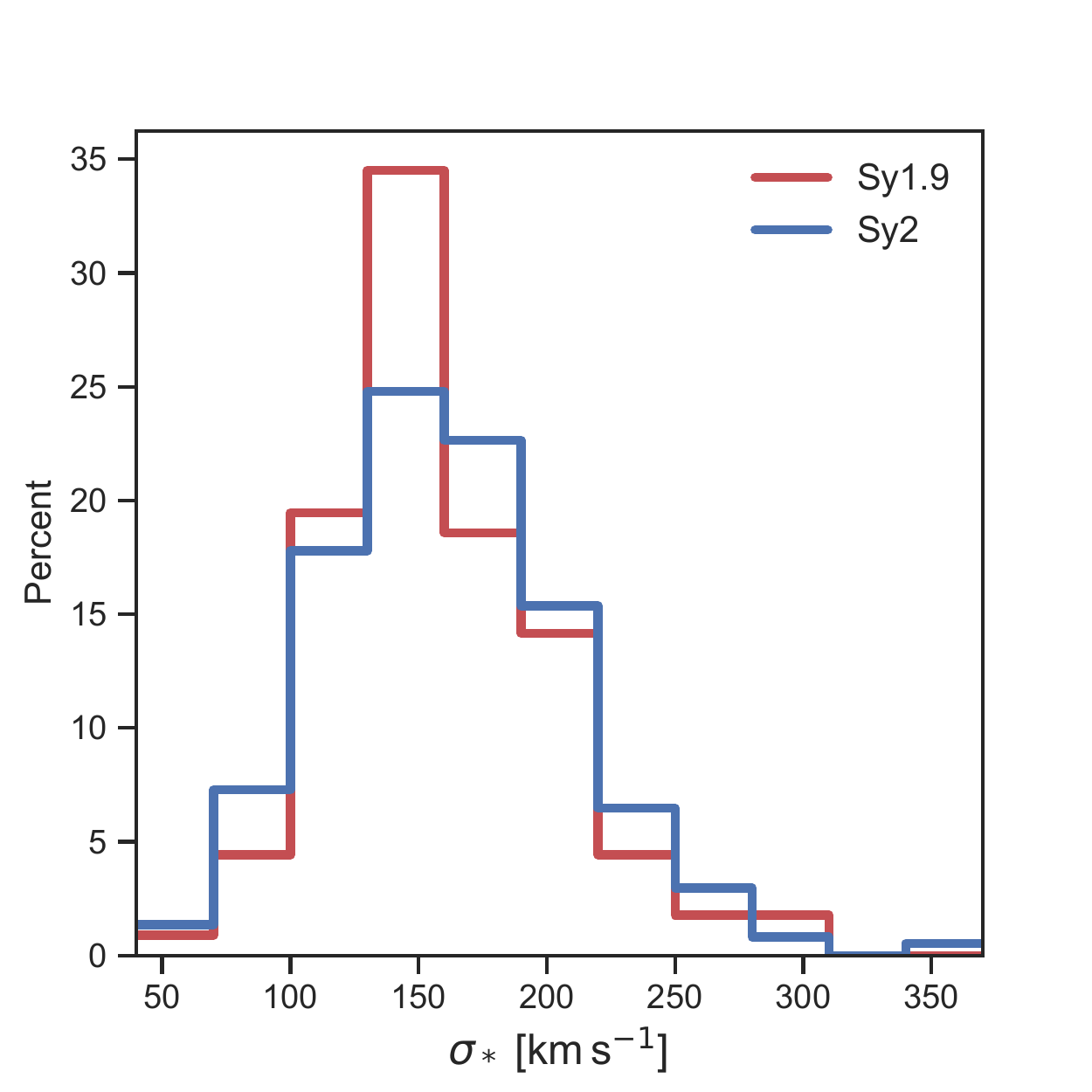}
\includegraphics[width=0.49\textwidth]{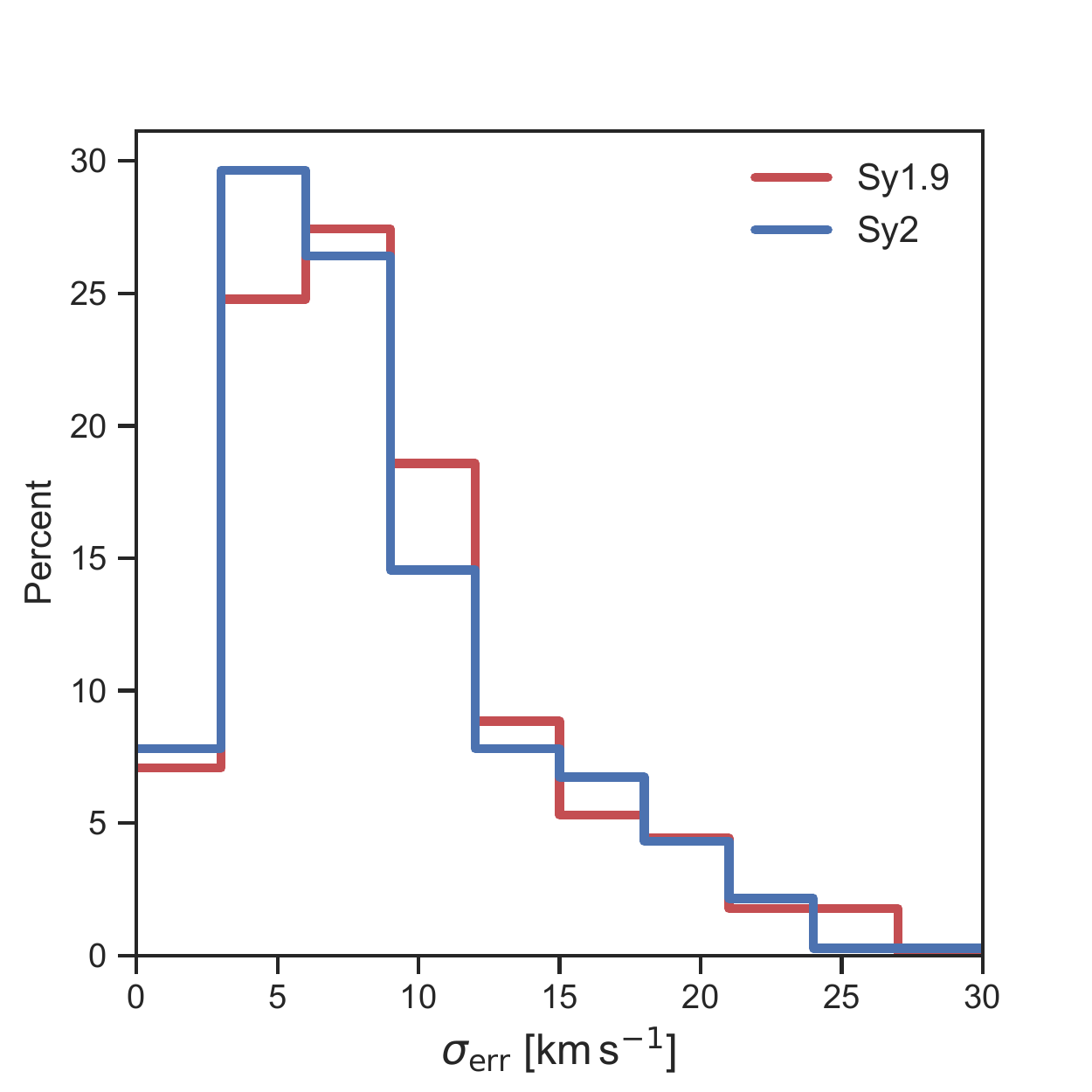}
\caption{Distributions of velocity dispersions (left) and measurement errors (right) for Sy\,1.9 ($N=113$) and Sy\,2 ($N=371$).  The two subsamples are very similar, with a median of 153$\pm$5 and 157$\pm$3\,\kmps for Sy\,1.9 and Sy\,2, respectively. }
\label{fig:agntypecomp}
\end{figure*}

\begin{deluxetable*}{lllllllllll}
\tabletypesize{\scriptsize}
\tablecaption{Best Spectral Measurements \label{tab:veldispbest}}
\tablehead{
\colhead{BAT ID} &      \colhead{Galaxy} & \colhead{DR2 Type} &
\colhead{Tele./Inst.} &                          \colhead{Res. Blue} & \colhead{Res. Red} &  \colhead{Mask} &     \colhead{$z_{3880-5550\,\text{\AA}}$} & \colhead{$\sigma_{3880-5550\,\text{\AA}}$} & \colhead{$z_{\mathrm{CaT}}$} & \colhead{$\sigma_{\mathrm{CaT}}$}\\
\colhead{} &      \colhead{} &                        \colhead{} &  \colhead{} & \colhead{(\AA)} & \colhead{(\AA)} & \colhead{} &     \colhead{(\kmpssh)} & \colhead{(\kmpssh)} & \colhead{(\kmpssh)} & \colhead{(\kmpssh)}\\
\colhead{(1)}& \colhead{(2)}& \colhead{(3)}& \colhead{(4)}&\colhead{(5)}& \colhead{(6)}& \colhead{(7)}& \colhead{(8)}& \colhead{(9)}& \colhead{(10)}& \colhead{(11)}
}
\startdata
     1 &       2MASXJ00004876-0709117 &    Sy\,1.9 &       APO/SDSS &  2.9 &  \nodata &        n &   11561$\pm$14 &   126$\pm$6 &  11557$\pm$17 &      124$\pm$7 \\
     4 &       2MASXJ00032742+2739173 &      Sy\,2 &       APO/SDSS &  2.9 &  \nodata &        n &   12170$\pm$15 &   146$\pm$5 &  12139$\pm$18 &        \\
     7 &      SDSSJ000911.57-003654.7 &      Sy\,2 &       APO/SDSS &  3.0 &  \nodata &        n &   22497$\pm$22 &  205$\pm$12 &               &                \\
    10 &                     LEDA1348 &    Sy\,1.9 &   VLT/X-Shooter &  1.3 &      1.4 &        n &   29488$\pm$25 &  253$\pm$14 &               &                \\
    13 &                   LEDA136991 &      Sy\,2 &   Palomar/DBSP &  2.3 &      1.8 &        n &                &             &   3837$\pm$15 &      131$\pm$8 \\
    17 &                     ESO112-6 &      Sy\,2 &      SOAR/GMAN &  2.7 &  \nodata &        w &                &             &   8911$\pm$16 &      147$\pm$7 \\
    20 &       2MASXJ00343284-0424117 &      Sy\,2 &      Keck/LRIS &  3.9 &      4.7 &        n &   65639$\pm$31 &  294$\pm$22 &               &                \\
    24 &                       MRK344 &      Sy\,2 &   Palomar/DBSP &  4.1 &      4.9 &        n &    7768$\pm$18 &  152$\pm$10 &   7658$\pm$14 &      121$\pm$7 \\
    25 &        SWIFTJ004039.9+244539 &    Sy\,1.9 &   Palomar/DBSP &  2.0 &      1.3 &        n &   24018$\pm$16 &   134$\pm$6 &  23952$\pm$23 &     177$\pm$14 \\
    28 &                      NGC235A &    Sy\,1.9 &      SOAR/GMAN &  2.7 &  \nodata &        w &                &             &   6817$\pm$13 &      200$\pm$3 \\
    29 &       2MASXJ00430184+3017195 &      Sy\,2 &   Palomar/DBSP &  4.1 &      4.9 &        n &                &             &  15310$\pm$21 &     135$\pm$17 \\
    31 &                   MCG-2-2-95 &      Sy\,2 &   VLT/X-Shooter &  1.3 &      1.4 &        n &    6008$\pm$11 &    83$\pm$1 &   6010$\pm$11 &       99$\pm$2 \\
    32 &                     ESP39607 &      Sy\,2 &   VLT/X-Shooter &  1.3 &      1.4 &        n &   61938$\pm$43 &  269$\pm$29 &               &                \\
    33 &                       Mrk348 &    Sy\,1.9 &   Palomar/DBSP &  4.1 &      4.9 &        w &                &             &   4676$\pm$25 &      82$\pm$21 \\
    37 &       2MASXJ00520383-2723488 &      Sy\,2 &   VLT/X-Shooter &  1.3 &      1.4 &        n &   23599$\pm$18 &  195$\pm$10 &  23626$\pm$17 &     182$\pm$15 \\
    44 &       2MASXJ01003490-4752033 &      Sy\,2 &      SOAR/GMAN &  2.7 &  \nodata &        w &                &             &  14706$\pm$15 &      160$\pm$7 \\
    49 &                    MCG-7-3-7 &      Sy\,2 &      SOAR/GMAN &  2.7 &  \nodata &        w &                &             &   9211$\pm$14 &      190$\pm$7 \\
    50 &                    ESO243-26 &      Sy\,2 &   VLT/X-Shooter &  1.3 &      1.4 &        n &    5915$\pm$16 &    93$\pm$8 &   5937$\pm$15 &      104$\pm$5 \\
    53 &                         UM85 &      Sy\,2 &   VLT/X-Shooter &  1.3 &      1.4 &        n &   12524$\pm$15 &   140$\pm$5 &  12510$\pm$15 &      145$\pm$6 \\
    55 &   2MASXJ01073963-1139117 &      Sy\,2 &   Palomar/DBSP &  4.1 &      4.9 &        n &   14353$\pm$18 &   151$\pm$9 &  14315$\pm$19 &     181$\pm$15 \\
    57 &                         3C33 &      Sy\,2 &   VLT/X-Shooter &  1.3 &      1.4 &        n &   18289$\pm$16 &   239$\pm$7 &  18274$\pm$16 &     248$\pm$11 \\
\enddata
\tablecomments{The columns are as follows.  (1) BAT 70 month survey catalog ID.\footnote{\url{https://swift.gsfc.nasa.gov/results/bs70mon/}} (2) Host galaxy. (3) AGN type based on optical spectroscopy including Sy\,1.9 (narrow \hbeta\ and broad \halpha) and Sy\,2 (narrow \hbeta\ and \halpha) from \citet{Koss_DR2_catalog}. (4) Observatory and instrument used. (5) and (6) Instrumental resolution FWHM in \AA\ from the DR2 for the blue (3880-5550\,\AA) and/or red (CaT) region.  (7) Emission line mask used for measurements, where 'n' is normal mask and 'w' is more extensive line list to cover weak emission lines. (8) and (10) Redshift measurements for the 3880--5550\,\AA\ and CaT region based on the template fits and associated (1$\sigma$) error.  A 10\,\kmps systematic error has been added due to wavelength calibration uncertainty.  (9) and (11) Velocity dispersion measurements for the 3880--5550\,\AA\ and CaT region based and associated (1$\sigma$) error.  \autoref{tab:veldispbest} is available in its entirety in machine-readable format.  A portion is shown here for guidance regarding its form and content.}
\end{deluxetable*}

\subsubsection{Comparisons between Measurements \label{sec:meascomp}}

In \autoref{fig:dr2_vdispcomp}, we compare the \sigs\ measurements of AGN with multiple spectra within BASS (i.e., best and secondary), further split by the two fitting regions.  
We find that, in general, there is good agreement among the 3880--5550\,\AA\ region measurements, with a scatter of 27\,\kmps (i.e., rms,  of $\sigma_{\mathrm{Best}}-\sigma_{\mathrm{Sec}}$), and 7/69 (10\%) sources, where the difference between the measurements exceeds 30\,\kmps\ beyond the 1$\sigma$ error bars.  For the CaT region, the scatter between measurements is slightly larger, 31\,\kmpssh, and there is indeed a larger number of sources with significant discrepancies (11\%, 11/100).  It is unclear if the slightly higher fraction of outliers compared to 3880--5550\,\AA\ is related to the smaller fitting region or possible undetected issues with telluric features.  The corresponding median absolute deviation (MAD), which is a robust estimator that is more resilient to outliers than the standard deviation, is merely 20 and 14\,\kmpssh, for the 3880--5550\,\AA\ and CaT regions, respectively.

We next compare our best \sigs\ values to literature measurements of BASS AGN from other studies (\autoref{fig:dr2_litcomp}) to understand the importance of varying instrumental setups, slit sizes and apertures, fitting regions and software, and templates.  We focus on comparisons with large studies of velocity dispersions for comparison with significant sample sizes (N$>$50) with several overlapping AGN in our sample (N$>$5).  This includes an early study of 85 AGN \citep{Nelson:1995:67} with low spectral resolution ($\sigma_{\mathrm{inst}}$=80--230\,\kmpssh), an atlas of CaT fits \citep{Garcia-Rissmann:2005:765} for 78 AGN  ($\sigma_{\mathrm{inst}}$=60\,\kmpssh), and 428 dwarf seyfert nuclei \citep{Ho:2009:398} from a Palomar Survey  ($\sigma_{\mathrm{inst}}$=42--118\,\kmpssh).  We also include overlapping AGN in the Hobby-Eberly Telescope Massive Galaxy Survey \citep[HETMGS,][]{vandenBosch:2015:10}, which is an optical long-slit spectroscopic survey of 1022 massive galaxies (and includes AGN) using the 10m Hobby-Eberly Telescope at McDonald Observatory ($\sigma_{\mathrm{inst}}$=108--180\,\kmpssh).  We limit our comparison to measurements of similar quality ($\Delta\sigs<20$\,\kmpssh) and if an AGN was observed by more than one of these studies, we use the one with the lowest error in velocity dispersion.   The final list includes 51 overlapping BASS AGN taken from these four studies.  The literature sample tends to have somewhat larger errors (other than the 10m HETMGS sample), whereby the BASS median error is 4\,\kmps\ compared to 9\,\kmps\ in the corresponding literature sample.  In 43/51 cases, the BASS measurements have equivalent or lower estimated errors.

We find a scatter of 28\,\kmps (again, in an rms sense), which is very similar to previously published comparisons between sets of \sigs\ measurements \citep[e.g., 28\,\kmpssh,][]{Ho:2009:1}.  The corresponding MAD we find is, again, smaller, at 16\,\kmpssh.  There are three (ID=573, 245, 548) significant outliers (e.g. $>$30\,\kmps\ offset after including 1$\sigma$ error bars) compared to a large study of massive galaxies by \citet{vandenBosch:2015:10}.  Two of the three outliers (ID=573 and 548) have some of the largest velocity dispersion errors (19 and 21\,\kmpssh, respectively) of the BASS sample.  The most extreme outlier is ID=548 (NGC 3718), with an offset of 91\,\kmps\ from the reported literature value.  In terms of outliers, when comparing their survey of dwarf AGN to the literature, \citet{Ho:2009:1} found 2\% (5/286) with offsets of more than 80\,\kmpssh, while our sample has only ID=548 2\% (1/51), so the number of outliers is quite similar percentagewise.

A summary of the comparisons between different overlapping samples of \sigs\ measurements is provided in \autoref{tab:vdispcomp}.  
In addition to the comparisons with the secondary spectra, within the best spectra, and with the literature, we also look at a subsample of the best spectra in the top three quartiles of the best measurements, with $\Delta \sigs{<}11$\,\kmpssh, to better exclude unreliable outliers.  In general, we do not find significant systematic offsets between samples, with the medians differing by at most $8$\,\kmps or ${<}1$\,\kmps when using the best spectra.

Given the large number of duplicate velocity dispersion measurements it is possible to estimate the systematic scatter beyond the errors we report for each measurement to check whether the computed uncertainties are underestimated or overestimated. While there are several possible functional forms for the measurement error distribution and the intrinsic scatter in measurements of \sigs, past studies have generally found that the error distribution does not significantly change the scatter \citep[see, e.g.  Appendix B in][]{Gultekin:2009:198}; we therefore prefer the Gaussian error distribution with Gaussian scatter, as it is straightforward.  With this assumption, we have run a bootstrap simulation assuming the 1$\sigma$ equivalent uncertainties of our \sigs\ measurements, provided by the \ppxf\ (resampling) procedure, to generate a distribution of 1000 Gaussian errors within the two samples and estimated the corresponding rms and MAD. We then compared the simulated rms and MAD, to the actual rms and MAD, and found that the intrinsic scatter is larger than reported based on individual measurements.  We can then subtract the simulated rms (or MAD) from the measured rms (or MAD) in quadrature to determine the additional intrinsic scatter.  We find that our errors between samples are underestimated by 16-31\,\kmps in rms and 8-18\,\kmpssh.  The underestimate in error scales with the predicted error, with the top three quartiles in the best spectra showing the smallest underestimates of the error at 14\,\kmps in rms and 8\,\kmps in MAD.  
This systematic scatter is likely driven by a combination of several factors but is not, however, due to \ppxf-based error estimation, as the comparisons include different observations, spectral regions, galaxy apertures, and spectral templates.  However, we stress that, when making a comparison, it is important to consider this increased scatter (on the order of 8-18 or 16-31\,\kmps\ rms when considering outliers).

A further comparison with the \sigs\ measurements in BASS DR1 is provided in Appendix D.

\begin{deluxetable*}{llrrrrrrrrr}
\tabletypesize{\footnotesize}
\tablecaption{Velocity Dispersion Comparisons \label{tab:vdispcomp}}
\tablehead{
\colhead{S1} & \colhead{S2} & \colhead{N} &       \colhead{$\langle\Delta \sigs\rangle$} & \colhead{rms} & \colhead{rms Pred.}&
\colhead{rms$+$} & \colhead{Med.} & \colhead{MAD} &       \colhead{MAD Pred.} & \colhead{MAD$+$}\\
\colhead{} & \colhead{} & \colhead{} &       \colhead{(\kmpssh)} & \colhead{(\kmpssh)} & \colhead{(\kmpssh)}&
\colhead{(\kmpssh)} & \colhead{(\kmpssh)} & \colhead{(\kmpssh)} &       \colhead{(\kmpssh)} & \colhead{(\kmpssh)}\\
\colhead{(1)}& \colhead{(2)}& \colhead{(3)}& \colhead{(4)}&\colhead{(5)}& \colhead{(6)}& \colhead{(7)}& \colhead{(8)}& \colhead{(9)}& \colhead{(10)}& \colhead{(11)}
}
\startdata
3880--5550 Best&3880--5550 Secondary&78&-7.5&27.1&14.5&22.9&-6.0&19.5&8.0&17.8\\
CaT Best&CaT Secondary&101&-2.3&31.2&13.8&28.0&0.0&14.0&7.1&12.1\\
3880--5550 Best&CaT Best&266&-2.1&26.7&12.9&23.5&-0.5&13.5&6.1&12.1\\
3880--5550 Best Q1--Q3&CaT Best Q1--Q3&162&0.2&16.3&8.0&14.1&1.0&9.5&4.4&8.4\\
Composite Best&Literature&51&-5.1&28.4&13.2&25.1&1.0&16.0&6.9&14.5\\
\enddata
\tablecomments{The columns are as follows. (1) Primary measurement sample used from the best spectra for comparison; Q1-Q3 indicates the upper 3 quartiles in measurement error in the best spectra that also have a secondary measurement region. (2) Measurement sample used for comparison. (3) Number of measurements in each sample.  (4) Mean offset between sample measurements. (5)--(7) The rms of the sample, predicted rms based on individual errors using bootstrapping, and additional rms based on subtraction of the predicted rms from the rms.  (8) Median offset.  (9)--(11) Median absolute deviation (MAD) of the sample, predicted MAD based on individual errors using bootstrapping, and additional MAD based on subtraction of the predicted MAD from the MAD.}
\end{deluxetable*}

\begin{figure*} 
\centering
\includegraphics[width=0.49\textwidth]{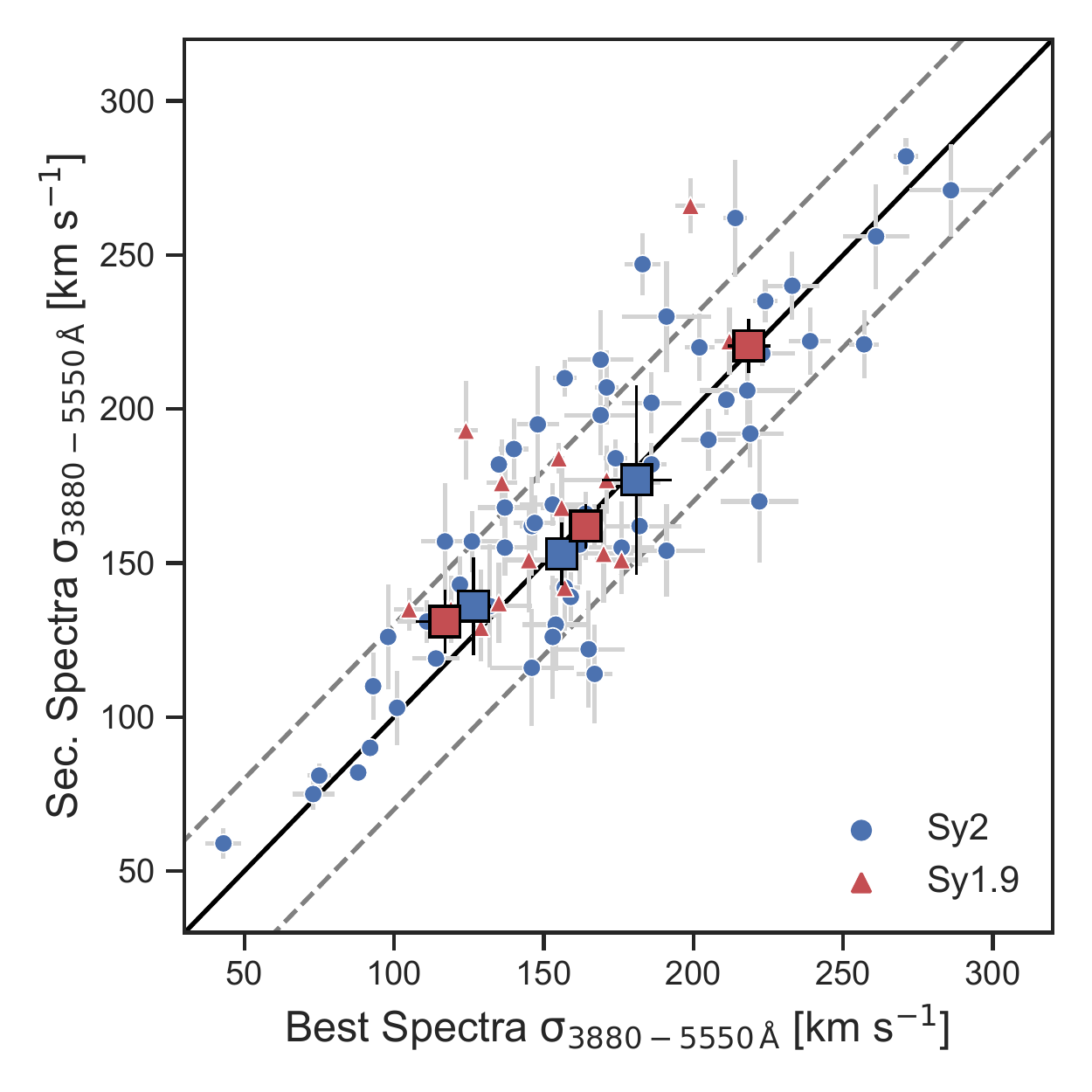}
\includegraphics[width=0.49\textwidth]{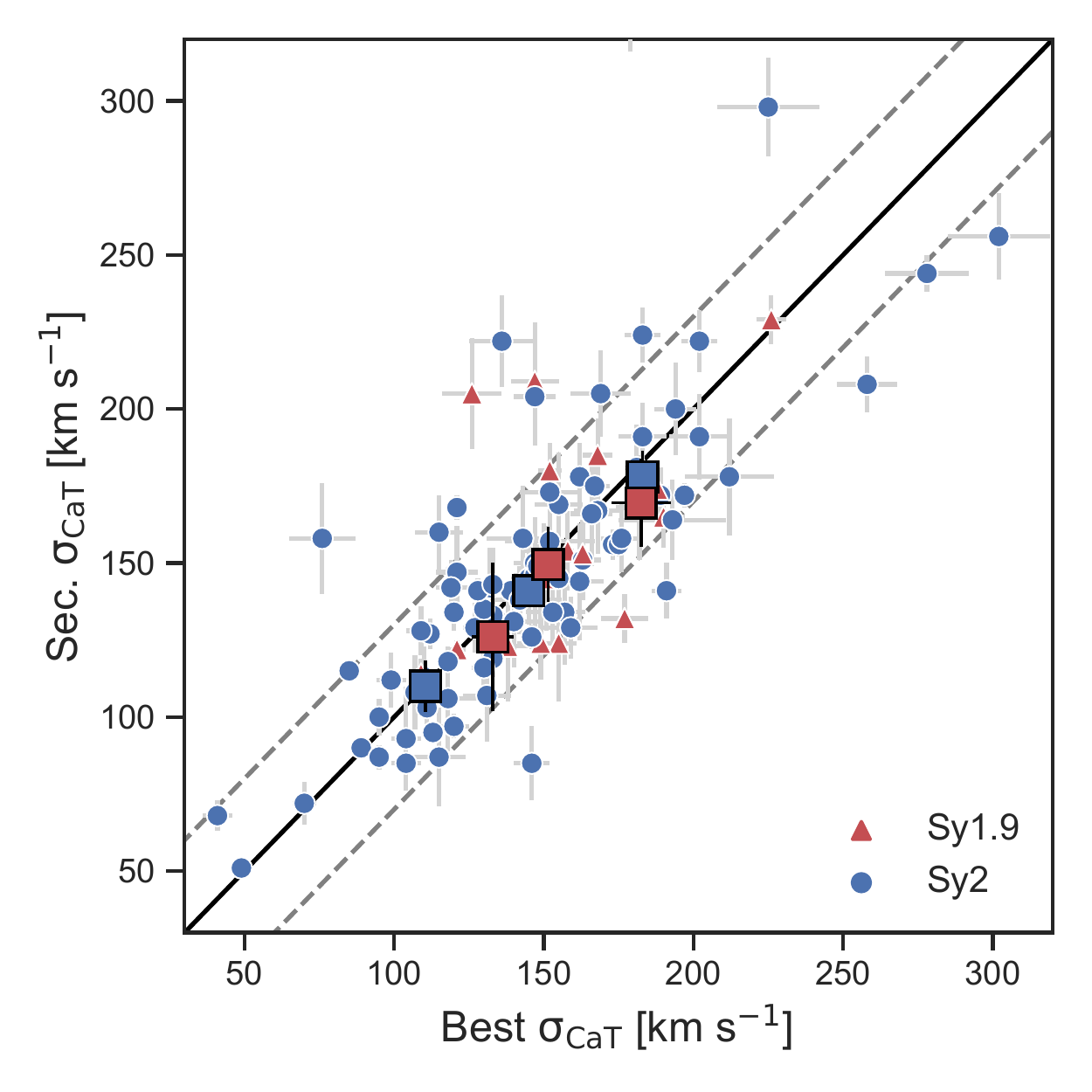}
\includegraphics[width=8cm]{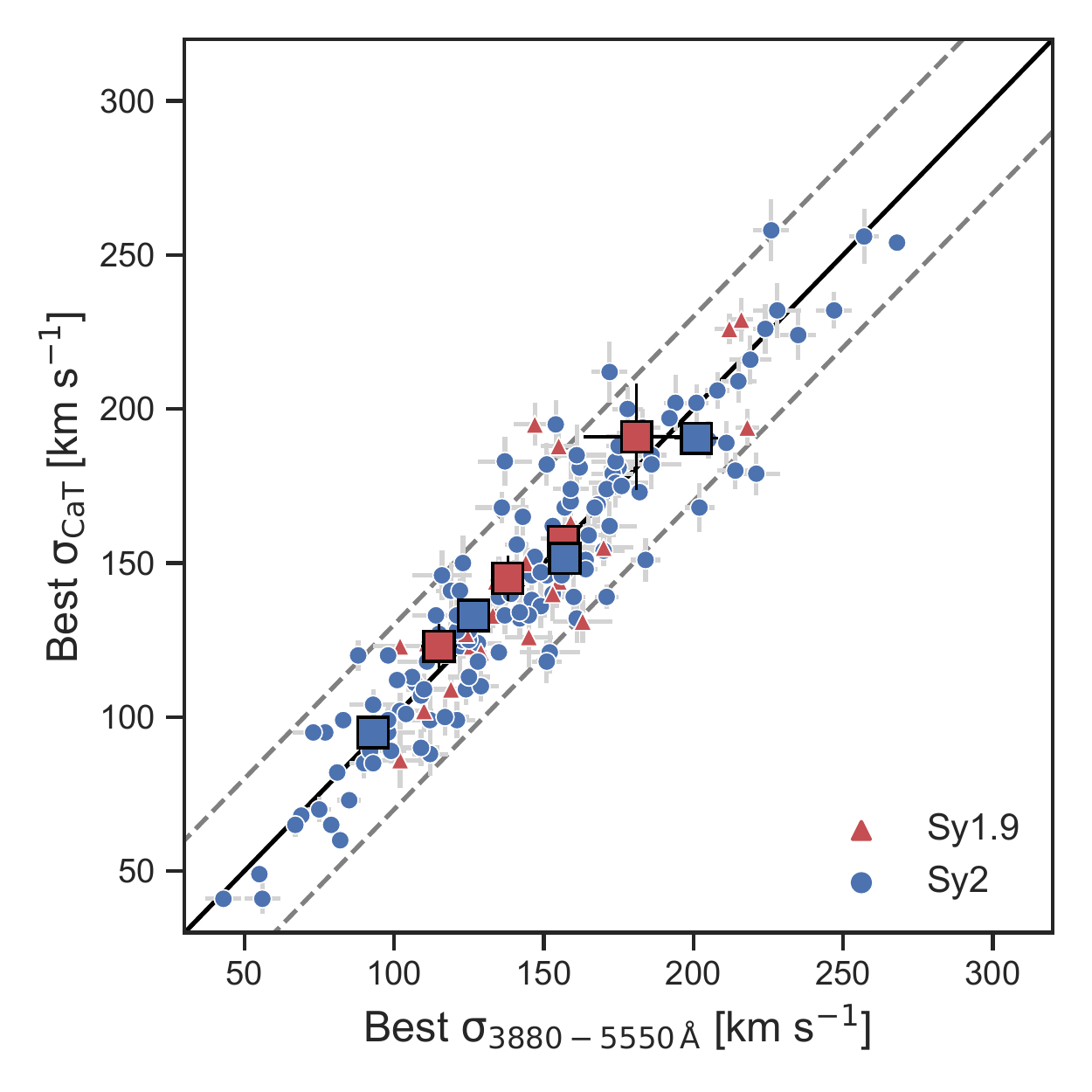}
\caption{{\em Top Left}: velocity dispersion measurements from the best spectra compared to a lower-quality different secondary spectra in the blue region (3880-5550\,\AA).  Here Sy\,1.9 (with broad \halpha) is shown by red triangles and Sy\,2 is shown by blue circles.  Error bars (at $1\sigma$) are shown in gray.    A solid black line indicates the one-to-one relation with dashed gray lines indicating offsets of 30\,\kmpssh.   The large squares indicate the binned medians for each subclass.  Error bars on the plotted median values are equivalent to 1$\sigma$ and calculated based on a bootstrap procedure with 100 realizations. The bin sizes were constructed to have equal numbers of sources in each bin.   {\em Top Right}: comparison between the velocity dispersion measurements from the best spectra and the secondary spectra in the CaT region (8350--8730\,\AA). {\em Bottom Right}:  comparison between the 3880--5550\,\AA\ and CaT within the single same best spectrum.    }
\label{fig:dr2_vdispcomp}
\end{figure*}

\begin{figure*} 
\centering
\includegraphics[width=0.49\textwidth]{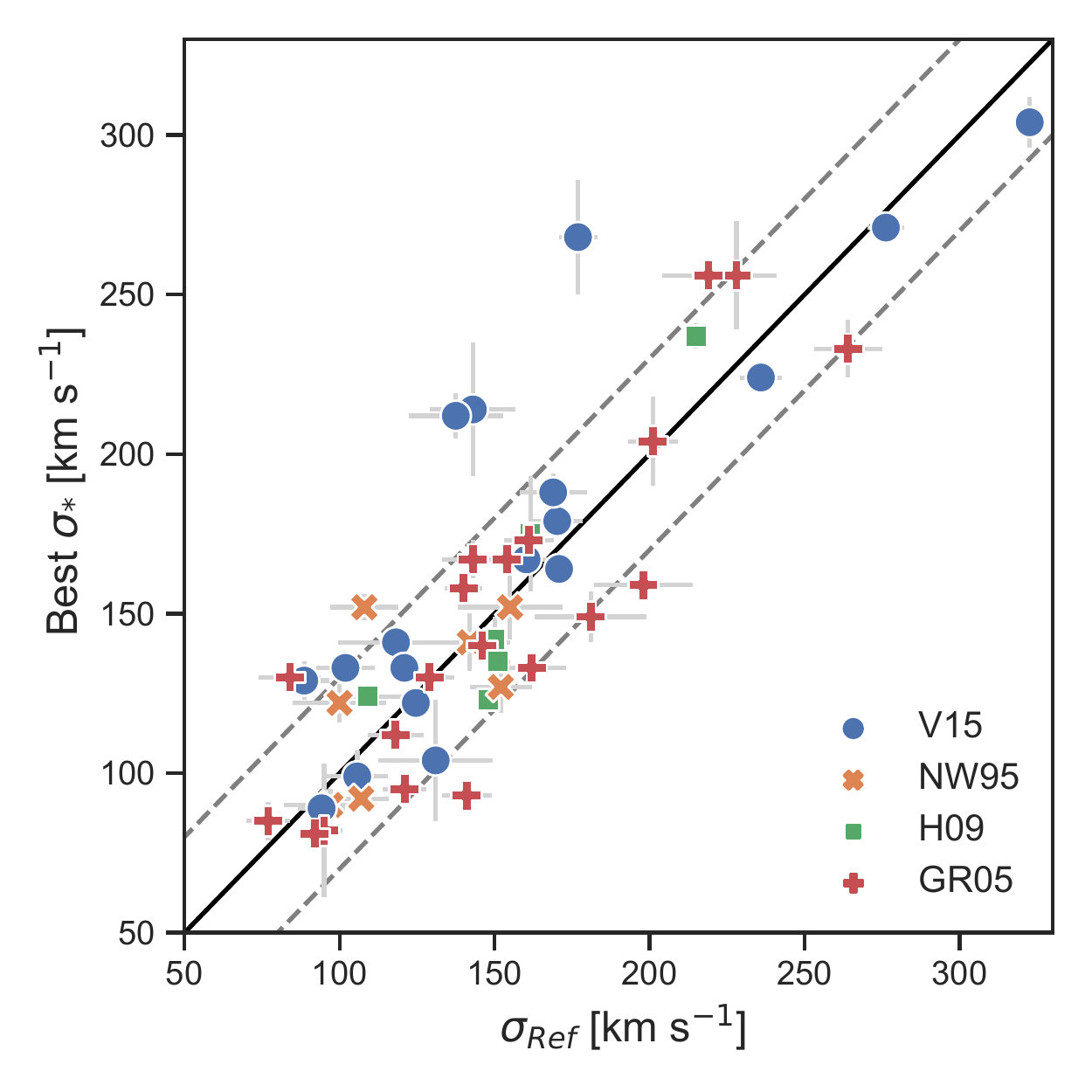}
\caption{Comparison between the velocity dispersion measurements from the best spectra and the literature values for the same AGN.  Error bars (at $1\sigma$) are shown in gray.    A black line indicates the one-to-one relation, with dashed gray lines indicating offsets of 30\,\kmpssh.  The comparison samples are as follows: H09=\citet{Ho:2009:1}; GR05=\citet{Garcia-Rissmann:2005:765}; V15=\citet{vandenBosch:2015:10}; and NW95=\citet{Nelson:1995:67}.   }
\label{fig:dr2_litcomp}
\end{figure*}


\subsection{Sample (In)completeness}
\label{sec:completeness}

The set of BASS DR2 AGN with \sigs\ measurements is completely free from biases related to inadequate instrumental resolution. This is thanks to our strategy of repeated targeting with higher spectral resolution setups in both the northern and southern hemispheres (with Palomar/DBSP in the north and VLT/X-Shooter or SOAR/Goodman in the south), providing an instrumental resolution of $\sigma_{\rm inst}=19-27$\,\kmpssh.  
Still, about 9\% of the Sy\,1.9/Sy\,2 sources in BASS DR2  (34/393) lack a stellar velocity dispersion measurement.\footnote{This excludes the \Nbonusvdisp\ AGN in the bonus sample, drawn from the  105 month BAT catalog, where follow-up optical spectroscopy is ongoing.}
A complete list of the sources that lack a \sigs\ measurement is provided in Appendix E (\autoref{tab:vdispfail}).  
The main contribution to this (minor) incompleteness  comes from AGN observed within the Galactic plane, where high optical extinction makes high-quality spectroscopy and \sigs\ measurements more difficult.  

Focusing on the subset of BASS DR2 sources that are seen outside the Galactic plane ($\lvert b\rvert{>}10^{\circ}$), the \sigs\ completion rate rises to 97\% (326/335).  Further focusing on $z<0.1$ AGN, the completion rate is 99\% (298/301), with only three AGN without high-quality measurements.  For somewhat higher-redshift sources ($0.1{<}z{<}0.59$) the completion rate drops to 82\% (28/34), driven by the limited S/N achievable within our longest exposures and the inability to use the CaT region due to telluric features.  
Among lower-redshift AGN outside the Galactic plane, the three obscured AGN without \sigs\ are largely mergers or emission line-dominated AGN. To demonstrate the challenges imposed by such systems, consider the dual AGN in Mrk 463E, which is completely dominated by emission lines with strong outflows \citep{Treister:2018:83}.

\subsection{Black Hole Mass Estimates}
\label{sec:mbh}

A key goal of the BASS project is to generate a set of BH mass (\mbh) measurements for the entire sample of BASS AGN. Only a small number of high-quality literature measurements from reverberation mapping, spatially resolved gas or stellar dynamics, and/or H$_2$O megamasers exist for the BASS AGN.  As part of this study, we therefore provide estimates of the BH masses of the obscured AGN where velocity dispersion measurements can be used to estimate the BH mass following the well-established SMBH--host correlations.

For simplicity and consistency with the BASS DR1, we calculated \mbh\ using the best-fit $\mbh{-}\sigs$ relation from \citet{Kormendy:2013:511}:
\begin{equation}\label{eq:mbh_sigs}
\log\left(\mbh/\Msun\right) = 4.38\times \log\left(\sigs/200\,\kms\right) + 8.49 \,\, .
\end{equation}
The slope of this relation is consistent with that found by \cite{Gultekin:2009:198}, and both are considerably shallower than the slope of the relation derived by \citet{McConnell:2013:184}, who reported a value of 5.64.  The effect of using different scaling relations for \mbh\ is on order of 0.1-0.3 dex at the median velocity dispersion \citep[e.g., \lmbh=8 vs. \lmbh=7.7--7.9, \sigs=155\,\kmpssh,][]{McConnell:2013:184,Woo:2013:49}. The difference is somewhat larger at lower velocity dispersions \citep[e.g., at 100\,\kmps \lmbh=7.2 vs. \lmbh=6.7,][]{McConnell:2013:184}.  Alternatively, using the more recent relations from \citet{Greene:2020:257} for all galaxies including limits, the corresponding \mbh\ value is larger at 100\,\kmps corresponding to \lmbh=7.5.

We provide a full list of the resulting \Nvdisp\ \mbh\ estimates from the best \sigs\ measurements in \autoref{tab:mbhbest}.  For those AGN where acceptable fits for both the 3880--5550\,\AA\ and CaT spectral regions are available, we also provide a weighted (mean) \sigs\ and corresponding weighted \mbh\ measurement.
Using our large sample of measurements, we can understand how the real uncertainties on the \sigs\ measurements contribute to the error in inferring \mbh.  This is important because the highly nonlinear nature of the $\mbh-\sigs$ scaling relation makes it difficult to simply add (in quadrature) the 8-18\,\kmps systematic error, as the same kilometer per second error will be much larger for low velocity dispersions.

We summarize several comparisons of inferred SMBH masses---between spectral regions, between best and secondary spectra, and to literature values---in \autoref{tab:mbhcomp}.  We find very small systematic offsets in the median ($\leq$0.07 dex) \mbh\ measured between different samples and somewhat larger offsets in the means due to outliers, though still $\leq$0.1 dex.  In order to understand the scatter, we performed a simulation similar to the one done for velocity dispersions but using the lognormal distributions of \mbh\ and errors inferred from \sigs\ and associated errors.  We used similar Gaussian distributions based on the 1$\sigma$ errors to generate a grid of 1000 distributions within the two samples and estimated the rms and MAD.   We find that there is additional scatter on the order of 0.2--0.4 dex in rms or 0.11--0.21 in MAD. 
We note that these are consistent with, and indeed somewhat smaller than, the intrinsic scatter around even the tightest observed $\mbh-\sigs$ relations \cite[e.g.,][]{Gultekin:2009:198,Kormendy:2013:511}.

To calculate AGN bolometric luminosities (\Lbol) and Eddington ratios ($\log L/L_{\rm Edd}$, \lledd), we follow the prescription used throughout the BASS DR2  \citep[see][]{Koss_DR2_catalog}. In brief, \Lbol\ is calculated from the intrinsic luminosity in the 14--150\,keV range, which is derived in \citet[][see their Table 12]{Ricci:2017:17}.  This calculation was made using the 70 month average Swift-BAT 14--195\,keV spectra plus data below 10\,keV from Swift XRT, XMM-Newton, Chandra, Suzaku, and/or ASCA fit to detailed spectral models.  We follow the prescription used in other BASS DR2 publications \citep{Koss_DR2_catalog} and use the 14--150\,keV emission with a bolometric correction of 8 which is consistent with past studies \cite[e.g.,][]{Vasudevan:2009:1124}. 
The \lledd\ is then calculated as $\lledd=\Lbol / (1.5{\times}10^{38}[\mbh/\Msun])$.
More elaborate bolometric corrections are beyond the scope of the present work, as 
\Lbol\ and \lledd\ are used simply to illustrate the part of parameter space occupied by obscured BASS AGN.

The distributions of SMBH masses and Eddington ratios for the Sy\,1.9 ($N=113$) and Sy\,2 ($N=371$) sources in BASS are shown in \autoref{fig:agntypembh}.  The SMBH mass distributions for the two AGN subclasses are very similar, with medians of $\log(\mbh/\Msun)=8.01\pm0.05$  and $8.00\pm0.03$ for Sy\,1.9 and Sy\,2, respectively (standard deviations of 0.56 and 0.59 dex, respectively).  A Kolmogorov-Smirnov (K-S) test indicates that the samples are not significantly different in their \mbh\ ($p$=0.45).  For Eddington ratios, the distributions are also very similar with $\log\lledd=-1.65\pm0.10$ and $-1.72\pm0.03$ for Sy\,1.9 and Sy\,2, respectively (standard deviations of 0.77 and 0.64 dex, respectively).  A K-S test indicates that the samples are not significantly different in their Eddington ratios ($p$=0.12).


\autoref{fig:edd_ratio_limits} shows the distribution of our AGN in the \sigs\- (or \Mbh) vs. redshift parameter space.  The lowest \sigs\ (and corresponding \Mbh) that we could have measured, given our instrumental setups, is highlighted. 
Importantly, we also illustrate the limiting values of \Mbh\ (and \sigs) corresponding to various values of \lledd, given the BAT flux limit, to show the range of \lledd\ that our study is sensitive to as a function of redshift.  
This figure is interesting for both the distribution of sources and the parameter space where sources are not detected.  While BASS DR2 has the instrumental resolution to detect low-mass, super-Eddington SMBHs (i.e., $\lesssim10^{5}$\,\Msun), we do not find such systems despite having a "complete" sample of AGN, including obscured sources (perhaps missing only highly CT AGN, $\NH{>}10^{25}$\,\nhunit).  Additionally, there is only a very small volume where the BAT survey detects SMBHs with extremely low Eddington ratios, $\lesssim 10^{-5}$ (i.e., at $z{<}0.007$). The detection of such objects would allow the study of the emission properties of advection-dominated accretion flows  \cite[][]{YuanNarayan14}.  However, these radiatively inefficient accretion flows result in broadband spectral energy distributions that are thought to be markedly different from those characterizing standard, thin-disk accretion and are expected to be intrinsically fainter, making them even harder to detect with Swift/BAT (see, e.g., \citealt{Ryan:2017:199}).

An elaborate investigation of the distributions of luminosities, BH masses, and Eddington ratios of all BASS DR2 AGN, relying in part on the \sigs\ measurements presented here and correcting for various selection effects that are present in the survey, is described in a companion paper \citep{Ananna_DR2_XLF_BHMF_ERDF}.

\begin{deluxetable*}{llllll}
\tabletypesize{\scriptsize}
\tablecaption{Best Velocity Dispersion and \mbh\ Measurements \label{tab:mbhbest}}
\tablehead{
\colhead{BAT ID} & \colhead{Region} & \colhead{\sigs} &       \colhead{$\log$ \mbh} & \colhead{\sigs\ Weighted} & \colhead{\mbh\ Weighted}\\
\colhead{} & \colhead{} & \colhead{(\kmpssh)} &       \colhead{($\log$ \Msun)} & \colhead{(\kmpssh)} & \colhead{($\log$ \Msun)}\\
\colhead{(1)}& \colhead{(2)}& \colhead{(3)}& \colhead{(4)}&\colhead{(5)}& \colhead{(6)}
}
\startdata
     1 &   3880--5550 &     126$\pm$6 &  7.61$\pm$0.09 &        125$\pm$5 &    7.60$\pm$0.07 \\
     4 &   3880--5550 &     146$\pm$6 &  7.89$\pm$0.07 &        123$\pm$4 &   7.56$\pm$0.06 \\
     7 &   3880--5550 &    205$\pm$12 &  8.54$\pm$0.11 &                  &                 \\
    10 &   3880--5550 &    253$\pm$14 &   8.94$\pm$0.10 &                  &                 \\
    13 &     CaT &     131$\pm$8 &  7.68$\pm$0.12 &                  &                 \\
    17 &     CaT &     147$\pm$7 &   7.90$\pm$0.09 &                  &                 \\
    20 &   3880--5550 &    294$\pm$22 &  9.22$\pm$0.14 &                  &                 \\
    24 &     CaT &     121$\pm$7 &  7.54$\pm$0.11 &        132$\pm$6 &    7.7$\pm$0.08 \\
    25 &   3880--5550 &     134$\pm$6 &  7.73$\pm$0.08 &        141$\pm$6 &   7.83$\pm$0.07 \\
    28 &     CaT &     200$\pm$3 &  8.49$\pm$0.03 &                  &                 \\
    29 &     CaT &    135$\pm$17 &  7.74$\pm$0.23 &                  &                 \\
    31 &   3880--5550 &      83$\pm$1 &  6.82$\pm$0.03 &         88$\pm$1 &   6.94$\pm$0.02 \\
    32 &   3880--5550 &    269$\pm$29 &  9.05$\pm$0.19 &                  &                 \\
    33 &     CaT &     82$\pm$21 &   6.80$\pm$0.44 &                  &                 \\
    37 &   3880--5550 &    195$\pm$10 &  8.45$\pm$0.09 &        191$\pm$8 &   8.41$\pm$0.08 \\
    44 &     CaT &     160$\pm$7 &  8.06$\pm$0.08 &                  &                 \\
    49 &     CaT &     190$\pm$7 &  8.39$\pm$0.07 &                  &                 \\
    50 &     CaT &     104$\pm$5 &  7.24$\pm$0.09 &        101$\pm$4 &   7.18$\pm$0.08 \\
    53 &   3880--5550 &     140$\pm$5 &  7.82$\pm$0.07 &        143$\pm$4 &   7.85$\pm$0.05 \\
    55 &   3880--5550 &     151$\pm$9 &  7.96$\pm$0.12 &        159$\pm$8 &   8.06$\pm$0.09 \\
    57 &   3880--5550 &     239$\pm$7 &  8.83$\pm$0.06 &        242$\pm$6 &   8.85$\pm$0.05 \\
    58 &     CaT &     118$\pm$5 &  7.49$\pm$0.08 &                  &                 \\
    62 &     CaT &     131$\pm$9 &  7.68$\pm$0.13 &                  &                 \\
    63 &   3880--5550 &     106$\pm$1 &  7.29$\pm$0.02 &        109$\pm$1 &   7.33$\pm$0.01 \\
    64 &     CaT &     168$\pm$5 &  8.16$\pm$0.05 &        160$\pm$4 &   8.07$\pm$0.05 \\
    65 &   3880--5550 &    123$\pm$16 &  7.56$\pm$0.23 &                  &                 \\
    70 &     CaT &     121$\pm$7 &  7.54$\pm$0.11 &                  &                 \\
\enddata
\tablecomments{The columns are as follows. (1) Catalog ID in the BAT survey.\footnote{\url{https://swift.gsfc.nasa.gov/results/bs70mon/}} (2) Region having the lowest error in velocity dispersion measurement. (3)   Best velocity dispersion measurement and associated (1$\sigma$) error. (4) Inferred \mbh\ based on \sigs\ using \citealt{Kormendy:2013:511} \mbh--\sigs\ relation and associated error.  (5)  Weighted velocity dispersion as the average of the 3880--5550\,\AA\ and CaT dispersions, weighted by their respective error bars. \autoref{tab:mbhbest} is published in its entirety in machine-readable format.  A portion is shown here for guidance regarding its form and content.}
\end{deluxetable*}

\begin{deluxetable*}{llrrrrrrrrr}
\tabletypesize{\footnotesize}
\tablecaption{\mbh\ Comparison Using Different \sigs\ Measurements \label{tab:mbhcomp}}
\tablehead{
\colhead{S1} & \colhead{S2} & \colhead{N} &       \colhead{$\langle\Delta\log\mbh\rangle$} & \colhead{rms} & \colhead{rms Pred.}&
\colhead{rms$+$} & \colhead{Med.} & \colhead{MAD} &       \colhead{MAD Pred.} & \colhead{MAD$+$}}
\startdata
3880--5550 Best&3880--5550 Secondary&78&-0.09&0.31&0.17&0.26&-0.07&0.23&0.09&0.21\\
CaT Best&CaT Secondary&101&-0.02&0.37&0.18&0.33&0.00&0.20&0.09&0.17\\
3880--5550 Best&CaT Best&266&-0.02&0.33&0.16&0.28&0.00&0.18&0.08&0.16\\
3880--5550 Best Q1--Q3&CaT Best Q1--Q3&162&0.01&0.23&0.11&0.20&0.01&0.13&0.06&0.11\\
Composite Best&Literature&51&0.05&0.36&0.18&0.31&-0.02&0.22&0.09&0.20\\
\enddata
\tablecomments{Column descriptions same as \autoref{tab:vdispcomp}, unless otherwise noted. All \mbh-related quantities are listed in dex.}
\end{deluxetable*}

\begin{figure*} 
\centering
\includegraphics[width=0.49\textwidth]{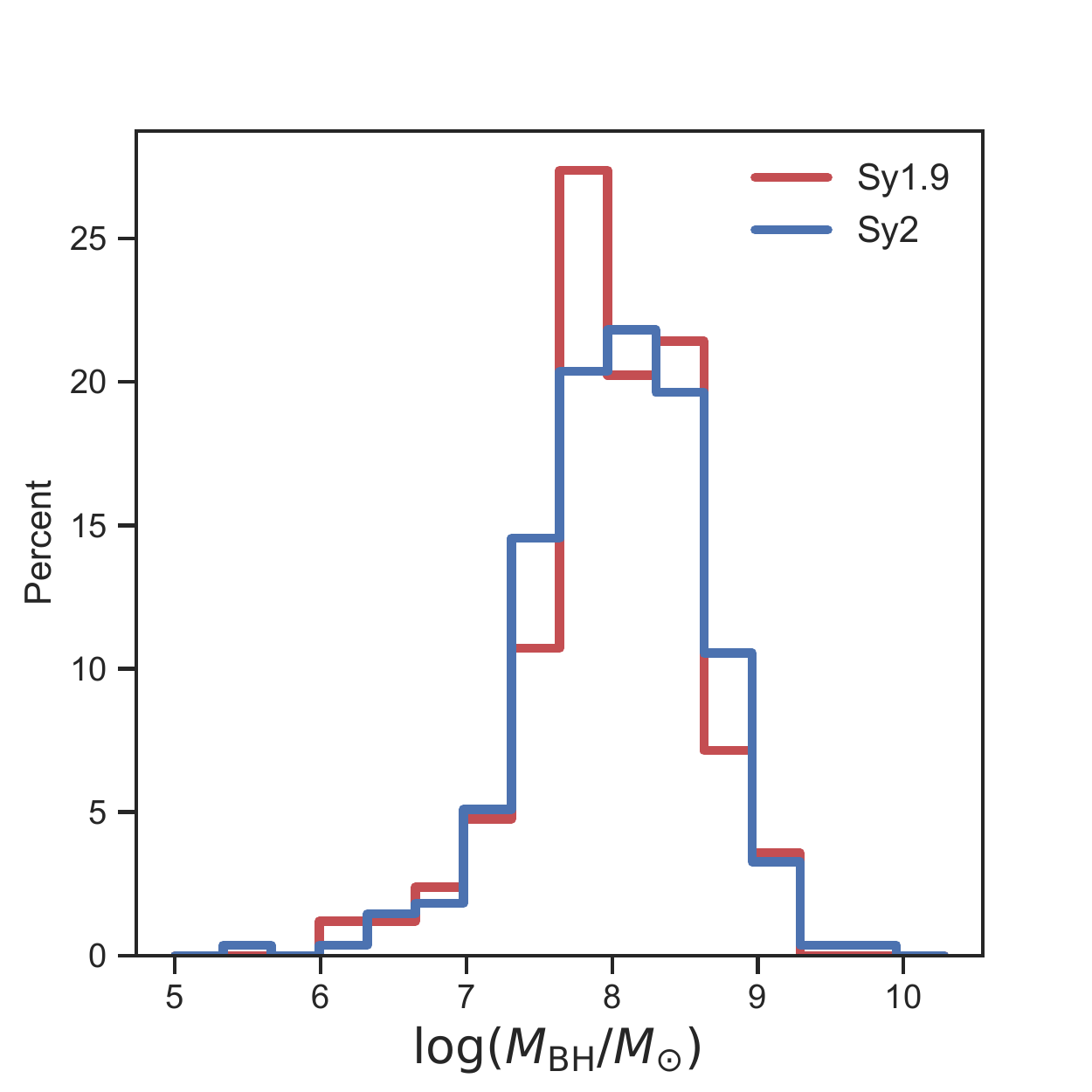}
\includegraphics[width=0.49\textwidth]{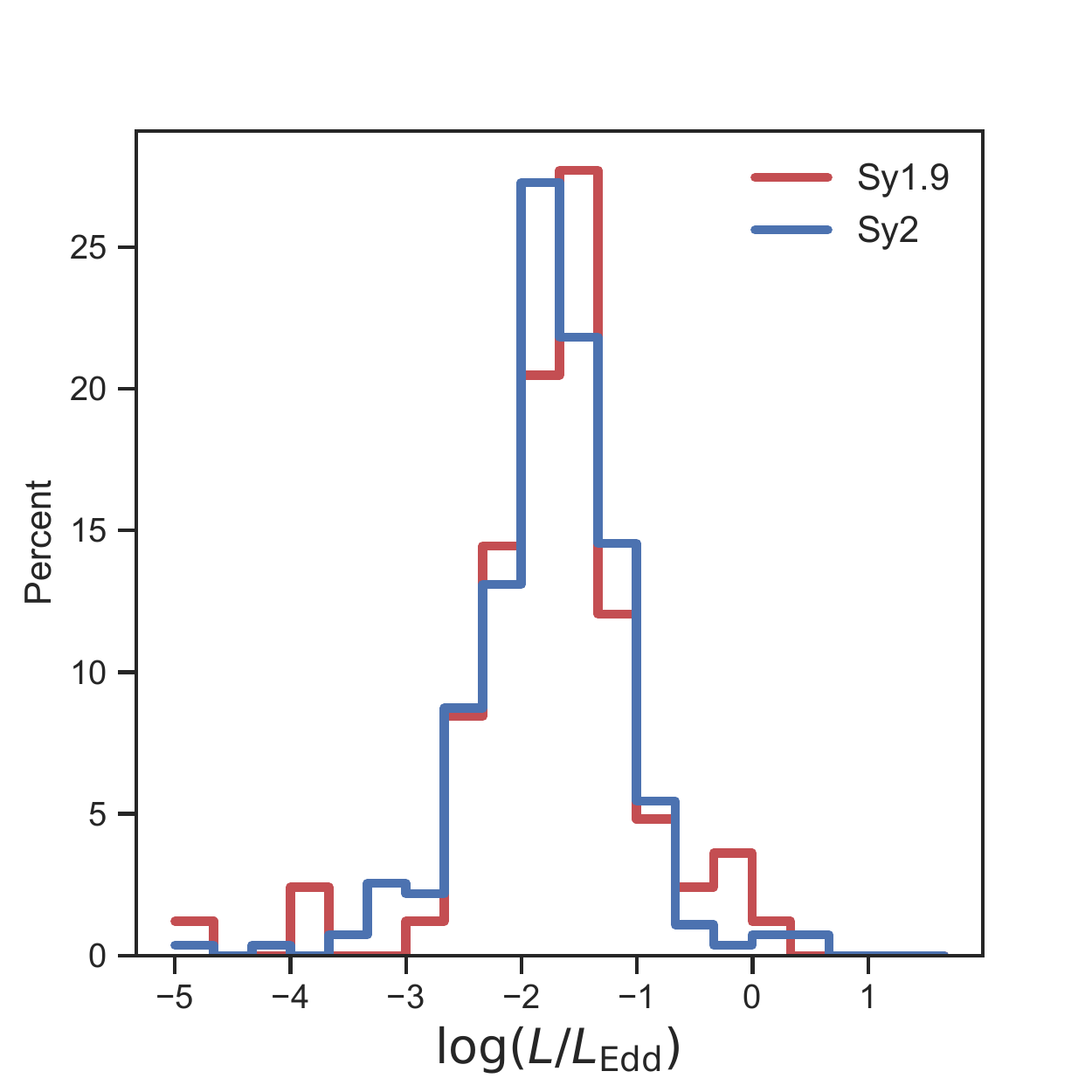}
\caption{Frequency distributions of SMBH masses and Eddington ratios for the Sy\,1.9 ($N$=113) and Sy\,2 ($N$=371).}
\label{fig:agntypembh}
\end{figure*}

\begin{figure*} 
\centering
\includegraphics[width=1\textwidth]{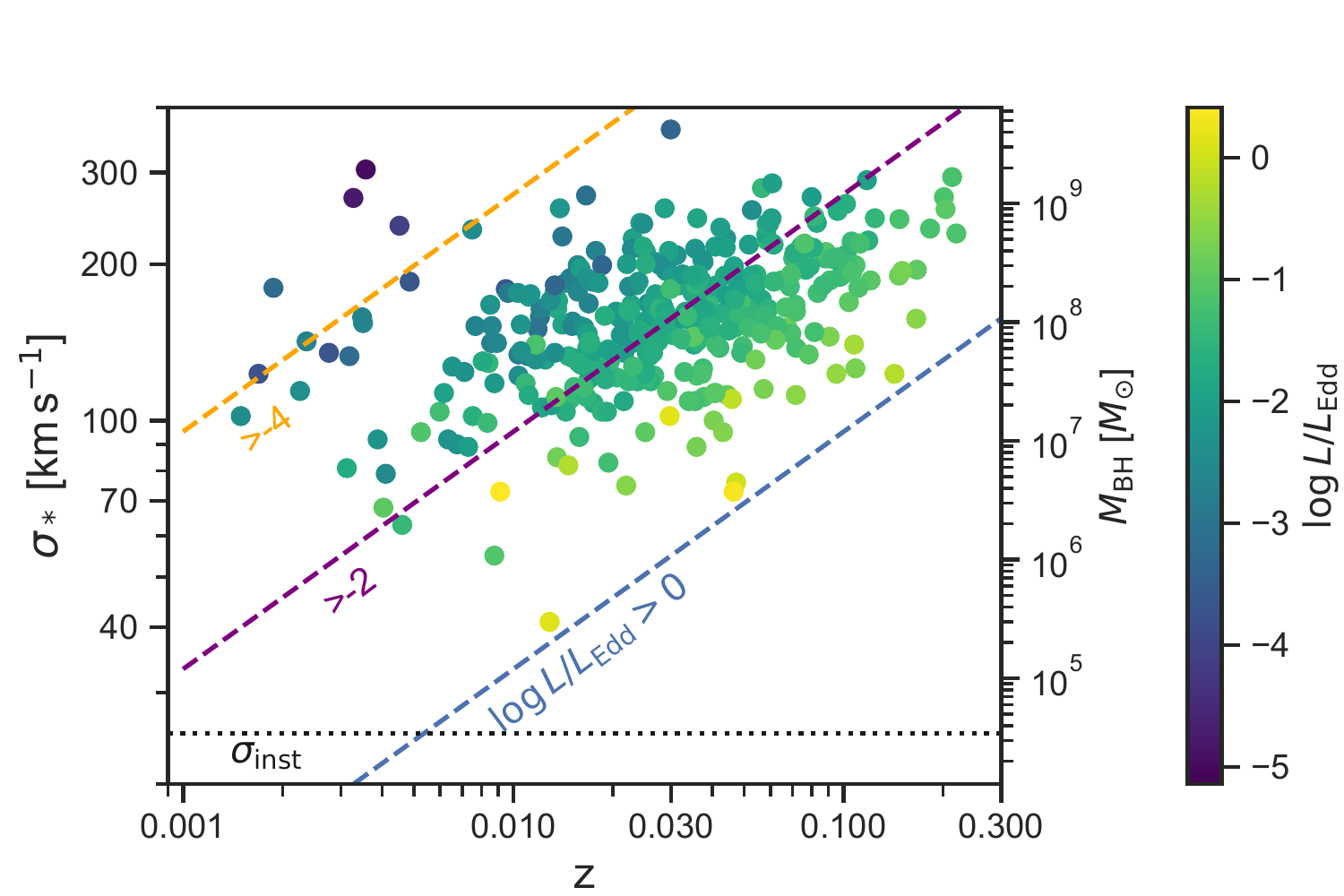}
\caption{Plot of the distribution of \sigs\ and \mbh\ (based on the \citealt{Kormendy:2013:511} relation) with redshift.  The color bar indicates the Eddington ratio (\lledd) of the AGN.  Error bars in \sigs\ span 1-29\,\kmpssh, with a median of 7\,\kmpssh. The effective sensitivity limits on the ER due to the survey flux limits are illustrated with diagonal dashed lines.  For instance, the flux limit at $z{=}0.01$ (\lbol${=}1.8{\times}10^{43}$\ergps), corresponds to an \lledd=1 (blue dashed line) source with \mbh=$1.5\times10^5$\,\Msun\ or \sigs=35\,\kmpssh.    Therefore, any sources detected at z=0.01 with \sigs{<}35\,\kmps will always be super-Eddington because of the survey flux limits.  The black dotted line indicates the instrumental limits of the survey for velocity dispersion measurements.  One higher-redshift AGN, BAT ID=1204 (\sigs=364\,\kmpssh, ER=0.20, $z=0.597$), is not shown.}
\label{fig:edd_ratio_limits}
\end{figure*}


\section{Discussion}
\label{sec:discussion}

As a nearly obscuration-free tracer of SMBH accretion up to CT levels, the 14-195 keV ultrahard X-rays surveyed by BAT/BASS provide a useful tracer of nearby obscured AGN activity over the whole sky.  The ${>}$99\% completeness of our spectroscopic coverage, combined with very high spectral resolution ($ \sigma_{\rm inst}\sim25$\,\kmpssh) to effectively resolve even the narrowest absorption features for systems with small BHs, make it a unique legacy sample for future AGN studies, where a similar sensitivity, completeness, and/or resolution may be achievable.  Here we review BASS in comparison to other AGN surveys, provide comparisons with direct measurements of SMBH masses, and investigate the feasibility of future programs of dynamical measurements and modeling to better measure the SMBH distribution of these AGN.

\subsection{Comparison to Other Surveys }

In \autoref{fig:comp_SDSS}, we compare our BASS AGN \sigs\ measurements to those measured for spectroscopically selected star-forming galaxies (SFGs) and narrow-line AGN in the SDSS drawn from the OSSY catalog \citep{Oh:2011:13}, all restricted to $z{<}0.1$.  
The identification of (narrow-line) AGN and SFGs in SDSS is based on strong emission line ratio diagnostics \citep[e.g., \OIII/\Hbeta\ vs.~\NII/\Halpha][]{Baldwin:1981:5,Veilleux:1987:295,Kewley:2001:37}.  The majority of SFGs and a significant fraction of the AGN in the SDSS have \sigs\ that are below the nominal instrumental resolution of SDSS; thus, their true distribution below 70\,\kmps is highly uncertain.  Nearly all of the BASS AGN have velocity dispersions that are larger than the average for SDSS-selected AGN (by $\sim1.5\times$, i.e. 150\,\kmps vs. 100\,\kmpssh). This is consistent with the fact that the SDSS narrow-line AGN population leans toward the more numerous, lower stellar mass galaxies compared to the BAT-selected AGN \citep[see, e.g.,  Figure 12 in][]{Koss:2011:57}.  Interestingly, the higher velocity dispersions of the BASS AGN (e.g. $\sim$150\,\kmpssh) are more consistent with the much more luminous type 2 quasars identified in the SDSS  via strong emission line ratio diagnostics  and luminous \OIII\ emission at somewhat higher redshifts  \citep[$z\sim$0.3, $\sim$160\,\kmpssh,][]{Kong:2018:116}.

One surprising result is that, despite the SFGs being roughly a factor 220 more numerous in a similar redshift range, and having low velocity dispersions ($\sigs {\lesssim}70$\,\kmpssh), and thus potentially hosting lower-mass BHs, we do not detect their active counterparts within the Swift/BAT survey. If a significant population of super-Eddington low-\mbh\ AGN exists, they either lie below the flux limits of the BAT survey or are not hard X-ray sources. The $\sim$150\,\kmps\ is more consistent with SFGs at higher redshifts ($0.6<{z}<1$) such as those observed in COSMOS \citep[][]{Straatman_2018}.


%
Looking into the low-\sigs, low-mass end of the distribution for our BASS AGN in more detail, we note that our BASS DR2 sample has six low velocity dispersion AGN that have \sigs\ lower than the SDSS instrumental resolution of 70\,\kmpssh, which translates to BH masses $\mbh\lesssim10^{6.5}$\,\Msun\ (see Eq.~\ref{eq:mbh_sigs}) that are also robustly measured through our highest spectral resolution setups ($\sigma_{\rm inst}\sim25$\,\kmpssh, which translates to $10^{4.5}$\,\Msun).
One of these six sources, NGC 7314, has been the focus of many X-ray variability studies \citep[e.g.][]{Emmanoulopoulos:2016:2413} motivated by the hope of identifying smaller SMBHs, suggested by the extremely low central stellar velocity dispersion. 
We finally compare our sources to some of the lowest-mass dwarf AGN selected from weak broad \Halpha\ line emission identified within the SDSS by \cite{Baldassare:2020:L3}. As \autoref{fig:dwarf_comp} shows, our sources are typically located at lower redshifts ($z{<}0.03$) and/or with higher \sigs. The small corresponding BH masses of the dwarf AGN, if detected above the BAT flux limit, would imply super-Eddington accretion, which may not necessarily be detected in the BAT band. 


Considering the high-\sigs, high-mass end of the (active) galaxy population, in \autoref{fig:massive_comp} we compare the BASS AGN \sigs\ distribution to that of the HETMGS.  The HETMGS survey was designed to find nearby galaxies where the SMBH sphere of influence (SOI) can be resolved by seeing-limited spectroscopy.  This selection scheme means HETMGS includes some small but nearby galaxies that appear sufficiently large on the sky (i.e., due to their small distance). 
We limit our comparison between HETMGS and BASS AGN to $0.02{<}z{<}0.04$, adopting  $z{>}0.02$ to focus only on the most massive galaxies and  $z{<}0.04$ as the HETMGS survey does not extend further than this.  
Interestingly, the typical velocity dispersions in the HETMGS (sub)sample are significantly larger than those of the BASS AGN (mean values of $\sim$275 versus $\sim$150\,\kmpssh, respectively).  This highlights the fact that the BASS AGN are not hosted in the largest bulges known in the nearby universe, which are predominantly in passive galaxies, but rather in an intermediate population with likely lower masses.   
Combining the largest velocity dispersions observed within HETMGS of $\approx$400\,\kmpssh, and the BAT survey flux limit (see \autoref{fig:edd_ratio_limits}), we conclude that BASS is not sensitive enough to detect SMBHs accreting at extremely low rates, $\lledd\lesssim 10^{-5}$, and hence probe the advection-dominated regime of SMBH accretion.

In summary, the BASS AGN tend to avoid both the highest- and lowest-mass systems, as well as ones with super-Eddington accretion, and the selection function is complicated.  For deeper understanding of the intrinsic \mbh\ and \lledd\ distributions, we again refer to the companion dedicated study by \cite{Ananna_DR2_XLF_BHMF_ERDF}.

\subsection{Comparison with Higher Precision Direct Measurements}

The velocity dispersion measurements studied here provide a useful proxy to \mbh\ for large statistical studies of obscured AGN.  However, direct measurements provide much higher accuracy and naturally avoid the usage of empirical scaling relations. For obscured AGN, the relevant direct \mbh\ measurement methods include dynamical gas and/or stellar modeling \citep[e.g.,][]{Kormendy:2013:511} and H$_2$O megamasers \citep[e.g.,][]{Staveley-Smith:1992:725}.
A list of 64 direct measurements for the 70 month BAT AGN was compiled as part of the BASS DR1 \citep{Koss:2017:74} and more recently for the DR2 \citep{Koss_DR2_catalog}, though the majority of masses in these lists were derived from reverberation mapping campaigns of unobscured (broad-line) AGN \citep[e.g.,][]{Bentz:2009:199}, which are not part of the present sample of host-dominated, narrow-line AGN.  
The updated DR2 list in \cite{Koss_DR2_catalog} matches the BASS AGN to compilations of megamaser measurements \citep[e.g.,][]{Kuo:2011:20} and the latest large samples of direct measurements of early- and late-type spiral galaxies \citep[e.g., 145 galaxies in][]{Sahu:2019:155}. Despite the BASS AGN being the brightest X-ray-selected AGN across the sky, there are only 18 of these measurements, including 10 megamasers, four dynamical gas measurements, and four stellar dynamical measurements that overlap with our velocity dispersion sample.    The BAT-detected AGN at the center of the Perseus cluster (Perseus A; aka NGC 1275) also does not have a successful \sigs\ measurement, due to the strong emission lines present in the spectrum. This source does have a dynamical (molecular) gas measurement reported by \cite{Scharwachter:2013:2315} although that paper cautions that the gas mass in the core may be nonnegligible; thus, their measurement could be considered as an upper limit.


In \autoref{fig:directcomp}, we compare the available direct \mbh\ measurements to those derived here through our \sigs\ measurements and the $\mbh{-}\sigs$ relation.  There are clearly large discrepancies among megamasers, where the \sigs-based estimates seem to significantly overpredict \mbh\ (median offset of 0.7 dex).  In contrast, our \sigs-based measurements show better agreement with stellar and/or gas dynamical measurements, although the latter are only available for a small sample.  
These findings are consistent with studies of much larger samples of spiral and early-type galaxies, which have found that spiral galaxies with nonmaser dynamical \mbh\ measurements do not appear to show an offset, but megamasers are significantly offset \citep[e.g., $0.6\pm0.14$ dex;][]{Greene:2016:L32}.  It is still unclear whether this offset in megamasers is because they trace the true \mbh\ distribution, while the nonmaser \mbh\ population is biased because of the inability to resolve the SOI of small SMBHs (${<}10^{7}$\,\Msun).  Or the maser population maybe somewhat offset because of a stronger rotational contribution to the stellar velocity dispersion \citep[e.g.,][]{Caglar:2020:A114}.  Alternatively, the nonmaser distribution could be the true one.  We leave further discussion to works with much larger samples of direct measurements, where these differences can be addressed in a much more comprehensive way.

\subsection{Feasibility of Additional Direct Measurements}

Given the extremely small sample of high-quality megamaser and dynamical measurements (i.e., only 3.7\%, or 18/\Nvdisp\ AGN) and large offsets compared to the few megamaser measurements, it is worth considering how many more AGN-dominated galaxy centers could potentially be spatially resolved to yield BH masses derived from dynamical modeling of gas and/or stars.
For these studies, facilities such as NIR integral field units, assisted by adaptive optics (AO; e.g., VLT/MUSE or Keck/OSIRIS) or submillimeter interferometers (e.g., ALMA or NOEMA), could be considered.

In these dynamical models, to detect the influence of the BH, the region within which the BH gravity dominates over that of the host must be spatially resolved. The size of this region, referred to as the SOI, is commonly given by $r_{\rm SOI} = G\mbh\sigs^{-2}$.  Given a distance, $D$, the apparent size projected on the sky is $\theta_{\rm SOI}=r_{\rm SOI} \, D^{-1}$.
This simple prescription is derived assuming virialized, spherically symetric, and isotropic gas motion in the target galaxy nucleus \citep[see, e.g.,][]{vandenBosch:2015:10}.  Plugging in our $\mbh{-}\sigs$ relation (see \autoref{eq:mbh_sigs}) and rescaling to a size in parsecs results in
\begin{equation}\label{eq:r_soi}
r_{\rm SOI} =  33\,\mathrm{pc} \left(\frac{\sigs}{200\,\kmps}\right)^{2.38} \,\, . 
\end{equation}
As an example, an SMBH with $\mbh=10^{8}$\,\Msun\ (or $\sigs=155$\,\kmpssh) would have $r_{\rm SOI}=18$\,pc, and at the median redshift of the BASS AGN ($z$=0.03),  would have $\theta_{\rm SOI}=$0\farcs03.



We use this definition to determine whether resolving the SOI is feasible for any given galaxy.  There can be significant additional challenges to obtaining measurements for dynamical modeling (e.g., stellar mass-to-light ratios and/or nonaxisymmetric structures).  This is combined with instrumental and observational challenges, such as the requirement of AO instruments to have a bright nearby tip-tilt star and the limited spectral range in the $K$ band ($\sim$2 \micron) for CO-bandhead observations of redshifted sources.  Given all of these challenges, our following attempt to illustrate potential sources for follow-up observations should be considered as a first step to identify initial candidates.  

The calculated $r_{\rm SOI}$ (in parsecs) for our sample of BASS AGN with quality \sigs\ measurements are shown in \autoref{fig:soi}.   Clearly, there is a large number of candidate AGN that could possibly be resolved using direct measurements.  Existing facilities could resolve the stellar dynamics in  128/\Nvdisp\ of our sources (27\%, or $128-17=111$ where such measurements do not yet exist).  For gas dynamics, the picture may be even more positive, given the higher-resolution that can be achieved with submillimeter interferometers. Specifically, a resolution of  0\farcs03, achievable with ALMA over much of the sky (e.g., $\delta < 25^{\circ}$), could potentially provide direct \mbh\ measurements for 325/\Nvdisp\ of our obscured AGN (68\%, or $325-18=307$ with no prior measurements).  We note that gas dynamical measurements require Keplerian motions for a relatively large reservoir of circumnuclear gas, as well as high-resolution optical imaging (with the Hubble Space Telescope; FWHM=0\farcs08) to estimate the contribution of stellar mass in the region near the SMBH.  However, the most recent of these gas dynamical measurements have been done at similar spatial scales (e.g., $\theta_{\mathrm{SOI}}$=0\farcs07, \citealt{Smith:2021:5984}) and with gas detections in the inner regions of similarly small regions near the SMBH (e.g., $r_{\mathrm{SOI}}$=13 pc; \citealt{Davis:2017:4675}).  The best available ALMA resolution will also increase again in late 2022, thanks to the high-frequency, long-baseline capabilities in the highest bands (e.g., $\sim$0\farcs01).

Increasing the sample size in \mbh--\sigs\ to better understand the relation is a critical goal within BASS, as it is sensitive to the strongest accreting SMBHs in the nearby universe, where there may be more gas available for direct imaging measurements.  A number of these ALMA programs to enlarge the sample of high spatial resolution observations among BASS AGN are just beginning or already ongoing. For gas dynamical measurements, an ALMA program to obtain CO(2--1) measurements at $\approx$100 pc resolution for 33 nearby and luminous AGN was recently completed, with additional programs ongoing, and generally, more than 100 BASS AGN now have ALMA observations at CO(2--1) regardless of resolution.  Additionally, single-dish observations to measure whether the AGN lie in galaxies that are sufficiently gas-rich to facilitate high-resolution molecular gas observations have been carried out with APEX \citep{Koss:2021:29} for 213 nearby AGN  ($0.01{<}z{<}0.05$), although naturally higher-resolutions will be needed to confirm the presence of sufficient gas in the inner regions of these galaxies.  


\begin{figure*} 
\centering
\includegraphics[width=0.49\textwidth]{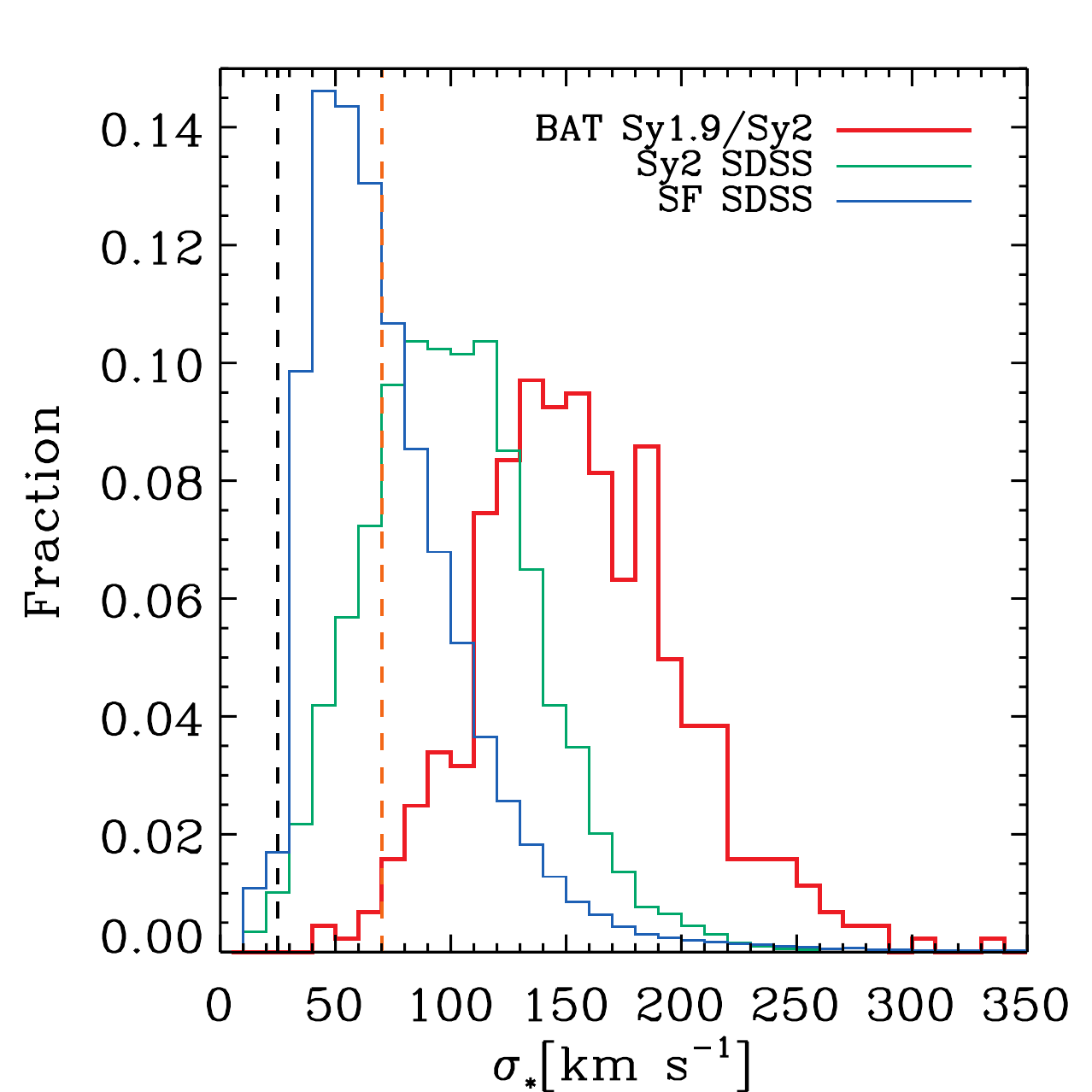}
\caption{Comparison of nearby obscured BASS AGN (Sy\,1.9 and Sy\,2, $z<0.1$, $N$=444) to optically selected nearby ($z<0.1$) narrow-line Seyferts ($N$=4926) and SF galaxies ($N$=97,216) in the SDSS DR7 from the OSSY catalog \citep{Oh:2011:13}.  The orange dashed line indicates the instrumental limit of the SDSS ($\sim$70\,\kmpssh) compared to the instrumental limits for BASS AGN, shown with a black dashed line ($\sim$25\,\kmpssh).  }
\label{fig:comp_SDSS}
\end{figure*}

\begin{figure*} 
\centering
\includegraphics[width=0.6\textwidth]{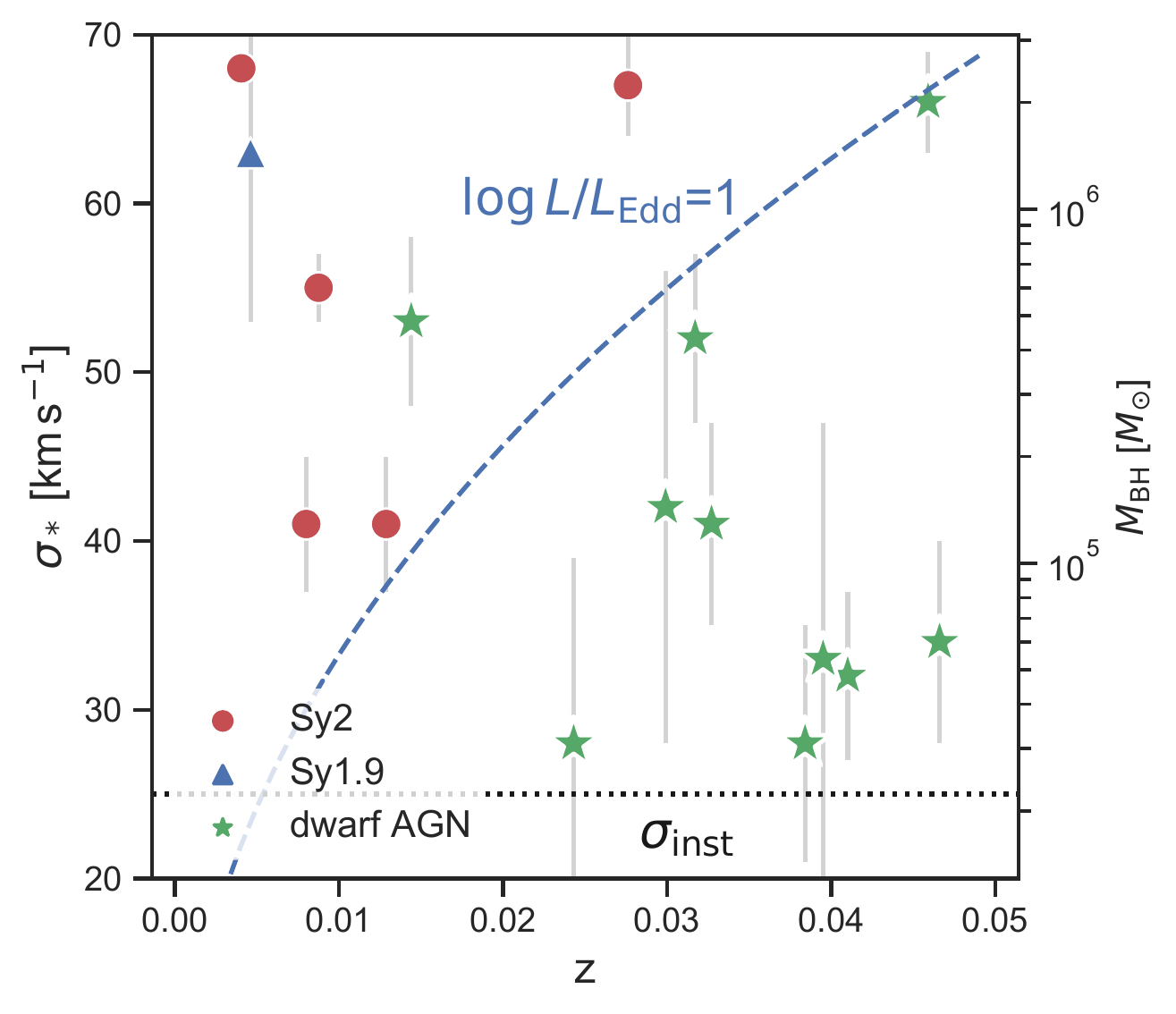}
\caption{Comparison of the BASS AGN to the velocity dispersion of dwarf galaxies hosting AGN \citep{Baldassare:2020:L3}.  This dwarf AGN sample was identified in the SDSS at $z<0.055$ based on emission line diagnostics and broad lines in \Halpha\ \citep{Reines:2013:116}.  The dashed blue line indicates the sky sensitivity limit for a source at the Eddington limit (\lledd=1) at the BAT survey all-sky flux limits, so any Sy\,1.9 and Sy\,2 detected by BAT below this line would be super-Eddington.  The black dotted line indicates the BASS instrumental limits for velocity dispersion measurements  ($\sigma_{\mathrm{inst}}{\sim}25$\,\kmpssh) which is the similar to the the Keck study ($\sigma_{\mathrm{inst}}{=}23$\,\kmpssh).}
\label{fig:dwarf_comp}
\end{figure*}

\begin{figure*} 
\centering
\includegraphics[width=0.6\textwidth]{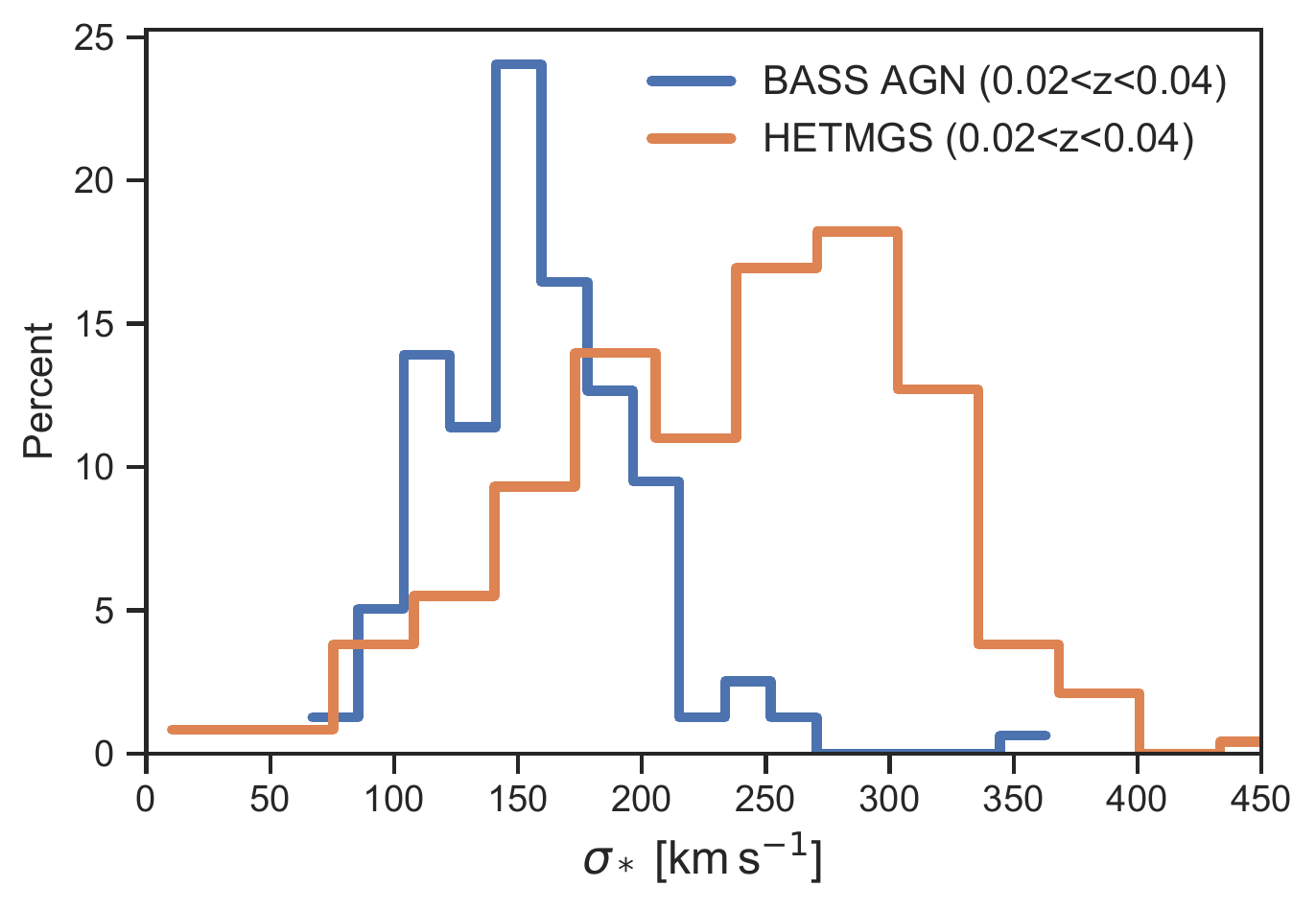}
\caption{Comparison of the BASS AGN to the velocity dispersions of the HETMGS survey \citep{vandenBosch:2015:10}, which was designed to select the most massive BHs (and highest velocity dispersions) at all redshifts that could be dynamically resolved.  For both samples, the redshift range is limited to $0.02<z<0.04$, to avoid the low-mass galaxies at low redshifts in HETMGS that can be resolved.}
\label{fig:massive_comp}
\end{figure*}


\begin{figure*} 
\centering
\includegraphics[width=0.49\textwidth]{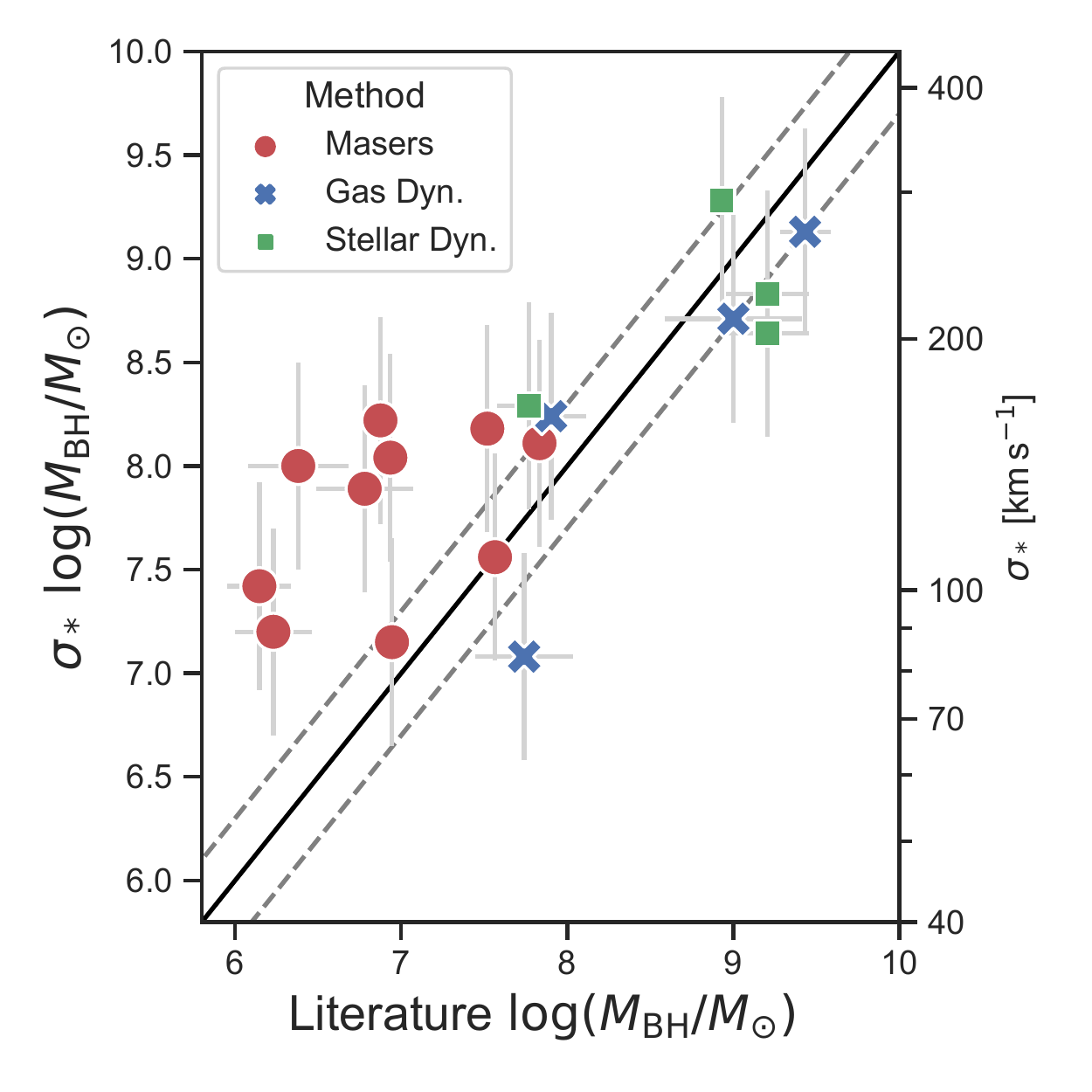}
\caption{Comparison of the inferred \mbh\ mass estimate from \sigs\ compared to direct measurements using megamaser compilations \citep[][]{Greene:2016:L32}, as well as stellar \citep{Neumayer:2010:449,Medling:2011:32,Walsh:2012:79} and gas dynamics \citep{Tadhunter:2003:861,Capetti:2005:465,Wold:2006:449a} measurements.  Error bars in \mbh\ inferred from \sigs\ have been set to 0.5 dex based on past comparisons with direct measurements \citep[e.g., 0.44 dex;][]{Gultekin:2009:198}.}
\label{fig:directcomp}
\end{figure*}

\begin{figure*} 
\centering
\includegraphics[width=0.49\textwidth]{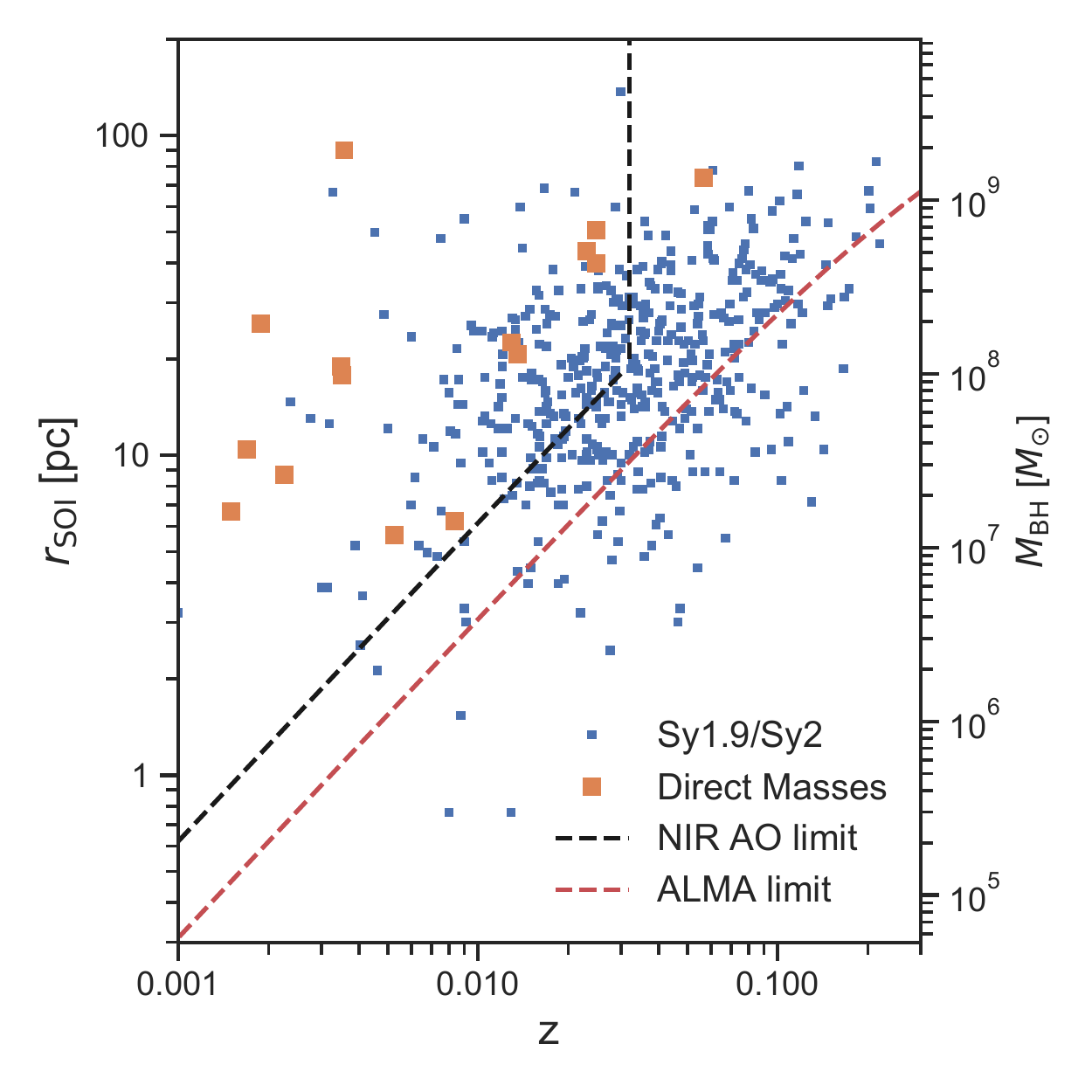}
\caption{Estimated SOI ($r_{\mathrm{SOI}}$) in parsecs for obscured AGN in this sample based on the velocity dispersion measurements.  The dashed lines indicate the ability to detect the dynamical influence of the SMBH, assuming the observations have a minimum spatial resolution of twice the $r_{\mathrm{SOI}}$ \citep{Davis:2014:911}.   The black dashed lines indicate the limiting resolution for AO observations of stellar dynamics using the CO bandheads in the $K$ band assuming a good tip-tilt star is available (0\farcs06 FWHM). The vertical line is due to these bands shifting out of the $K$ band (at $z=0.032$). The red dashed line indicates the capabilities of ALMA to resolve the gas dynamics for the CO 2--1 line at higher spatial resolutions (0\farcs03 FWHM).  The large orange squares indicate sources with dynamical and/or megamaser measurements. }
\label{fig:soi}
\end{figure*}

\section{Summary}
\label{sec:summary}


We present new measurements of central stellar velocity dispersions for \Nvdisp\ Sy\,1.9 and Sy\,2 luminous AGN drawn  from the BASS DR2.  We have developed a set of high-resolution spectral templates ($R$=10,000) from the VLT/X-Shooter library to take advantage of the very high spectral resolutions of many of the spectra we used ( $R\simeq5000$, $\sigma_{\rm inst}$=25\,\kmpssh).  This constitutes the largest study of X-ray-selected obscured AGN with velocity dispersion measurements and includes \Nvdispmeas\ independent measurements of the spectral regions relevant for the \Cahk\ and \mgb\ (3880--5550\,\AA) and the CaT (8350--8730\,\AA) absorption features from \Nvdispspec\ spectra. 
Our measurements span a wide range, \sigs=40--360\,\kmpssh, corresponding to 4--5 orders of magnitude in SMBH mass ($\mbh\sim10^{5.5-9.6}$\,\Msun). Combined with the wide range in bolometric luminosity probed by the BAT all-sky survey ($\lbol\sim10^{42-46}$\,\ergps), we can explore an unprecedentedly broad range in Eddington ratios, \lledd$\sim10^{-5}$ to 2.  
Thus, this BASS DR2 Sy\,1.9/Sy\,2 sample represents a significant advance with a nearly complete census of velocity dispersions of hard X-ray-selected obscured AGN in the local universe ($z{<}0.2$), covering 99\% of nearby AGN ($z{<}0.1$) and 82\% at higher redshifts ($0.1{<}z<0.6$; outside the Galactic plane).    

Using this large survey of central stellar velocity dispersions, we draw the following insights,
\begin{itemize}
\item With a sample of \Nvdispmeas\ velocity dispersion measurements with significant duplications and two spectral regions (3880--5550 and CaT), we find that there is no significant offset between the two regions.  However, there is larger intrinsic scatter than expected (e.g. 8-18\,\kmps MAD) based on fitting errors or due to the presence of outliers (e.g., 14-28\,\kmps\ rms).  This leads to additional uncertainties in the scatter of inferred SMBH measurements of 0.11--0.21 dex in MAD or 0.2--0.4 dex in rms.
\item  The obscured BASS AGN occupy a unique space with their velocity dispersion properties, having much higher velocity dispersions (i.e., 150 vs. 100\,\kmpssh) than the more numerous, optically selected narrow-line AGN (i.e. drawn from the SDSS) but not having a significant population of the highest velocity dispersions (i.e., $>$250\,\kmpssh) of the nearby universe.
\item Despite having sufficient spectral resolution to resolve very small BHs, we do not find a significant population of low-redshift, low-\mbh\ ($\lesssim10^6\,\Msun$), super-Eddington sources among BAT ultrahard X-ray-selected AGN.
\item Based on preliminary estimates of the SOI for the SMBHs in our sample, existing facilities can obtain direct \mbh\ measurements using stellar and gas dynamics for a considerably larger number of AGN than what is currently available.
\end{itemize}
To summarize, the BASS DR2 catalog of stellar velocity dispersion measurements and the implied BH masses provide an extremely useful resource for studies of the low-redshift population of actively accreting SMBHs and their host galaxies.

\clearpage

\begin{acknowledgments}
We thank Vivian Baldassare and Amy Reines for their assistance with dwarf AGN masses.  
We acknowledge support from NASA through ADAP award NNH16CT03C (M.K.); 
the Israel Science Foundation through grant No. 1849/19 (B.T.); 
the European Research Council (ERC) under the European Union's Horizon 2020 research and innovation program, through grant agreement No. 950533 (B.T.);
FONDECYT Regular 1190818 (E.T., F.E.B.) and 1200495 (E.T., F.E.B); FONDECYT Postdoctoral Fellowship 3210157 (A.R.); 
ANID grants CATA-Basal AFB-170002 (E.T., F.E.B.), ACE210002 (E.T., F.E.B.) and FB210003 (C.R., E.T., F.E.B.); ANID Anillo ACT172033 and Millennium Nucleus NCN19\_058 (E.T.); and Millennium Science Initiative Program  – ICN12\_009 (F.E.B.); an ESO fellowship (M.H.,J.M.); Fondecyt Iniciacion grant 11190831 (C.R.); the National Research Foundation of Korea grant NRF-2020R1C1C1005462 and the Japan Society for the Promotion of Science ID: 17321 (K.O.); Comunidad de Madrid through the Atracción de Talento Investigador Grant 2018-T1/TIC-11035 (I.L.); YCAA Prize Postdoctoral Fellowship (M.B.); Conselho Nacional de Desenvolvimento Cient\'{i}fico e Tecnol\'ogico  ( CNPq, Proj. 311223/2020-6,  304927/2017-1 and  400352/2016-8), Funda\c{c}\~ao de amparo 'a pesquisa do Rio Grande do Sul (FAPERGS, Proj. 16/2551-0000251-7 and 19/1750-2), Coordena\c{c}\~ao de Aperfei\c{c}oamento de Pessoal de N\'{i}vel Superior (CAPES, Proj. 0001, R.R.). This work was performed in part at the Aspen Center for Physics, which is supported by National Science Foundation grant PHY-1607611.  We acknowledge the work done by the Swift/BAT team and the 50+ BASS scientists  to make this project possible.

Some of the optical spectra were taken with DoubleSpec at Palomar via Yale (PI: M. Powell; 2017--2019, 16 nights), as well as Caltech (PI: F. Harrison) and JPL (PI: D. Stern) from programs from 2013--2021.

This work made use of observations collected at the European Southern Observatory under ESO programs 0101.A-0765(A), 0101.B-0456(B), 0102.A-0433(A),  0102.B-0048, 0103.A-0521(A),  0103.A-0777(A), 0104.A-0353(A), 090.A-0830(A), 092.B-0083(A), 093.A-0766(A), 095.B-0059(A), 098.A-0635(B), 099.A-0403(A), 099.A-0403(B), 094.B-0321(A), 098.B-0551(A),  60.A-9421(A), and 60.A-9100(J).  Based on observations from six CNTAC programs:  CN2018A-104, CN2018B-83, CN2019A-70, CN2019B-77, CN2020A-90,  and CN2020B-48 (PI C. Ricci).  Based on NOIRLab proposals (2012A-0463, PI: M. Trippe).  Based on observations obtained at the Southern Astrophysical Research (SOAR) telescope, which is a joint project of the Minist\'{e}rio da Ci\^{e}ncia, Tecnologia e Inova\c{c}\~{o}es (MCTI/LNA) do Brasil, the US National Science Foundation's NOIRLab, the University of North Carolina at Chapel Hill (UNC), and Michigan State University (MSU).

Some of the data presented herein were obtained at the W.M. Keck Observatory, which is operated as a scientific partnership among the California Institute of Technology, the University of California, and the National Aeronautics and Space Administration. The Observatory was made possible by the generous financial support of the W. M. Keck Foundation.  The authors wish to recognize and acknowledge the very significant cultural role and reverence that the summit of Maunakea has always had within the indigenous Hawaiian community.  We are most fortunate to have the opportunity to conduct observations from this mountain. 

This research has made use of NASA's ADS Service. 
This research has made use of the NASA/ IPAC Infrared Science Archive, which is operated by the Jet Propulsion Laboratory, California Institute of Technology, under contract with the National Aeronautics and Space Administration.

Funding for SDSS-III has been provided by the Alfred P. Sloan Foundation, the Participating Institutions, the National Science Foundation, and the U.S. Department of Energy Office of Science. The SDSS-III website is http://www.sdss3.org/. The SDSS-III is managed by the Astrophysical Research Consortium for the Participating Institutions of the SDSS-III Collaboration, including the University of Arizona, the Brazilian Participation Group, Brookhaven National Laboratory, Carnegie Mellon University, University of Florida, the French Participation Group, the German Participation Group, Harvard University, the Instituto de Astrofisica de Canarias, the Michigan State/Notre Dame/JINA Participation Group, Johns Hopkins University, Lawrence Berkeley National Laboratory, Max Planck Institute for Astrophysics, Max Planck Institute for Extraterrestrial Physics, New Mexico State University, New York University, Ohio State University, Pennsylvania State University, University of Portsmouth, Princeton University, the Spanish Participation Group, University of Tokyo, University of Utah, Vanderbilt University, University of Virginia, University of Washington, and Yale University.
 \end{acknowledgments}


\vspace{5mm}
\facilities{IRSA, Keck:I (LRIS), Magellan:Clay, Hale (Doublespec),  Swift (BAT), VLT:Kueyen (X-Shooter), VLT:Antu (FORS2), SOAR (Goodman)}

\software{astropy \citep{Collaboration:2013:A33}, 
          Matplotlib \citep{Hunter:2007:90}, 
          Numpy \citep{vanderWalt:2011:22}
          }.

\clearpage
\appendix

\section{Emission Lines Masked \label{sec:emlinesmask}}
A list of the masked emission lines is given in \autoref{tab:emlinesmask}.

\begin{table}
\begin{center}
\caption{Emission lines masked in \ppxf\ host galaxy fitting} 
\begin{tabular}{ll}
\tableline
\tableline
Emission line&Wavelength [\AA]  \\
\tableline
H8 + He I &3889.1\\
\neiii&3967.41\\
\sii&4071.24\\
H$\delta$&4101.76\\
$\left[{\rm Fe}\,\textsc{v}\right]$&4229\\
H$\gamma$&4340.47\\
\oiii &4363.21\\
HeII&4686\\
H$\beta$&4861.33\\
\oiii&4958.92\\
\oiii&5006.84\\
$\left[{\rm N}\,\textsc{i}\right]$&5200\\
\tableline Mask for Weaker Lines\\ \tableline
He I&4026\\
$\left[{\rm Ar}\,\textsc{iv}\right]$ &4712\\
$\left[{\rm Ar}\,\textsc{iv}\right]$&4740\\
$\left[{\rm Fe}\,\textsc{vi}\right]$ &5146\\
$\left[{\rm Fe}\,\textsc{vii}\right]$ &5159\\
$\left[{\rm Fe}\,\textsc{viii}\right]$ &5176\\
$\left[{\rm Ca}\,\textsc{v}\right]$ &5309\\
$\left[{\rm Fe}\,\textsc{vi}\right]$ &5485\\
$\left[{\rm Cl}\,\textsc{iii}\right]$ &5518\\
$\left[{\rm Cl}\,\textsc{iii}\right]$ &5538\\
${\rm O}\,\textsc{i}$&8446\\
$\left[{\rm Cl}\,\textsc{ii}\right]$&8578.7\\
$\left[{\rm Fe}\,\textsc{ii}\right]$&8617\\
Pa12&8750\\
Pa11&8863\\
\tableline
\end{tabular}
\label{tab:emlinesmask}
\end{center}
\end{table}

\section{Excluded Templates from Hierarchical Clustering}
\label{sec:exclude_temp}
Examples of excluded templates from clustering are provided in \autoref{fig:excl_uvb}.

\begin{figure*} 
\centering
\includegraphics[width=0.49\textwidth]{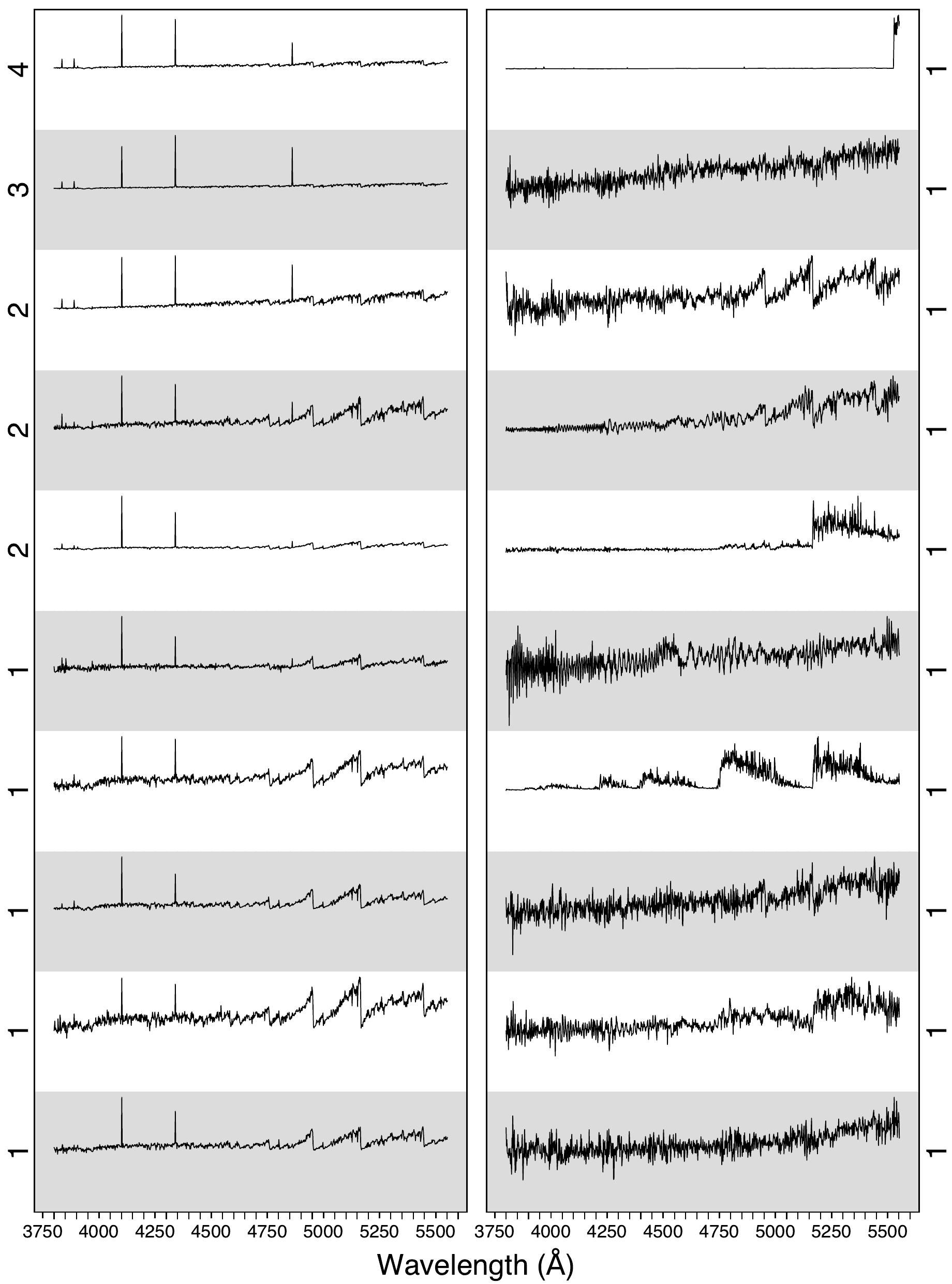}
\includegraphics[width=0.49\textwidth]{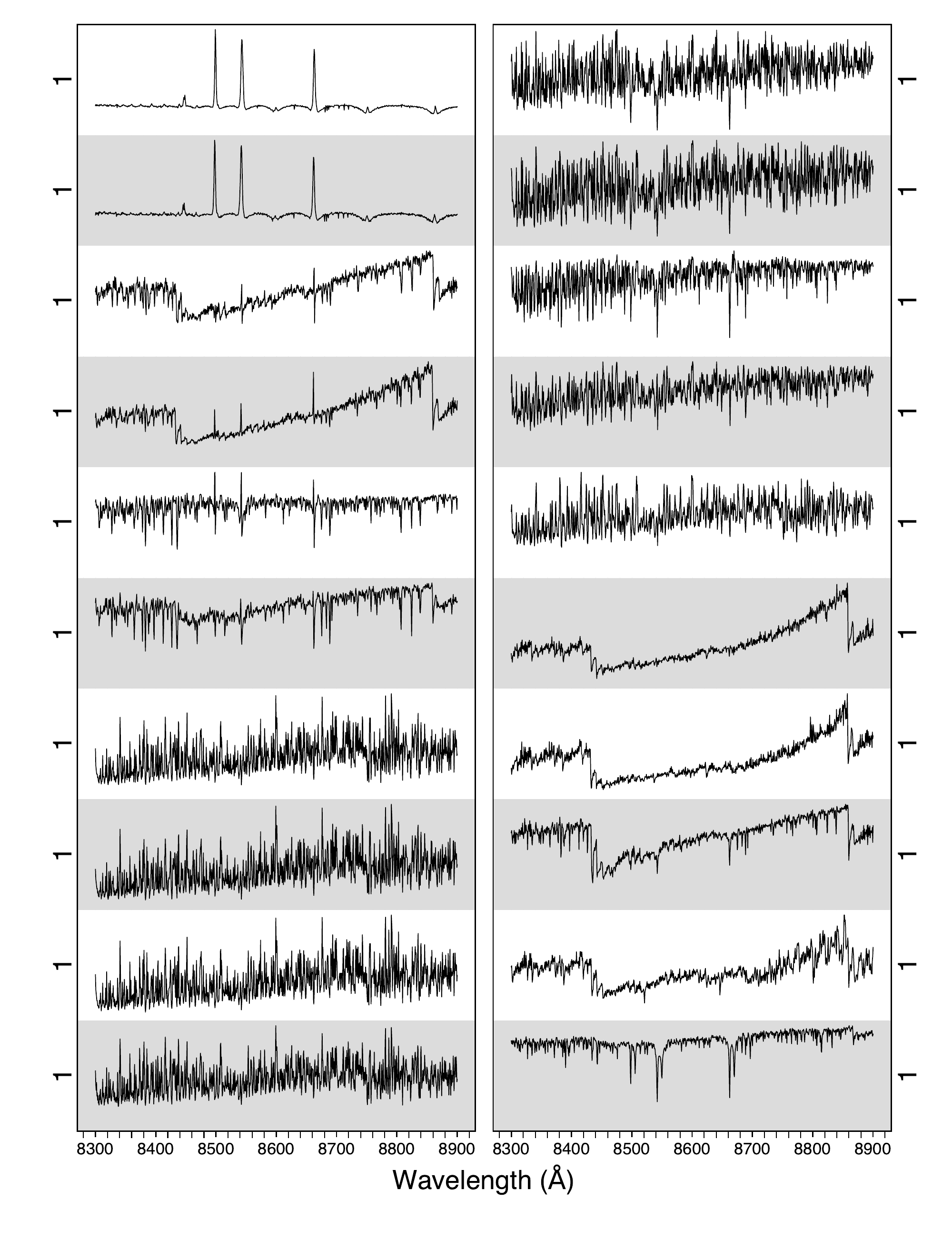}
\caption{Examples of excluded X-Shooter templates from hierarchical clustering analysis.  {\it Left}: UVB templates that were excluded due to emission line features.  {\it Left-middle}: UVB templates that were excluded because of low S/N or because they were only found in one spectrum.  The number on the left and right side of the figure indicates the number of spectra the features were found in the parent sample in the clustering analysis.  {\it Right-middle and right}: Examples of excluded X-Shooter VIS templates from hierarchical clustering analysis.  }
\label{fig:excl_uvb}
\end{figure*}

\section{Secondary Velocity Dispersion Measurements}
\label{sec:sec_meas}
Here we provide secondary measurements in \autoref{tab:veldispsec}.

\begin{deluxetable*}{lllllllllll}
\tabletypesize{\scriptsize}
\tablecaption{Secondary Spectral Measurements \label{tab:veldispsec}}
\tablehead{
\colhead{BAT ID} &      \colhead{Galaxy} &                        \colhead{DR2 Type} &  \colhead{Tele./Inst.} & \colhead{Res. Blue} & \colhead{Res. Red} & \colhead{Mask} &     \colhead{$z_{3880-5550\,\text{\AA}}$} & \colhead{$\sigma_{3880-5550\,\text{\AA}}$} & \colhead{$z_{\mathrm{CaT}}$} & \colhead{$\sigma_{\mathrm{CaT}}$}\\
\colhead{} &      \colhead{} &                        \colhead{} &  \colhead{} & \colhead{\AA} & \colhead{\AA} & \colhead{} &     \colhead{\kmps} & \colhead{\kmps} & \colhead{\kmps} & \colhead{\kmps}
}
\startdata
  13 &                LEDA136991 &      Sy\,2 &   Palomar/DBSP &  4.1 &      4.9 &        n &               &             &   3827$\pm$11 &     107$\pm$15 \\
    17 &                  ESO112-6 &      Sy\,2 &   VLT/X-Shooter &  1.3 &      1.4 &        n &               &             &   8897$\pm$10 &     150$\pm$13 \\
    28 &                   NGC235A &    Sy\,1.9 &    Gemini/GMOS &  4.8 &  \nodata &        n &    6869$\pm$5 &   188$\pm$6 &               &                \\
    50 &                 ESO243-26 &      Sy\,2 &      SOAR/GMAN &  2.7 &  \nodata &        w &               &             &    5968$\pm$4 &       85$\pm$6 \\
    53 &                      UM85 &      Sy\,2 &   Palomar/DBSP &  4.1 &      4.9 &        w &  12536$\pm$10 &  187$\pm$10 &  12516$\pm$11 &     145$\pm$15 \\
    57 &                      3C33 &      Sy\,2 &   Palomar/DBSP &  2.3 &      1.8 &        n &  18302$\pm$14 &  222$\pm$11 &               &                \\
    63 &                   NGC454E &      Sy\,2 &      SOAR/GMAN &  2.7 &  \nodata &        w &               &             &    3730$\pm$2 &       95$\pm$3 \\
    70 &                MCG+8-3-18 &      Sy\,2 &   Palomar/DBSP &  4.8 &      6.5 &        w &               &             &   6351$\pm$10 &     147$\pm$15 \\
    72 &                   NGC526A &    Sy\,1.9 &      SOAR/GMAN &  2.7 &  \nodata &        w &               &             &    5789$\pm$4 &      153$\pm$4 \\
    81 &                 ESO244-30 &      Sy\,2 &      SOAR/GMAN &  2.7 &  \nodata &        w &               &             &    7878$\pm$6 &      100$\pm$6 \\
    83 &                  ESO353-9 &      Sy\,2 &      SOAR/GMAN &  2.7 &  \nodata &        w &               &             &    5077$\pm$2 &      126$\pm$4 \\
    84 &                    NGC612 &      Sy\,2 &      SOAR/GMAN &  2.7 &  \nodata &        w &               &             &   9137$\pm$12 &     336$\pm$20 \\
    87 &                 ESO297-18 &      Sy\,2 &      SOAR/GMAN &  2.7 &  \nodata &        w &               &             &    7864$\pm$8 &      141$\pm$9 \\
    88 &                LEDA138434 &    Sy\,1.9 &   Palomar/DBSP &  4.1 &      4.9 &        n &               &             &  22206$\pm$13 &     123$\pm$18 \\
    92 &               LEDA1656658 &    Sy\,1.9 &   Palomar/DBSP &  4.1 &      4.9 &        n &               &             &  21281$\pm$18 &     209$\pm$19 \\
    96 &                MCG-1-5-47 &      Sy\,2 &   Palomar/DBSP &  4.8 &      3.4 &        n &               &             &    5662$\pm$9 &      85$\pm$12 \\
   101 &                  UGC01479 &      Sy\,2 &   Palomar/DBSP &  4.1 &      4.9 &        n &               &             &    5029$\pm$6 &      142$\pm$9 \\
   112 &                    ARP318 &      Sy\,2 &       APO/SDSS &  3.0 &  \nodata &        n &    3963$\pm$2 &   182$\pm$3 &    3971$\pm$3 &      168$\pm$4 \\
   123 &                  VIIZw232 &      Sy\,2 &   Palomar/DBSP &  4.8 &      6.5 &        n &  18937$\pm$16 &  206$\pm$17 &               &                \\
   140 &                   NGC1052 &      Sy\,2 &    Gemini/GMOS &  4.8 &  \nodata &        n &    1560$\pm$3 &   218$\pm$4 &               &                \\
   145 &                2MFGC02171 &      Sy\,2 &   Palomar/DBSP &  4.8 &      6.5 &        n &               &             &  10720$\pm$11 &     131$\pm$15 \\
   149 &                 LEDA89928 &      Sy\,2 &   Palomar/DBSP &  2.3 &      1.8 &        n &  17803$\pm$17 &  170$\pm$20 &  17767$\pm$14 &     256$\pm$14 \\
   151 &                LEDA166445 &      Sy\,2 &   Palomar/DBSP &  4.1 &      4.9 &        n &               &             &    4690$\pm$8 &     160$\pm$12 \\
   153 &                   NGC1125 &      Sy\,2 &   Palomar/DBSP &  4.1 &      4.9 &        n &    3435$\pm$9 &  103$\pm$12 &    3334$\pm$5 &      127$\pm$5 \\
\enddata
\tablecomments{\autoref{tab:veldispsec} is published in its entirety in the machine-readable format.  A portion is shown here for guidance regarding its form and content.  Column descriptions same as \autoref{tab:veldispbest}.}
\end{deluxetable*}

\section{Comparison of the DR1 and DR2 Velocity Dispersions}
\label{sec:dr1comp}
In comparing with the DR1, we limit our sample to only the sources with $\Delta\sigs<20$\,\kmpssh, whereas the DR1 included fits up to 50\,\kmps\ to better understand any systematic offsets.  Outside of the SDSS sample and a small number of Palomar/DBSP and Gemini/GMOS spectra that are included in this study, we note that nearly all of the DR1 sources were reobserved as part of the higher spectral resolution DR2 survey.  This limits the sample to 29 Sy\,1.9 and 89 Sy\,2 galaxies that were measured in both the DR1 and the current sample.  Overall, we find that there is a median offset in the DR1 of 9.6 and 6.6\,\kmps\ for Sy\,1.9 and Sy\,2, respectively (e.g. $\vert \sigma_{\mathrm{DR1}}-\sigma_{\mathrm{Best}}\vert$).  These offsets correspond to median offsets of 0.11 and 0.08 dex for Sy\,1.9 and Sy\,2, in measured SMBH, respectively (or a median of 0.09 dex for the combined sample).  This systematic difference is relatively small compared to the expected typical systematic error (e.g. 0.3-0.5 dex), so we do not expect it to affect any specific DR1 studies that were based on the entire sample.  However, there are outliers in specific samples and the Sy\,1.9, tend to show increased offsets, so we explore this further.

A useful test of the effect of the new templates, changed fitting regions, and updated \ppxf\ software applied is provided in the overlapping spectra in the SDSS.  There are 46 Sy\,2 and 17 Sy\,1.9 overlapping within the DR1 and DR2. A comparison of the velocity dispersion measurements from the DR1 and DR2 for the SDSS samples is shown in \autoref{fig:dr1_sdsscomp}. The DR1 fit the entire 3900--7000\,\AA\ region for the SDSS spectra, whereas the DR2 was limited to $3880-5550$\,\AA. There is a systematic shift in both Sy\,2 and Sy\,1.9, with the velocity dispersion measurements being somewhat larger in the DR1 than the DR2 and Sy\,1.9 showing larger offsets and scatter.  Some of the larger scatter is likely from the fact that Sy\,1.9 has larger typical measurement errors than Sy\,2 (e.g., for the DR1, the median $1\sigma$ error is 10 and 6\kmpssh, respectively).  Among Sy\,2, the median offset ($\vert \sigma_{\mathrm{DR1}}-\sigma_{\mathrm{best}}\vert$) is 4.0\,\kmpssh. Among Sy\,1.9, the median offset is larger at 14\,\kmpssh.

We then expand our comparison of our current work to the other DR1 samples that are not in the current study.  This includes spectra taken from the Perkins 1.8m telescope and DeVeny Spectrograph at the Lowell Observatory and CTIO 1.5m RC spectrograph, as well as archival optical spectra obtained as part of the final data release for the 6dF Galaxy Survey \citep[][]{Jones:2009:683}.  There are six systems observed with other telescopes that were significant outliers (e.g. ${>}$30\,\kmps beyond 1$\sigma$ error bars) between the DR1 and DR2, which includes four Sy\,1.9 and two Sy\,2.  The outliers showed differences in velocity dispersion between 38 and 85\,\kmps corresponding to offsets in SMBH mass of 0.2-0.6 dex.  All of the six outliers were at higher redshifts ($z{>}0.059$) and had high AGN luminosities (\Lbol${>}10^{45}$\,\ergps) and larger errors (12-20\,\kmpssh) than typical of the sample. The Sy\,1.9 outliers all show evidence of extremely broad lines that were not properly masked in the original DR1 fitting procedure, which excluded regions of 3200\,\kmps around \Halpha.  Three of the systems show double-peaked and/or asymmetric broad lines (ID 522, 715, and 817), which are rare in the BAT sample.  The final system (ID=785) shows a very broad line (\Halpha\ FWHM=7800\,\kmpssh).  The two outlier Sy\,2 systems are likewise unique, with the presence of strong outflows or double-peaked narrow emission lines.  Recent, optical integral field spectroscopy with the Gemini North Telescope \citep{Couto:2020:5103} for 4C+29.30 (BAT ID=426) indicated a large southern knot 1\arcsec\ south of the nucleus (and hence in the 3\arcsec\ SDSS fiber), which also presents high velocity dispersions ($\sim$250\,\kmpssh) attributed to an outflow.  The other Sy\,2 outlier, SDSS J000911.58-003654.7 (BAT ID=7), shows double-peaked narrow emission lines, and five nearby galaxies within 30\arcsec\ at similar redshift ($<500$\,\kmpssh), indicative of a galaxy group.

Finally, we examine a DR1 sample from the 2.1m telescope at the Kitt Peak (KP) national observatory with the GoldCam spectrograph.  It is clear that the KP sample specifically shows ${>}$30\,\kmps\ outliers for the majority of the sample.  The outliers all appear above  200\,\kmpssh, so a possible reason is a significant underestimation of the instrumental resolution.  The median offset is 55\,\kmps\ corresponding to 0.41 dex in SMBH mass.

In summary, while the a median systematic offset with the DR1 is small (e.g. 9.6 and 6.6\,\kmps\ for Sy\,1.9 and Sy\,2, respectively), there is a population of outliers that bias the distributions. We find that the Sy\,1.9 show larger systematic offsets in the DR1 likely due to AGN contamination.  This is seen in the outliers tending to have very broad lines and be double-peaked sources.  As the AGN broad \Halpha\ emission and continuum emission scales with the hard X-ray emission \citep[e.g.,][]{Mejia_Broadlines}, the host galaxy absorption features are likely diluted in these systems, leading to erroneously large measurements compared to the DR2 measurements which did not fit the \Halpha\ region.  The small offset among Sy\,2 is puzzling but may be related to the lower spectral resolution of the DR1 sample.  Finally, with the higher-quality observations, we can see that the KP sample has large systematic errors and should not be used.  As the DR2 observations presented here are of much higher S/N and spectral resolution, they should be used in place of these measurements.

\begin{figure*} 
\centering
\includegraphics[width=6cm]{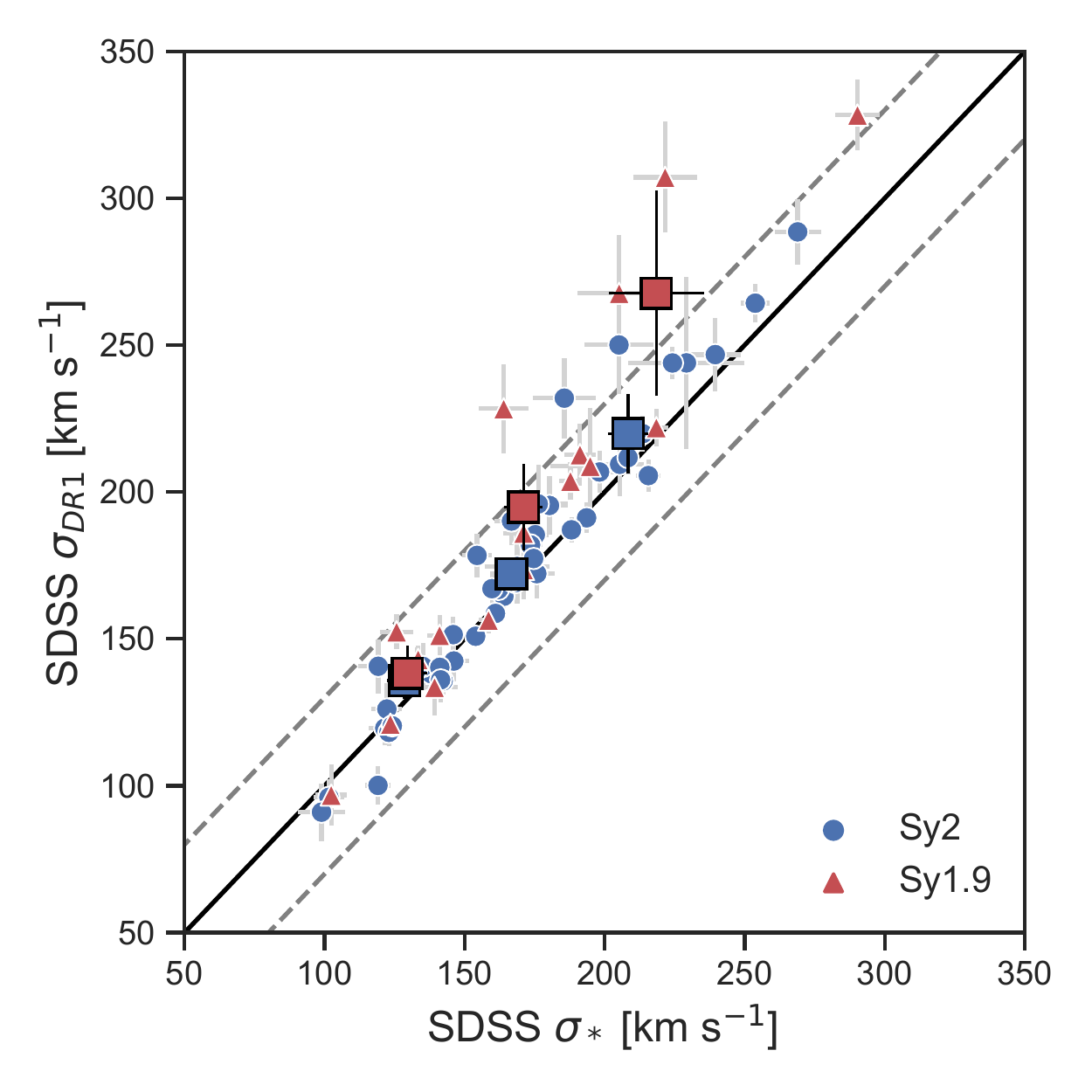}
\includegraphics[width=9cm]{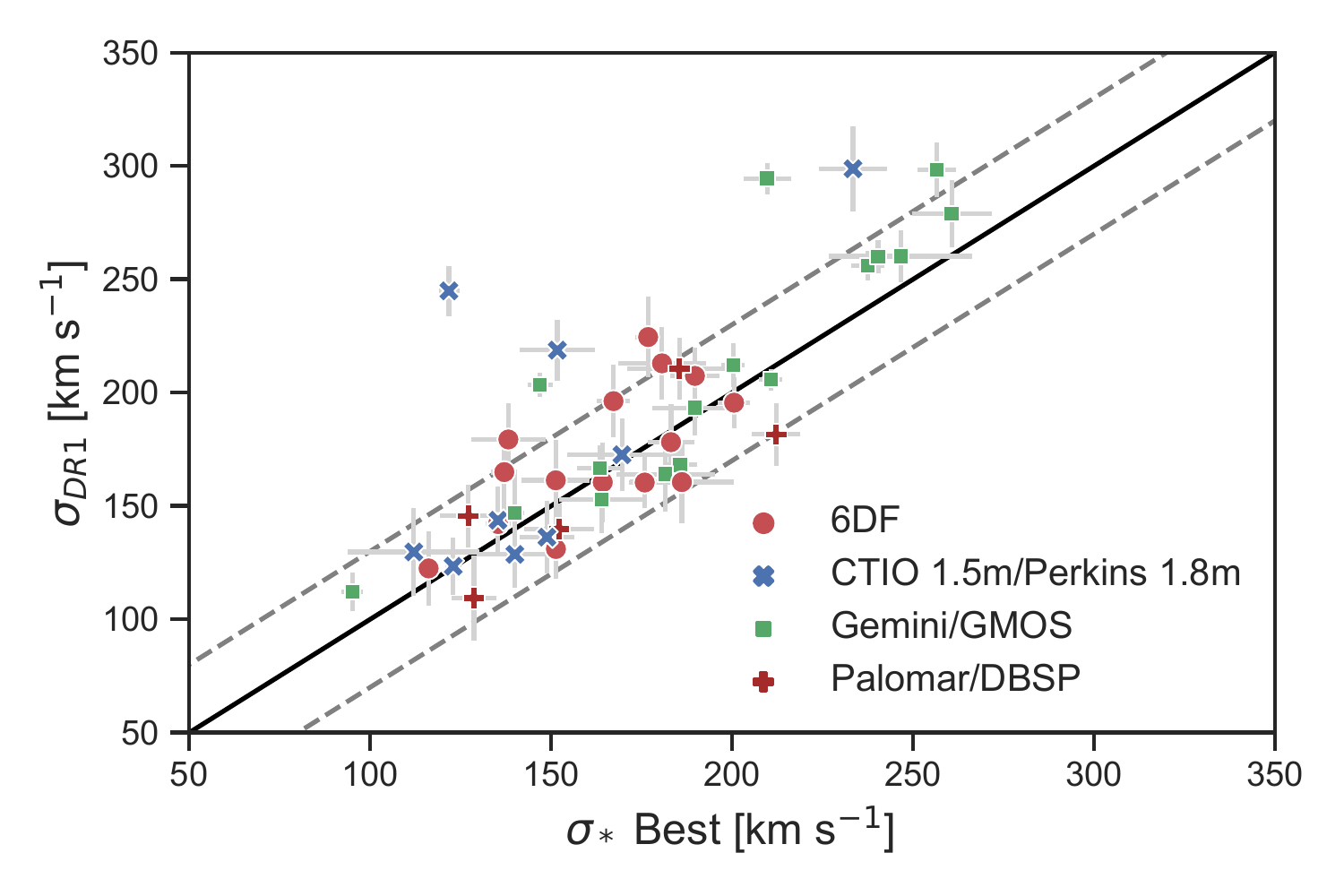}
\includegraphics[width=8cm]{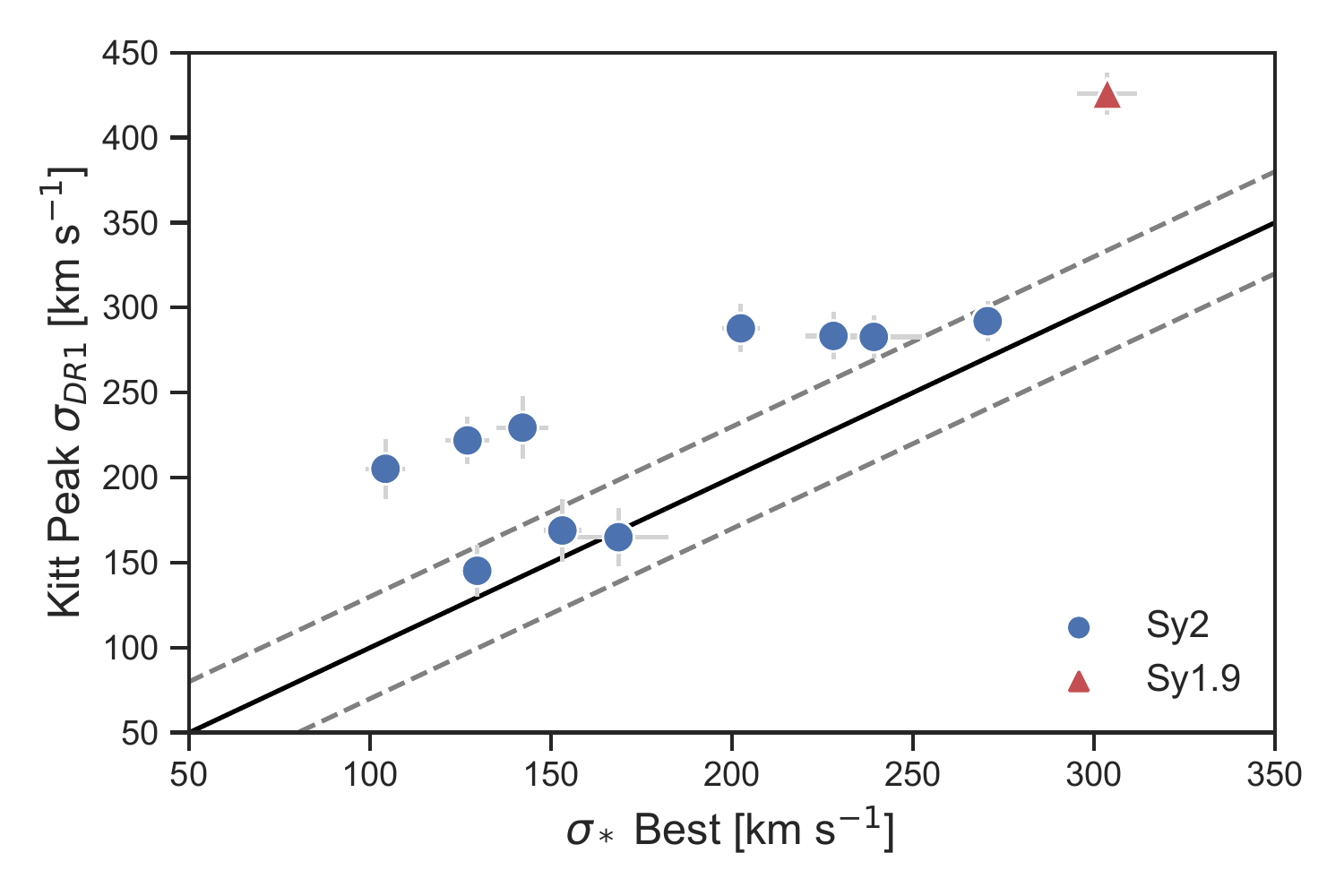}
\caption{Comparison between different velocity dispersion measurements.  In all panels, Sy\,1.9 (with broad \halpha) is shown with red triangles and Sy\,2 is shown with blue circles.  Error bars (at $1\sigma$) are shown in gray.  A solid black line indicates the one-to-one relatio,n with dashed gray lines indicating offsets of 30\,\kmpssh.  {\em Top left}: comparison between the velocity dispersion measurements from the SDSS for the DR1 and DR2 for DR1 measurements with $\Delta\sigs{<}20$\,\kmpssh.  {\em Top right}: comparison between the velocity dispersion measurements from other DR1 samples and the current best measurements. {\em Bottom}:  comparison between the velocity dispersion measurements from the KP DR1 samples where there is a significant fraction of outliers.   }
\label{fig:dr1_sdsscomp}
\end{figure*}

\section{Velocity Dispersion Failures in the DR2}
\label{sec:vdispfail}

A list of the AGN for which we were unable to measure velocity dispersions is provided in \autoref{tab:vdispfail}.

\begin{deluxetable*}{lllllll}
\tabletypesize{\scriptsize}
\tablecaption{DR2 Velocity Dispersion Failures \label{tab:vdispfail}}
\tablehead{
\colhead{BAT ID} & \colhead{Counterpart} & \colhead{Reason} &       \colhead{$A_V$} & \colhead{DR2 Type} & \colhead{z}&
\colhead{Best Spectra}\\
\colhead{}& \colhead{}& \colhead{}& \colhead{(mag)}&\colhead{}& \colhead{}& \colhead{}\\
\colhead{(1)}& \colhead{(2)}& \colhead{(3)}& \colhead{(4)}&\colhead{(5)}& \colhead{(6)}& \colhead{(7)}
}
\startdata
5&2MASXJ00040192+7019185&GalExt&2.67&Sy\,1.9&0.0957&Palomar/DBSP\\
18&2MASXJ00331831+6127433&GalExt&3.72&Sy\,1.9&0.1042&Palomar/DBSP\\
119&2MASSJ02162672+5125251&GalExt&0.59&Sy\,2&0.4223&Palomar/DBSP\\
172&2MASXJ03181899+6829322&GalExt&2.29&Sy\,1.9&0.0906&Palomar/DBSP\\
173&NGC1275&Emission&0.5&Sy\,1.9&0.0168&Palomar/DBSP\\
249&LEDA1797736&GalExt&3.5&Sy\,2&0.061&Palomar/DBSP\\
285&2MASXJ05325752+1345092&GalExt&2.51&Sy\,1.9&0.024&VLT/X-Shooter\\
315&IRAS05581+0006&GalExt&2.26&Sy\,1.9&0.1144&Magellan/MagE\\
360&2MASXJ07091407-3601216&GalExt&2.2&Sy\,2&0.1107&VLT/X-Shooter\\
367&1RXSJ072352.4-080623&GalExt&0.93&Sy\,1.9&0.1449&Palomar/DBSP\\
372&1RXSJ072720.8-240629&GalExt&3.1&Sy\,1.9&0.1219&VLT/X-Shooter\\
381&3C184.1&lowSN&0.09&Sy\,1.9&0.1183&Palomar/DBSP\\
433&SWIFTJ085429.35-082428.6&lowSN&0.09&Sy\,2&0.1884&VLT/X-Shooter\\
441&2MASXJ09023729-4813339&GalExt&5.18&Sy\,2&0.0392&VLT/X-Shooter\\
476&CXOJ095220.1-623234&GalExt&0.93&Sy\,1.9&0.2521&VLT/X-Shooter\\
494&SDSSJ102103.08-023642.6&lowSN&0.12&Sy\,2&0.2936&VLT/X-Shooter\\
505&SDSSJ103315.71+525217.8&lowSN&0.06&Sy\,2&0.1404&Palomar/DBSP\\
516&2MASSJ10445192-6025115&GalExt&9.55&Sy\,2&0.047&Nospec\\
533&2MASXJ11140245+2023140&lowSN&0.06&Sy\,2&0.0271&Palomar/DBSP\\
639&2MASXJ12475784-5829599&GalExt&1.86&Sy\,2&0.0276&VLT/X-Shooter\\
661&2MASXJ13103701-5626551&GalExt&1.98&Sy\,2&0.1142&VLT/X-Shooter\\
692&4U1344-60&GalExt&9.02&Sy\,1.9&0.0128&VLT/X-Shooter\\
703&Mrk463E&Emission&0.09&Sy\,1.9&0.0501&Palomar/DBSP\\
745&2MASXJ14545815+8554589&lowSN&0.53&Sy\,2&0.112&Palomar/DBSP\\
747&LEDA3085605&GalExt&2.54&Sy\,2&0.0187&VLT/X-Shooter\\
896&1RXSJ173728.0-290759&GalExt&7.1&Sy\,1.9&0.0218&VLT/X-Shooter\\
929&2MASSJ17485512-3254521&GalExt&5.24&Sy\,1.9&0.0207&Nospec\\
976&CXOJ182557.5-071022&GalExt&6.14&Sy\,2&0.037&VLT/X-Shooter\\
1073&2MASXJ20183871+4041003&GalExt&10.6&Sy\,2&0.0145&Palomar/DBSP\\
1075&1RXSJ202400.8-024527&lowSN&0.22&Sy\,1.9&0.1375&VLT/X-Shooter\\
1096&SWIFTJ210001.06+430209.6&GalExt&4.09&Sy\,2&0.066&Palomar/DBSP\\
1110&4C+50.55&GalExt&7.53&Sy\,1.9&0.0154&Palomar/DBSP\\
1191&2MASXJ23203662+6430452&GalExt&5.15&Sy\,1.9&0.0729&Palomar/DBSP\\
1207&CXOJ235221.9+584531&GalExt&3.97&Sy\,2&0.1629&Palomar/DBSP\\
\enddata
\tablecomments{The columns are as follows. (1) Catalog ID from BAT survey. (2) Host galaxy. (3) Reason for failure.  GalExt=Failure due to being within the Galactic plane and having high optical extinction, lowSN=low S/N typically because at $z>0.1$, Emission=Galaxy spectra was dominated by strong and broad emission lines that contaminated the absorption lines, Nospec=No suitable high-resolution spectra were available due to the very high levels of extinction. (4) Visual extinction due to Milky Way foreground dust using the maps of \citet{Schlegel:1998:525} and the extinction law derived by \citet{Cardelli:1989:245}. (5) AGN type based on optical spectroscopy including Sy\,1.9 (narrow \hbeta\ and broad \halpha) and Sy\,2 (narrow \hbeta\ and \halpha) from \citet{Koss_DR2_catalog}. (6) Redshift based on emission lines from \citet{Koss_DR2_catalog}. (7) Best available spectra in the DR2 based on telescope diameter and spectral resolution.}
\end{deluxetable*}

\bibliography{bibfinal,bib_dr2papers,bib_add_here}{}
\bibliographystyle{aasjournal}

\end{document}

%% file: mycommands.tex
\newcommand{\todo}{\ifmmode \text{\color{red}\Huge{\(\bullet\)}} \else {\color{red}{\Huge$\bullet$}}\fi}
\newcommand{\tido}{\ifmmode {{\color{red}\bullet}} \else {\color{red}$\bullet$}\fi}

\newcommand{\E        }[1]{\ifmmode 10^{#1} \else $10^{#1}$\fi}
\newcommand{\tE        }[1]{\ifmmode \times10^{#1} \else $\times10^{#1}$\fi}
\newcommand{\til}{\ifmmode \sim \else $\sim$\fi}
\renewcommand{\~} {\ifmmode \sim \else $\sim$\fi}

\newcommand{\logNH }{\ifmmode \log (N_{\rm H}/{\rm cm}^{-2}) \else $\log (N_{\rm H}/{\rm cm}^{-2})$\fi}

\newcommand{\Mbh   }{\ifmmode M_{\rm BH} \else $M_{\rm BH}$\fi}

\newcommand{\pc}	{\ifmmode {\rm pc} \else pc\fi}
\newcommand{\ld}	{\ifmmode {\rm l.d.} \else l.d.\fi}
\newcommand{\kms}	{\ifmmode {\rm km\,s}^{-1} \else km\,s$^{-1}$\fi}
\newcommand{\cc}	{\ifmmode {\rm cm}^{-3}    \else cm$^{-3}$\fi}
\newcommand{\cmii}	{\ifmmode {\rm cm}^{-2}    \else cm$^{-2}$\fi}
\newcommand{\ergs}	{\ifmmode {\rm erg\,s}^{-1} \else erg s$^{-1}$\fi}
\newcommand{\ergcms}	{\ifmmode {\rm erg\,cm}^{-2}\,{\rm s}^{-1} \else erg\,cm$^{-2}$\,s$^{-1}$\fi}
\newcommand{\ergcmsA}	{\ifmmode {\rm erg\,cm}^{-2}\,{\rm s}^{-1}\,{\rm\AA}^{-1}
\else erg\,cm$^{-2}$\,s$^{-1}$\,\AA$^{-1}$\fi}
\newcommand{  \ergcmsHz  }{\ifmmode{\rm erg\,cm}^{-2}\,{\rm s}^{-1}\,{\rm Hz}^{-1}
                       \else ergs\,cm$^{-2}$\,s$^{-1}$\,Hz$^{-1}$\fi}
\newcommand{\kev}	{\ifmmode {\rm keV} \else keV\fi}

\newcommand{\mic}	{\ifmmode {\rm \mu m} \else $\mu$m\fi}
\newcommand{\vFWHM}	{\ifmmode v_{\mbox{\tiny FWHM}} \else $v_{\mbox{\tiny FWHM}}$\fi}
\newcommand{\vBLR}	{\ifmmode v_{\mbox{\tiny BLR}} \else $v_{\mbox{\tiny BLR}}$\fi}
\newcommand{\sigBLR}	{\ifmmode \sigma_{\mbox{\tiny BLR}} \else $\sigma_{\mbox{\tiny BLR}}$\fi}
\newcommand{\vNLR}	{\ifmmode v_{\mbox{\tiny NLR}} \else $v_{\mbox{\tiny NLR}}$\fi}
\newcommand{\tauBLR}	{\ifmmode \tau_{\mbox{\tiny BLR}} \else $\tau_{\mbox{\tiny BLR}}$\fi}

\newcommand{\Hubble}	{\ifmmode {\rm km\,s}^{-1}\,{\rm Mpc}^{-1} \else km\,s$^{-1}$\,Mpc$^{-1}$\fi}
\newcommand{\NDunit}	{\ifmmode {\rm Mpc}^{-3} \else Mpc$^{-3}$\fi}
\newcommand{\LFunit}	{\ifmmode {\rm Mpc}^{-3}\,{\rm mag}^{-1} \else Mpc$^{-3}$\,mag$^{-1}$\fi}
\newcommand{\MFunit}	{\ifmmode {\rm Mpc}^{-3}\,{\rm dex}^{-1} \else Mpc$^{-3}$\,dex$^{-1}$\fi}

\newcommand{\Msun}{\ifmmode M_{\odot} \else $M_{\odot}$\fi}
\newcommand{\Lsun}{\ifmmode L_{\odot} \else $L_{\odot}$\fi}
\newcommand{\Zsun}{\ifmmode Z_{\odot} \else $Z_{\odot}$\fi}
\newcommand{\mpyr}{\ifmmode \Msun\,{\rm yr}^{-1} \else $\Msun\,{\rm yr}^{-1}$\fi}

\newcommand{\qnote}{\ifmmode q_{0} \else $q_{0}$\fi}
\newcommand{\Hnote}{\ifmmode H_{0} \else $H_{0}$\fi}
\newcommand{\hnote}{\ifmmode h_{0} \else $h_{0}$\fi}
\newcommand{\anote}{\ifmmode a_{0} \else $a_{0}$\fi}

\newcommand{  \Halpha   }{\ifmmode {\rm H}\alpha \else H$\alpha$\fi}
\newcommand{  \ha   	}{\ifmmode {\rm H}\alpha \else H$\alpha$\fi}
\newcommand{  \Hbeta    }{\ifmmode {\rm H}\beta \else H$\beta$\fi}
\newcommand{  \hb    	}{\ifmmode {\rm H}\beta \else H$\beta$\fi}
\newcommand{  \Hgamma   }{\ifmmode {\rm H}\gamma \else H$\gamma$\fi}
\newcommand{  \Hdelta   }{\ifmmode {\rm H}\delta \else H$\delta$\fi}
\newcommand{  \Lya      }{\ifmmode {\rm Ly}\alpha \else Ly$\alpha$\fi}
\newcommand{  \Lyb      }{\ifmmode {\rm Ly}\beta \else Ly$\beta$\fi}
\newcommand{  \Pa       }{\ifmmode {\rm P}\alpha \else P$\alpha$\fi}
\newcommand{  \Pb       }{\ifmmode {\rm P}\beta \else P$\beta$\fi}
\newcommand{  \Bra      }{\ifmmode {\rm Br}\alpha \else Br$\alpha$\fi}
\newcommand{  \Brg      }{\ifmmode {\rm Br}\gamma \else Br$\gamma$\fi}
\newcommand{  \hii      }{\ifmmode {\rm H}\,\textsc{ii} \else H\,\textsc{ii}\fi}
\newcommand{  \hei      }{\ifmmode {\rm He}\,\textsc{i} \else He\,\textsc{i}\fi}
\newcommand{  \heii     }{\ifmmode {\rm He}\,\textsc{ii} \else He\,\textsc{ii}\fi}
\newcommand{  \HeIIuv   }{\ifmmode {\rm He}\,\textsc{ii}\,\lambda1640 \else He\,\textsc{ii}\,$\lambda1640$\fi}
\newcommand{  \HeIIop   }{\ifmmode {\rm He}\,\textsc{ii}\,\lambda4686 \else He\,\textsc{ii}\,$\lambda4686$\fi}
\newcommand{  \cii      }{\ifmmode {\rm C}\,\textsc{ii}  \else C\,\textsc{ii}\fi}
\newcommand{  \ciii     }{\ifmmode {\rm C}\,\textsc{iii}\right] \else C\,\textsc{iii}]\fi}
\newcommand{  \CIII     }{\ifmmode {\rm C}\,\textsc{iii}\right]\,\lambda1909 \else C\,\textsc{iii}]\,$\lambda1909$\fi}
\newcommand{  \civ      }{\ifmmode {\rm C}\,\textsc{iv}  \else C\,\textsc{iv}\fi}
\newcommand{  \CIV      }{\ifmmode {\rm C}\,\textsc{iv}\,\lambda1549 \else C\,\textsc{iv}\,$\lambda1549$\fi}
\newcommand{  \nii      }{\ifmmode [{\rm N}\,\textsc{ii}]  \else [N\,\textsc{ii}]\fi}
\newcommand{  \niii     }{\ifmmode {\rm N}\,\textsc{iii} \else N\,\textsc{iii}\fi}
\newcommand{  \niv      }{\ifmmode {\rm N}\,\textsc{iv}  \else N\,\textsc{iv}\fi}
\newcommand{  \NIVuv    }{\ifmmode {\rm N}\,\textsc{iv}\,\lambda1486 \else N\,\textsc{iv}\,$\lambda1486$\fi}
\newcommand{  \nv       }{\ifmmode {\rm N}\,\textsc{v}   \else N\,\textsc{v}\fi}
\newcommand{\oi}{\ifmmode \left[{\rm O}\,\textsc{i}\right] \else [O\,{\sc i}]\fi}
\newcommand{\OI}{\ifmmode \left[{\rm O}\,\textsc{i}\right]\,\lambda6300 \else [O\,{\sc i}]$\,\lambda6300$\fi}

\newcommand{\oii}{\ifmmode \left[{\rm O}\,\textsc{ii}\right] \else [O\,{\sc ii}]\fi}
\newcommand{\OII}{\ifmmode \left[{\rm O}\,\textsc{ii}\right]\,\lambda3727 \else [O\,{\sc ii}]\,$\lambda3727$\fi}
\newcommand{\oiii}{\ifmmode \left[{\rm O}\,\textsc{iii}\right] \else [O\,{\sc iii}]\fi}

\newcommand{\neiii}{\ifmmode \left[{\rm Ne}\,\textsc{iii}\right] \else [Ne\,{\sc iii}]\fi}

\newcommand{\OIII}{\ifmmode \left[{\rm O}\,\textsc{iii}\right]\,\lambda5007 \else [O\,{\sc iii}]\,$\lambda5007$\fi}

\newcommand{\NII}{\ifmmode \left[{\rm N}\,\textsc{ii}\right]\,\lambda6583 \else [N\,{\sc ii}]$\,\lambda6583$\fi}

\newcommand{\NeIII}{\ifmmode \left[{\rm Ne}\,\textsc{iii}\right]\,\lambda3968 \else [Ne\,{\sc iii}]$\,\lambda3968$\fi}

\newcommand{\NeV}{\ifmmode \left[{\rm Ne}\,\textsc{v}\right]\,\lambda3426 \else [Ne\,{\sc v}]$\,\lambda3426$\fi}

\newcommand{\HeII}{\ifmmode {\rm He}\,\textsc{ii}\,\lambda4686 \else He\,{\sc ii}$\,\lambda4686$\fi}

\newcommand{\sii}{\ifmmode \left[{\rm S}\,\textsc{ii}\right] \else [S\,{\sc ii}]\fi}

\newcommand{\SII}{\ifmmode \left[{\rm S}\,\textsc{ii}\right]\,\lambda6717,6731 \else [S\,{\sc ii}]$\,\lambda6717,6731$\fi}

\newcommand{  \OIIIuv   }{\ifmmode {\rm O}\,\textsc{iii}\,\lambda1663 \else O\,\textsc{iii}\,$\lambda1663$\fi}
\newcommand{  \oiv      }{\ifmmode {\rm O}\,\textsc{iv}  \else O\,\textsc{iv}\fi}
\newcommand{  \OIVuv    }{\ifmmode {\rm O}\,\textsc{iv}\,\lambda1402  \else O\,\textsc{iv}\,$\lambda1402$\fi}
\newcommand{  \OIVIR    }{\ifmmode {\rm O}\,\textsc{iv}\,25.9\,\mu {\rm m} \else O\,\textsc{iv}\,$25.9\,\mu$m\fi}
\newcommand{  \ovi      }{\ifmmode {\rm O}\,\textsc{vi}   \else O\,\textsc{vi}\fi}
\newcommand{  \Ovi      }{\ifmmode {\rm O}\,\textsc{vi}\,\lambda1035 \else O\,\textsc{vi}\,$\lambda1035$\fi}
\newcommand{  \nei      }{\ifmmode {\rm Ne}\,\textsc{i}   \else Ne\,\textsc{i}\fi}
\newcommand{  \neii     }{\ifmmode {\rm Ne}\,\textsc{ii}  \else Ne\,\textsc{ii}\fi}
\newcommand{  \NeiiIR   }{\ifmmode {\rm Ne}\,\textsc{ii}\,12.8\,\mu {\rm m} \else Ne\,\textsc{ii}\,$12.8\,\mu$m\fi}
\newcommand{  \neiv     }{\ifmmode {\rm Ne}\,\textsc{iv}  \else Ne\,\textsc{iv}\fi}
\newcommand{  \nev      }{\ifmmode {\rm Ne}\,\textsc{v}   \else Ne\,\textsc{v}\fi}
\newcommand{  \NevIR    }{\ifmmode {\rm Ne}\,\textsc{v}\,24.3\,\mu {\rm m} \else Ne\,\textsc{v}\,$24.3\,\mu$m\fi}
\newcommand{  \nevi     }{\ifmmode {\rm Ne}\,\textsc{vi}  \else Ne\,\textsc{vi}\fi}
\newcommand{  \mgi      }{\ifmmode {\rm Mg}\,\textsc{i}   \else Mg\,\textsc{i}\fi}
\newcommand{  \mgii     }{\ifmmode {\rm Mg}\,\textsc{ii}  \else Mg\,\textsc{ii}\fi}
\newcommand{  \MgII     }{\ifmmode {\rm Mg}\,\textsc{ii}\,\lambda2798 \else Mg\,\textsc{ii}\,$\lambda2798$\fi}

\newcommand{  \siii     }{\ifmmode {\rm S}\,\textsc{iii} \else S\,\textsc{iii}\fi}
\newcommand{  \siv      }{\ifmmode {\rm S}\,\textsc{iv}  \else S\,\textsc{iv}\fi}
\newcommand{  \sili     }{\ifmmode {\rm Si}\,\textsc{i}   \else Si\,\textsc{i}\fi}
\newcommand{  \silii    }{\ifmmode {\rm Si}\,\textsc{ii}  \else Si\,\textsc{ii}\fi}
\newcommand{  \Siliv    }{\ifmmode {\rm Si}\,\textsc{iv}  \else Si\,\textsc{iv}\fi}
\newcommand{  \SilIVuv  }{\ifmmode {\rm Si}\,\textsc{iv}\,\lambda1400  \else Si\,\textsc{iv}\,$\lambda1400$\fi}

\newcommand{  \caii     }{\ifmmode {\rm Ca}\,\textsc{ii}   \else Ca\,\textsc{ii}\fi}

\newcommand{  \feii     }{\ifmmode {\rm Fe}\,\textsc{ii}  \else Fe\,\textsc{ii}\fi}
\newcommand{  \feiii    }{\ifmmode {\rm Fe}\,\textsc{iii} \else Fe\,\textsc{iii}\fi}

\newcommand{ \Lhb   }{\ifmmode L\left(\hb\right) \else $L\left(\hb\right)$\fi}
\newcommand{ \fwhb  }{\ifmmode {\rm FWHM}\left(\hb\right) \else FWHM(\hb)\fi}
\newcommand{ \Lha   }{\ifmmode L\left(\ha\right) \else $L\left(\ha\right)$\fi}
\newcommand{ \fwha  }{\ifmmode {\rm FWHM}\left(\ha\right) \else FWHM(\ha)\fi}
\newcommand{ \Lmg   }{\ifmmode L\left(\mgii\right) \else $L\left(\mgii\right)$\fi}
\newcommand{ \fwmg  }{\ifmmode {\rm FWHM}\left(\mgii\right) \else FWHM(\mgii)\fi}
\newcommand{ \Lciv  }{\ifmmode L\left(\civ\right) \else $L\left(\civ\right)$\fi}
\newcommand{ \fwciv }{\ifmmode {\rm FWHM}\left(\civ\right) \else FWHM(\civ)\fi}
\newcommand{ \fwhm  }{\ifmmode {\rm FWHM} \else FWHM\fi} 
\newcommand{ \voff  }{\ifmmode v_{\rm off} \else $v_{\rm off}$\fi} 

\newcommand{ \mumg  }{\ifmmode \mu\left(\mgii\right) \else $\mu\left(\mgii\right)$\fi}
\newcommand{ \fmg   }{\ifmmode f\left(\mgii\right) \else $f\left(\mgii\right)$\fi}
\newcommand{ \muciv }{\ifmmode \mu\left(\civ\right) \else $\mu\left(\civ\right)$\fi}
\newcommand{ \fciv  }{\ifmmode f\left(\civ\right) \else $f\left(\civ\right)$\fi}

\newcommand{  \auvo     }{\ifmmode \alpha_{\nu,{\rm UVO}} \else $\alpha_{\nu,{\rm UVO}}$\fi}
\newcommand{  \Ledd     }{\ifmmode L_{\rm Edd} \else $L_{\rm Edd}$\fi}
\newcommand{  \lamLlam  }{\ifmmode \lambda L_{\lambda} \else $\lambda L_{\lambda}$\fi}
\newcommand{  \lLl      }{\ifmmode \lambda L_{\lambda} \else $\lambda L_{\lambda}$\fi}
\newcommand{  \nuLnu    }{\ifmmode \nu L_{\nu} \else $\nu L_{\nu}$\fi}
\newcommand{  \nLn      }{\ifmmode \nu L_{\nu} \else $\nu L_{\nu}$\fi}
\newcommand{  \Luv      }{\ifmmode L_{1450} \else $L_{1450}$\fi}
\newcommand{  \Lop      }{\ifmmode L_{5100} \else $L_{5100}$\fi}
\newcommand{  \lLop     }{\ifmmode \log\left(\Lop/\ergs\right) \else $\log\left(\Lop/\ergs\right)$\fi}
\newcommand{  \Lthree   }{\ifmmode L_{3000} \else $L_{3000}$\fi}
\newcommand{  \lLthree  }{\ifmmode \log\left(\Lthree/\ergs\right) \else $\log\left(\Lthree/\ergs\right)$\fi}

\newcommand{\Fthree}{\ifmmode F_{3000} \else $F_{3000}$\fi}
\newcommand{\fuv}{\ifmmode f_{\lambda}\left(1450{\rm \AA}\right) \else $f_{\lambda}\left(1450 {\rm \AA}\right)$\fi}
\newcommand{\fthree}{\ifmmode f_{\lambda}\left(3000{\rm \AA}\right) \else $f_{\lambda}\left(3000{\rm \AA}\right)$\fi}
\newcommand{\fH}{\ifmmode f_{\lambda}\left(1.65\micron\right) \else
$f_{\lambda}\left(1.65\micron\right)$\fi}

\newcommand{\fbol}{\ifmmode f_{\rm bol} \else $f_{\rm bol}$\fi}
\newcommand{\fbolwv}{\ifmmode f_{\rm bol}\left(\lambda\right) \else $f_{\rm bol}\left(\lambda\right)$\fi}
\newcommand{\fbolopt}{\ifmmode f_{\rm bol}\left(5100{\rm \AA}\right) \else $f_{\rm bol}\left(5100{\rm \AA}\right)$\fi}
\newcommand{\fbolthree}{\ifmmode f_{\rm bol}\left(3000{\rm \AA}\right) \else $f_{\rm bol}\left(3000{\rm \AA}\right)$\fi}
\newcommand{\fboluv}{\ifmmode f_{\rm bol}\left(1450{\rm \AA}\right) \else $f_{\rm bol}\left(1450{\rm \AA}\right)$\fi}

\newcommand{  \mbh      }{\ifmmode M_{\rm BH} \else $M_{\rm BH}$\fi}
\newcommand{  \lmbh     }{\ifmmode \log\left(\mbh/\Msun\right) \else $\log\left(\mbh/\Msun\right)$\fi} 
\newcommand{  \lledd    }{\ifmmode L/L_{\rm Edd} \else $L/L_{\rm Edd}$\fi}
\newcommand{  \logledd    }{\ifmmode \log\left(L/L_{\rm Edd}\right) \else $\log\left(L/L_{\rm Edd}\right)$\fi}
\newcommand{  \Lbol     }{\ifmmode L_{\rm bol} \else $L_{\rm bol}$\fi}
\newcommand{  \lbol     }{\ifmmode L_{\rm bol} \else $L_{\rm bol}$\fi}
\newcommand{  \lLbol    }{\ifmmode \log\left(\Lbol/\ergs\right) \else $\log\left(\Lbol/\ergs\right)$\fi} 
\newcommand{  \Lagn     }{\ifmmode L_{\rm AGN} \else $L_{\rm AGN}$\fi}
\newcommand{  \lagn     }{\ifmmode L_{\rm AGN} \else $L_{\rm AGN}$\fi}

\newcommand{  \tgrow     }{\ifmmode t_{\rm growth} \else $t_{\rm growth}$\fi}
\newcommand{  \tUni      }{\ifmmode t_{\rm Universe} \else $t_{\rm Universe}$\fi}

\newcommand{  \Mindot	}{\ifmmode \dot{M}_{\rm infall} \else $\dot{M}_{\rm infall}$\fi}
\newcommand{  \Mbhdot	}{\ifmmode \dot{M}_{\rm BH} \else $\dot{M}_{\rm BH}$\fi}
\newcommand{  \Maddot	}{\ifmmode \dot{M}_{\rm AD} \else $\dot{M}_{\rm AD}$\fi}

\newcommand{  \as	}{\ifmmode a_{\rm *} 		\else $a_{\rm *}$\fi}
\newcommand{  \avec	}{\ifmmode \vec{a}_{\rm *} 	\else $\vec{a}_{\rm *}$\fi}
\newcommand{  \re	}{\ifmmode \eta      	\else $\eta$\fi}

\newcommand{  \mseed    }{\ifmmode M_{\rm seed} \else $M_{\rm seed}$\fi}
\newcommand{  \mbul     }{\ifmmode M_{\rm Bulge} \else $M_{\rm Bulge}$\fi} 
\newcommand{  \mstar    }{\ifmmode M_{*} \else $M_{*}$\fi} 
\newcommand{  \mgal     }{\ifmmode M_{*} \else $M_{*}$\fi} 
\newcommand{  \mhost    }{\ifmmode M_{\rm Host} \else $M_{\rm Host}$\fi}
\newcommand{  \mm       }{\ifmmode M_{*}/M_{\rm BH} \else $M_{*}/M_{\rm BH}$\fi}
\newcommand{  \mmsmall  }{\ifmmode M_{\rm BH}/M_{*} \else $M_{\rm BH}/M_{*}$\fi}
\newcommand{  \mmlarge  }{\ifmmode M_{*}/M_{\rm BH} \else $M_{*}/M_{\rm BH}$\fi}
\newcommand{  \mmwp     }{\ifmmode \left(M_{*}/M_{\rm BH}\right) \else $\left(M_{*}/M_{\rm BH}\right)$\fi}
\newcommand{  \ml       }{\ifmmode M_{*}/L_{*} \else $M_{*}/L_{*}$\fi}
\newcommand{  \mlwp     }{\ifmmode \left(M_{*}/L\right) \else $\left(M_{*}/L\right)$\fi}
\newcommand{  \mlk      }{\ifmmode \left(M_{*}/L_{K}\right) \else $\left(M_{*}/L_{K}\right)$\fi}
\newcommand{  \sigs     }{\ifmmode \sigma_{*} \else $\sigma_{*}$\fi}
\newcommand{  \Reff     }{\ifmmode R_{\rm e} \else $R_{\rm e}$\fi}

\def\kmps{\hbox{$\km\s^{-1}\,$}}

\newcommand{\bj}{\ifmmode b_{\rm J} \else $b_{\rm J}$\fi}

\newcommand{\iab}{\ifmmode i_{\rm AB} \else $i_{\rm AB}$\fi}

\newcommand{\jab}{\ifmmode J_{\rm AB} \else $J_{\rm AB}$\fi}
\newcommand{\hab}{\ifmmode H_{\rm AB} \else $H_{\rm AB}$\fi}
\newcommand{\kab}{\ifmmode K_{\rm AB} \else $K_{\rm AB}$\fi}

\newcommand{\jveg}{\ifmmode J_{\rm Vega} \else $J_{\rm Vega}$\fi}
\newcommand{\hveg}{\ifmmode H_{\rm Vega} \else $H_{\rm Vega}$\fi}
\newcommand{\kveg}{\ifmmode K_{\rm Vega} \else $K_{\rm Vega}$\fi}

\def\arcsec{\hbox{$^{\prime\prime}$}}

\newcommand{  \Chisq    }{\ifmmode \chi^{2} \else $\chi^{2}$}
\newcommand{  \nelec    }{\ifmmode n_{e} \else $n_{e}$\fi}     
\newcommand{  \nh       }{\ifmmode n_{H} \else $n_{H}$\fi}     
\newcommand{  \Ncol     }{\ifmmode N_{col} \else $N_{col}$\fi} 
\newcommand{  \NH       }{\ifmmode N_{H} \else $N_{\rm H}$\fi}     

\def\sun{\hbox{$\odot$}}

\def\arcsec{\hbox{$^{\prime\prime}$}}

\def\ion#1#2{#1$\;${\small\rm\@Roman{#2}}\relax}

\newcommand{\OIIIa}{\ifmmode \left[{\rm O}\,\textsc{iii}\right]\,\lambda4959 \else [O\,{\sc iii}]\,$\lambda4959$\fi}
\newcommand{\NIIa}{\ifmmode \left[{\rm N}\,\textsc{ii}\right]\,\lambda6548 \else [N\,{\sc ii}]\,$\lambda6548$\fi}
\newcommand{\SIIa}{\ifmmode \left[{\rm S}\,\textsc{ii}\right]\,\lambda6716 \else [S\,{\sc ii}]\,$\lambda6716$\fi}
\newcommand{\SIIb}{\ifmmode \left[{\rm S}\,\textsc{ii}\right]\,\lambda6732 \else [S\,{\sc ii}]\,$\lambda6731$\fi}
\newcommand{\NeVa}{\ifmmode \left[{\rm Ne}\,\textsc{v}\right]\,\lambda3346 \else [Ne\,{\sc v}]\,$\lambda3346$\fi}
\newcommand{\NeVb}{\ifmmode \left[{\rm Ne}\,\textsc{v}\right]\,\lambda3426 \else [Ne\,{\sc v}]\,$\lambda3426$\fi}
\newcommand{\NeIIIa}{\ifmmode \left[{\rm Ne}\,\textsc{iii}\right]\,\lambda3869 \else [Ne\,{\sc iii}]\,$\lambda3869$\fi}
\newcommand{\NeIIIb}{\ifmmode \left[{\rm Ne}\,\textsc{iii}\right]\,\lambda3968 \else [Ne\,{\sc iii}]\,$\lambda3968$\fi}

\newcommand{\mgb}{\ifmmode \left{\rm Mg\,\textsc{I}}\right\,$b$\,\lambda5183, 5172, 5167 \else Mg\,{\sc I} $b$$\,\lambda5183, 5172, 5167$\fi}

\newcommand{\Cahk}{\ifmmode \left[{\rm Ca\,\textsc{ii} H and K}\right\,\lambda3969,3934 \else Ca\,\textsc{ii} H and K$\,\lambda3969,3934$\fi}

\newcommand{\Catrip}{\ifmmode \left[{\rm Ca}\,\textsc{ii}\right\,\lambda8498, 8542, 8662 \else Ca\,\textsc{ii} $\,\lambda8498, 8542, 8662$\fi}

\def\kmpssh{\hbox{$\km\s^{-1}$}}

\def\arcsec{{\mbox{$^{\prime \prime}$}}}

\def\erg{{\rm\thinspace erg}}

\def\g{{\rm\thinspace g}}

\def\K{{\rm\thinspace K}}

\def\km{{\rm\thinspace km}}

\def\Lsun{\hbox{$\rm\thinspace L_{\odot}$}}
\def\m{{\rm\thinspace m}}

\def\pc{{\rm\thinspace pc}}

\def\s{{\rm\thinspace s}}

\newcommand{\halpha}{\Halpha}

\newcommand{\hbeta}{\Hbeta}

\newcommand{\HeIIir}{\ifmmode {\rm He}\,\textsc{ii}\,\lambda8237 \else He\,{\sc ii}$\,\lambda8237$\fi}
\newcommand{\HeIir}{\ifmmode {\rm He}\,\textsc{i}\,\lambda10830 \else He\,{\sc i}$\,\lambda10830$\fi}

\newcommand{\SIII}{\ifmmode \left[{\rm S}\,\textsc{iii}\right]\,\lambda9531 \else [S\,\textsc{ii}]\,$\lambda9531$\fi}

\newcommand {\Lsoftint} {\ifmmode L^{\rm in}_{\mathrm{2-10\ keV}} \else $L^{\rm in}_{\mathrm{2-10\ keV}}$\fi}

\newcommand {\ergpersec} {\ifmmode {\rm erg~s}^{-1} \else erg~s$^{-1}$ \fi}

\newcommand {\nhunit} {cm$^{-2}$\xspace}

\def\micron{{\mbox{$\mu{\rm m}$}}}
\def\arcsec{{\mbox{$^{\prime \prime}$}}}

\def\arcsec{{\mbox{$^{\prime \prime}$}}}

\def\erg{{\rm\thinspace erg}}

\def\g{{\rm\thinspace g}}

\def\K{{\rm\thinspace K}}

\def\km{{\rm\thinspace km}}

\def\Lsun{\hbox{$\rm\thinspace L_{\odot}$}}
\def\m{{\rm\thinspace m}}

\def\pc{{\rm\thinspace pc}}

\def\s{{\rm\thinspace s}}

\def\ergps{\hbox{$\erg\s^{-1}\,$}}

\def\kmps{\hbox{$\km\s^{-1}\,$}}

\def\apjl{\hbox{ApJL}}
\def\apjs{\hbox{ApJS}}

\def\aap{\hbox{AAP}}
\def\micron{{\mbox{$\mu{\rm m}$}}}
\def\arcsec{{\mbox{$^{\prime \prime}$}}}

\newcommand {\iraf}{{\sc IRAF}\xspace}
\newcommand {\molecfit}{\texttt{molecfit}}
\newcommand {\ppxf}{\texttt{pPXF}}

\newcommand{\nuvr}{\ifmmode {\rm NUV}-r \else NUV-$r$\fi}
\newcommand{\mh}{\ifmmode M_{\rm H_2} \else $M_{\rm H_2}$\fi}
\newcommand{\mhi}{\ifmmode M_{\rm HI} \else $M_{\rm HI}$\fi}

\newcommand{\must}{\ifmmode \mu_{\ast} \else $\mu_{\ast}$\fi}
\newcommand{\hmol}{\ifmmode H_2 \else $H_2$\fi}
\newcommand{\rmol}{\ifmmode R_{\rm mol} \else $R_{\rm mol}$\fi}
\newcommand{\tdep}{\ifmmode t_{\rm dep}({\rm H_2}) \else $t_{\rm dep}({\rm H_2})$\fi}
\newcommand{\tdepHI}{\ifmmode t_{\rm dep}({\rm HI}) \else $t_{\rm dep}({\rm HI})$\fi}
\newcommand{\fgas}{\ifmmode f_{\rm H_2} \else $f_{\rm H_2}$\fi}
\newcommand{\fhi}{\ifmmode f_{\rm HI} \else $f_{\rm HI}$\fi}
\newcommand{\xco}{\ifmmode \alpha_{\rm CO} \else $\alpha_{\rm CO}$\fi}

\newcommand{\SiX}{\ifmmode \left[{\rm Si}\,\textsc{x}\right]\,\lambda14300 \else [Si\,{\sc x}]\,$\lambda14300$\fi}
\newcommand{\SiVI}{\ifmmode \left[{\rm Si}\,\textsc{vi}\right]\,\lambda19640 \else [Si\,{\sc vi}]\,$\lambda19640$\fi}
\newcommand{\SXI}{\ifmmode \left[{\rm S}\,\textsc{xi}\right]\,\lambda19196 \else [S\,{\sc xi}]\,$\lambda19196$\fi}
\newcommand{\SVIII}{\ifmmode \left[{\rm S}\,\textsc{viii}\right]\,\lambda9915 \else [S\,{\sc viii}]\,$\lambda9915$\fi}
\newcommand{\SIX}{\ifmmode \left[{\rm S}\,\textsc{ix}\right]\,\lambda12520 \else [S\,{\sc ix}]\,$\lambda12520$\fi}
\newcommand{\FeXIII}{\ifmmode \left[{\rm Fe}\,\textsc{xiii}\right]\,\lambda10747 \else [Fe\,{\sc xiii}]\,$\lambda10747$\fi}
\newcommand{\SiXI}{\ifmmode \left[{\rm Si}\,\textsc{xi}\right]\,\lambda19320 \else [Si\,{\sc xi}]\,$\lambda19320$\fi}

\def\arcsec{\hbox{$^{\prime\prime}$}}

\def\nh{N_\textrm{H}}

\def\ergcms{erg~cm$^{-2}$~s$^{-1}$}
\def\ergs{erg~s$^{-1}$}

\def\oiii{[O\hspace*{1mm}\textsc{iii}]}
\def\nii{[N\hspace{1mm}\textsc{ii}]}
\def\sii{[S\hspace{1mm}\textsc{ii}]}
\def\hei{He\hspace{1mm}\textsc{i}}
\def\caii{Ca\hspace{1mm}\textsc{ii}}
